# On the probability distribution of long-term changes in the growth rate of the global economy: An outside view

David Roodman[1]

June 15, 2020

**Abstract**: Daniel Kahneman and Amos Tversky argued for challenging *inside* views (informed by contextual specifics) with *outside* views (based on historical "base rates" for certain event types). A reasonable inside view of the prospects for the global economy in this century is that growth will converge to 2.5%/year or less: population growth is expected to slow or halt by 2100; and as more countries approach the technological frontier, economic growth should slow as well. To test that view, this paper models gross world product (GWP) observed since 10,000 BCE or earlier, in order to estimate a base distribution for changes in the growth rate as a function of the GWP level. For econometric rigor, it casts a GWP series as a sample path in a stochastic diffusion whose specification is novel yet rooted in neoclassical growth theory. After estimation, most observations fall between the 40th and 60th percentiles of predicted distributions. The fit implies that GWP explosion is all but inevitable, in a median year of 2047. The friction between inside and outside views highlights two insights. First, accelerating growth is more easily explained by theory than is constant growth. Second, the world system may be less stable than traditional growth theory and the growth record of the last two centuries suggest.

[1] Open Philanthropy, 182 Howard St., #225, San Francisco, CA, 94105, david.roodman@openphilanthropy.org. ORCID 0000-0001-6020-1347. This paper has benefited immensely from comments on earlier versions from my Open Philanthropy colleagues, including Holden Karnofsky, Carl Shulman, and Tom Davidson; and from Benjamin Garfinkel and Charles Jones. The author alone is responsible for the views expressed herein.

The intense vibration that rattles our epoch is nothing more, and nothing less, than the crest of a wave issuing from the origin of life, which, moving at first with infinite slowness, now swells and rears. It is thus the whole evolutionary dynamic that weighs on the moment now reached in the adventure of living things. Our situation is unique in the annals of life, yet inscribed for all time in the logic of history.—François Meyer (1974, p. 101)[1]

The analyst should therefore make every effort to frame the forecasting problem so as to facilitate the utilization of all the distributional information that is available.—Daniel Kahneman and Amos Tversky (1979, p. 316).

The steady state is not a bad place for the theory of growth to start, but may be a dangerous place for it to end.—Robert Solow (2000, p. 7)

# Introduction

At least twice in the last 10,000 years, the growth of the human economic system has accelerated markedly—developments that beforehand were hardly foreseen and afterward were called the agricultural and industrial revolutions. If the growth rate is so changeable over the very long term, the question arises as to whether growth might accelerate again, or even decelerate markedly, or reverse. An *inside* view of the question, one built on the specifics of the case (Kahneman and Lovallo 1973, p. 25), is that another step change is not in the cards. Even under the high variant of the population projection of the United Nations, global population growth will slow from the present 1% per annum to 0.7% by 2100 (UN 2019). Meanwhile, in leading Western economies, gross domestic product (GDP) per capita has grown with remarkable steadiness for a long time: in the U.S., per-capita growth has averaged 1.7%/year since 1820, a record that could reasonably expected to continue. Other countries have grown faster, but such rapid expansion looks like transitional catch-up growth; it proved so in Japan. These observations combine to suggest that that the global economy will converge in this century toward an aggregate growth rate of 2.5%/year or less (e.g., Lucas 2000).

Yet the human system could diverge from its present momentum for reasons undreamt of in the insider's philosophy. The most likely causes of large *negative* shocks are anthropogenic, for they could play out on a human rather than geological time scale: climate change, nuclear winter, a bioengineered pandemic (Ord 2020). Another wild card is the progress of artificial intelligence. Within this century, humans may do for the mind what they have already done for muscles and bones by inventing mechanical propulsion. We may devise machines that, while resembling the human brain no more than a 747 resembles a seagull, will surpass the brain in formulating and executing effective plans in complex situations. Like other epoch-making breakthroughs, this one could bring major harm, but also major benefits. It might permanently alter the growth rate of the human system (Hanson 2001).

To shed some additional light on the probability of such divergence, this paper adopts a radically *outside* view, seeking a parsimonious and statistically coherent mathematical model that captures important aspects of growth history as embodied in a very long-term series on gross world product (GWP), in order to estimate historical base rates or base distributions for growth changes. In research on expert judgment, Kahneman and

[1] "L'intense vibration qui secoue notre époque n'est rien de plus, mais rien de moins, que le sommet d'une vague issue des origines et qui, traînant d'abord infiniment sa lenteur, s'enfle et se cabre. C'est ainsi toute la dynamique évolutive qui pèse sur le point aujourd'hui atteint par l'aventure des vivants. Cette situation qui est la nôtre est à la fois unique dans les annales de la vie, et cependant inscrite depuis toujours dans la logique de son histoire."

Tversky (1979) found that supplementing inside views with outside ones improves accuracy. The insight has spread to project management (Flyvbjerg 2006) and current affairs prediction (Tetlock and Gardner 2015). The approach embodies the Bayesian logic that posterior distributions are most accurate when incorporating all available information. But the blending of inside and outside views need not occur formally; often it is the informal friction between multiple perspectives that best serves the forecaster and decisionmaker. Thus to investigate an outside view of the human trajectory is not to presume that it should dominate other views, only that there is a chance of increasing insight into a topic that could hardly be weightier.

As we look back in time, we quickly lose certainty about how many people lived and how well they lived. Nevertheless, GWP has been estimated back to 1 CE (Maddison 2001, 2003) and even to 1 million BCE (De Long 1998). The estimates contain information. One fact that leaps from the data is that over the very long term, the human population and economy have expanded superexponentially. The growth rate has grown. GWP doubled between 2000 and 2019; humanity's earliest doublings perhaps took millennia. Von Foerster, Mora, and Amiot (1960) first noted that the differential equation $y = sy^{1+B}$, with $B > 0$, conforms remarkably well to very-long-term series quantifying the scale of the human system (in their case, population). We can rewrite that as $\dot{\ln y} = sy^B$, in which $\dot{\ln y}$ is the growth rate. This functional form succinctly posits an endogenous scale effect in the human enterprise, an elasticity of growth to level.

As is well known, when projected forward, the superexponential equation sends $y$ to infinity in finite time. That is why Von Foerster, Mora, and Amiot predicted a population "doomsday" for Friday the 13th of November, 2026. Figure 1 illustrates the paradox with respect to GWP. In the graph, both axes are logarithmic, with time measured in years till 2050. On these scales, a line corresponds to a solution of the superexponential equation. Here, a line fits GWP well enough that the agricultural and industrial revolutions, the most profound economic events since language evoved, shrink to gentle undulations. Yet if the line is extrapolated forward, it never reaches 2050, even as it keeps rising. Projected GWP explodes by 2050.

That outcome seems crazy. But it does not follow that the quality of the fit from that simple model and the accompanying projection contain no information relevant to the human prospect. This paper therefore works on theoretical and statistical foundations for exploring that outside view and quantifying its limitations. Starting with theory, the paper makes the point, not novel, that if asymmetries are removed from the neoclassical macroeconomic model, so that all factors are endogenous and none grows at a rate fixed outside the model, then steady, positive output growth essentially never occurs. In effect, steady growth only emerges from a neoclassical-type model when assumed into it.

The paper's main technical novelty is designed to improve statistical rigor in fitting a nonlinear dynamic stochastic model to data. A univariate stochastic model is introduced that is mathematical kin with the neoclassical economic model. The statistical version integrates deterministic components of growth—production, reinvestment, depreciation—with stochasticity. For intuition, think of the informal equation $\dot{\ln y} = sy^B + \delta + \epsilon$. The term $\delta$ embodies exogenous depreciation and appreciation. $\epsilon$ represents a random shock in each infinitesimal time step. When properly epxressed in the stochastic calculus, this model casts the observed GWP series as a sample path in a *diffusion*. One can imagine alternative histories under the same probability law, in which, say, an epidemic or climate shock delays the invention of the wheel by a thousand years. Such possibilities shape the probability distribution for GWP at any given time. Typically they fatten the upper tail.

This diffusion model is close to and inspired by models in finance, the branch of economics that has most exploited the stochastic calculus.[2] As a tool for representing growth history, the model possesses several virtues. Conditional on a starting value, the probability distribution for $y$ at any time can be expressed analytically, which facilitates maximum-likelihood fitting. Because the error process is rigorously defined, not adduced *ad hoc* as in nonlinear least squares, it grounds inference in a coherent data generating process. For example, using GWP figures back to 10,000 BCE, the preferred estimate of the scale effect $B$ developed here is 0.55, with a standard error of 0.05.

Another attraction of the stochastic model is that it can soften the paradox of infinity in superexponential growth, by casting explosion as possible but not inevitable. But that essentially does not happen here. Conditioning on the 2019 GWP value, the preferred estimate puts the probability of *no* eventual explosion at $\sim 10^{-69}$. The median predicted explosion year is 2047. This projection not only collides with the laws of physics; it also lies in tension with the relative *constancy* of per-capita growth over the merely long term. (See Figure 2 on U.S. growth.) Perhaps this is an example of an outside view constructively challenging an inside view, for arguably the puzzle lies in the second half of that conjunction. Endogenous growth theory easily explains superexponential growth over the very long term, through the nonrivalry of innovation (Romer 1990; Kremer 1993). Increases in all inputs, including technology, bring more than proportional increase in output. There is no theory as straightforward for why frontier per-capita growth has been nearly constant in the last 200 years. For, as noted, constant growth only emerges from neoclassical-style models when injected into them by assumption. Jones (2003) sensibly proposes population growth as the most plausible ultimate source of constant exponential growth, with its rooting in biology. Yet in the U.S., long a nation near the economic frontier, population growth has actually declined fourfold, from 2.9% per annum in the 1820s to 0.7% in the 2010s (Bolt et al. 2018; UN 2019). Without so natural a theory for steady growth, it is hard to completely rule out that the steady growth is a temporary, century-scale deviation from superexponential growth or a transition to subexponential growth.

The notion that the growth of the human system has been accelerating may also seem to clash with a more growth-pessimistic line of evidence—evidence that returns to R&D investment have been falling over the last century. In the phrasing of Bloom et al. (2020), ideas are getting harder to find. But this contradiction is somewhat more apparent than real. If conventional inputs to production such as capital and labor together enjoy constant returns to scale, then *all* inputs, including technology, enjoy increasing returns. In the model developed here, under at least superficially plausible parameter choices, the overall increasing returns to scale generate superexponential growth even as returns to investment in total factor productivity (TFP) fall.

If the superexponential prediction of infinity is not easily refuted by a compelling theory for steady growth in the industrial era, nor by signs of declining returns to R&D, what should we make of it? One reading is that the extrapolation is *directional*: output will not go infinite, but may yet greatly increase, the most plausible cause being AI. Or the human system could implode, if escalating economic activity undermines an essential input such as natural resources. A reading that embraces both possibilities is that over the long term the world system is less stable than traditional growth theory and the last two centuries of growth suggest.

[2] But see Nuño and Moll (2018) for an application to macroeconomics.

Section 1 of this paper reviews previous work and elaborates on the motivation for diffusion modeling in econometrics. Then, to bridge from familiar theory to the econometrics, section 2 develops a deterministic, multi-factor model of growth with Cobb-Douglass production like those in Lee (1988) and Kremer (1993). The section next observes that when explosion or implosion occurs, it does so simultaneously in all factors, so that the system comes to be well approximated by a collection of univariate differential equations. But in the multivariate model, the timing of this denouement is in general intractable. Section 3 proposes representing the deterministic but intractable behavior of this multivariate system with a random, univariate, and more tractable stochastic differential equation. The exposition, mostly deferred to an appendix, fills some gaps in existing presentations of the Feller (1951b)/Cox-Ingersoll-Ross (1985; CIR) diffusion. Section 4 constructs data series for population and gross world product since 1 million BCE, as well as for gross domestic product (GDP)/capita in France, as a proxy for productivity at the economic frontier. Section 5 fits the stochastic model to these series and checks for robustness and goodness of fit. Section 6 works to extract meaning from the outsider model's projection of infinity. Section 7 concludes.

## 1 Previous work

The literature modeling human development over the very long run is rather short. Meyer (1947) is perhaps the first to identify a "loi d'accélération evolutive" in natural and human history; Meyer's data from human history consist of four dates seen as marking developmental upswings, which themselves arrived *accelerando*: 4500 BCE, 550 BCE, 1100 CE, and 1750 CE. Meyer explains the acceleration with a pseudoscientific riff on Heisenberg's uncertainty principle.

In the fall of 1960, two more firmly grounded articles on the history of human population appeared. In *Scientific American*, Deevey (1960) presents a coarse human population series back to 1 million BCE. Deevey depicts the series on log-log scales as in Figure 1, with a rising series of waves for the toolmaking, agricultural, and industrial revolutions. The overall linearity on these scales implies a race to infinity somewhere around the present.

In *Science*, Von Foerster, Mora, and Amiot (1960) more fully surfaces the paradox of infinite extrapolation, providing both a microtheory for and evidence of superexponential growth in human population. The paper begins by considering the differential equation for exponential growth and decay:

$$\dot{y} = sy. \tag{1}$$

In human populations, the growth rate $s$ is determined by the balance between natality and mortality. Von Foerster, Mora, and Amiot observe that in exponential growth, the rate of expansion is not an emergent property, although they do not use that term. That is, if two subpopulations grow at rate $s$ then their union does too. As a consequence, a microtheory for exponential growth of a system need not posit interactions among constituents. But a realistic microtheory of growth of the human system *does* posit interactions. These may be inhibitory, as when individuals compete for limited resources, or synergistic, as when innovations are copied. Then, scale matters for it affects the quantity of such interactions.

Von Foerster, Mora, and Amiot therefore generalize (1) to what I write as

$$\dot{y} = sy^{1+B}. \tag{2}$$

When $B \neq 0$, the particular solution is

$$y = (y_0^{-B} - sBt)^{-1/B}, \tag{3}$$

in which $t$ is time and $y_0$ is the initial value. If $B > 0$, $y$ goes to infinity at time $t_c = y_0^{-B}/sB$. This tendency becomes clearer if we rewrite (3) as

$$y = \left(sB(t_c - t)\right)^{-1/B}. \tag{4}$$

Von Foerster, Mora, and Amiot fit (4) to a 2,000-year-long world population series using least squares—though how this nonlinear estimation was carried out in the age of the slide rule is unclear. Most likely $t_c$ was first estimated somewhat informally, at Friday the 13th of November 2026. Holding $t_c$ fixed, (4) is log-linear in $t_c - t$, and can be fit with ordinary least squares in logarithms.

More recently, Kapitza (1996), Varfolomeyev and Gurevich (2001), Korotayev (2007), Johansen and Sornette (2001), and Dolgonosov (2016) fit versions of (4) to very long-term series for population or GWP. The last two take data from De Long (1998), which is the first paper to venture a GWP series covering a million years. Typically in these papers, the differential equation is solved, and the solution fit to GWP as a function of time. The methods are not precisely described and evidently do not produce standard errors.

Also since 1960, authors have theorized mechanisms that accelerate population and economic growth. Kuznets (1960, pp. 328–29) points out that rising population increases the absolute number of "geniuses" whose discoveries can benefit all people. Arrow (1962) models learning by doing, in which accumulated gross investment drives labor productivity. Boserup (1965) focuses on technological change in agriculture induced by increasing population density. In the initial wave of research on endogenous growth, Romer (1986, 1990), Grossman and Helpman (1991), and Aghion and Howitt (1992) insert parallel ideas into the neoclassical tradition.

A related literature models the major economic transitions in history. Most of the papers seek to reproduce the industrial revolution with a structure featuring two production regimes, such as agriculture and manufacturing or research and final goods. Optimizing agents allocate a resource across these regimes, or else to investments whose productivity differs between the regimes, such as education of children. The allocations are influenced by, and sometimes influence, an aggregate trend such as productivity growth in manufacturing.[3] Jones (2001) goes farthest in calibrating such a model to historical data, fitting a system with endogenous technology and fertility to population and GWP/capita starting in 25,000 BCE. The "unified growth theory" of Galor and Weil (2000) is perhaps the most comprehensive; its interplay between fertility, education, technology, and land can explain the Malthusian era, the industrial revolution, and the demographic transition. In contrast with these two-regime models, Becker, Murphy and Tamura (1990) and Acemoglu and Zilibotti (1997) are more unitary, the first emphasizing the role of human capital accumulation, the second the difficulty of diversifying away investment risk in a pre-industrial economy. Notably, the Becker, Murphy and Tamura system possesses low- and high-income equilibria, which assigns "luck" a role in the path of history. Lagerlöf (2003a, b) contemplates stochasticity in the form of mortality shocks termed "epidemics."

[3] Goodfriend and McDermott (1995), Galor and Weil (2000), Laitner (2000), Fernández-Villaverde (2001), Jones (2001), Kögel and Prskawetz (2001), Galor and Moav (2002), Hansen and Prescott (2002), Hazan and Berdugo (2002), Tamura (2002), Lagerlöf (2003a, b), Doepke (2004).

*Sui generis* in the literature is Hanson (2000), which models GWP over 2 million years as the sum of exponential growth terms with different time constants. The terms help the model match the waves for the toolmaking, agricultural, and industrial revolutions in Deevey's plot. While the sum-of-exponentials state equation can be derived from a particular differential equation of motion—a first-order, multivariate, linear homogenous system—Hanson (2000) does not invoke this theoretical motivation. The economic model is that the exponential growth dynamics leading to all the major economic revolutions were present in our ape-like ancestors 2 million years ago but, because of different starting levels and growth constants, took different amounts of time to burst forth to measurable levels. After identifying the parameters shaping each of these components, Hanson (2000) studies their statistics, notes the tendency for acceleration, and projects the timing and magnitude of the *next* exponential growth mode. Hanson suggests that its doubling time will be measured in days. If this reasoning were pursued further, to a cascade of revolutions, a singularity would occur. In this sense, the Hanson approach, though built on exponential growth, produces superexponential.

Lee (1988) and Kremer (1993) most influence the modeling done here. Lee (1988) introduces a model of long-term human development with one fixed factor, natural resources, and two endogenous factors, population and technology. The three combine in Cobb-Douglas production to determine output. Population growth rises with output per capita, per Malthus. But population growth slows with technological advance, which can explain the worldwide fall in fertility since the 1950s. Meanwhile technology growth is increasing in population, per Kuznets, which allows for an economic takeoff.

Lee expresses these relationships in an unusual functional form. The dynamical system is

$$\dot{\ln \mathbf{y}} = \mathbf{B} \ln \mathbf{y} + \boldsymbol{\delta},$$

where $\mathbf{y}$ is a 2-vector consisting of population and technology, the natural logarithm is taken elementwise, $\mathbf{B}$ is $2 \times 2$, and $\boldsymbol{\delta}$ is $2 \times 1$. This is an inhomogeneous linear system in $\ln \mathbf{y}$, whose solution for $\ln \mathbf{y}$ is an affine combination of exponential growth terms.[4] The solutions for $\mathbf{y}$ are therefore *double* exponentials, which can grow ever faster, yet which, in formal contrast to (4), never reach a singularity.

Kremer (1993) adapts the model of Lee and then—uniquely in the literature—brings model to data using econometrics. Focusing on population, Kremer outdoes Von Foerster, Mora, and Amiot (1960) by assembling a population series reaching back a million years. The earliest observations come from Deevey (1960). Lacking a GWP series of comparable span, Kremer simplifies the theoretical model by (usually) assuming that income per person is fixed at a Malthusian equilibrium. A bivariate model for output and population becomes a univariate one for population. Kremer also roots the model in the neoclassical tradition, in which a central dynamic is the reinvestment of output into factors. Taking production as Cobb-Douglas produces an equation of motion for population of the form (2). In this way, the Kremer reformulation of Lee restores the mathematical potential for singularity.

---

[4] If $\lambda_1, \lambda_2$ are the eigenvalues of $\mathbf{B}$ and $\mathbf{v}_1, \mathbf{v}_2$ are corresponding eigenvectors, we have $\ln \mathbf{y} = \mathbf{v}_1 e^{\lambda_1 t} + \mathbf{v}_2 e^{\lambda_2 t} - \mathbf{B}^{-1}\boldsymbol{\delta}$.

In the econometrics, Kremer's dependent variable is the compound annual growth rate between observations of population. Estimation is by nonlinear least squares (NLS) and is from an econometric point of view dynamic: the estimator is challenged with explaining each observation conditional on the previous. A finite-difference analog of the equation of motion (2) is estimated rather than the solution (4). Applied to the population series through 1960, the cusp of the global fertility decline, Kremer (1993, Table VI, col. 2) estimates the scale effect $B$ at 1.22 (standard error 0.112).

A point of departure for the present paper is the observation that even in Kremer (1993), the econometric model is not based on a coherent data generating process (DGP). For intuition, note first that essentially all observation spacings in the Kremer data are multiples of five years. So we could take a model for the quinquennial DGP as the building block for DGPs for observations of any spacing. The five-year DGP specification corresponding to NLS is

$$\begin{aligned} \Delta \ln y_t &= s y_{t-5}^B + \delta + \epsilon_t \\ \mathrm{E}[\epsilon_t] &= 0 \\ \mathrm{Var}[\epsilon_t] &= \sigma^2 \end{aligned} \tag{5}$$

Given a realization of the random term $\epsilon_t$, we have $y_t = y_{t-5} + y_{t-5} \times (s y_{t-5}^B + \delta + \epsilon_t)$. Substituting the five-year lag of this formula into itself gives the implied model for decennially spaced observations, and so on. The algebra is complex because it expresses the way a shock in one period folds into subsequent nonlinear dynamics. It produces a random variable which, as a model for 10-year growth, differs from the 10-year analog of (5). Yet it is the 10-year analog of (5) that NLS brings to decadal observations. It is in this sense that NLS is internally inconsistent.

Note that while for intuition we imagined a data set in which some observations are spaced quinquennially or decadally, the logic applies even when observations are uniformly spaced. A dynamic NLS econometric model does not capture how moment-to-moment stochasticity interacts with nonlinear dynamics to shape the error distribution for each observation. Rather, it expediently tacks an i.i.d., finite-variance error onto a deterministic model. A plausible consequence is modeling a fat-tailed process with a thin-tailed DGP.

The present paper more rigorously addresses the evolution of stochastic, dynamic, nonlinear processes, using stochastic differential equations. By passing to the infinitesimal limit in time steps, the stochastic calculus produces models that are internally consistent in the sense just mentioned. This should allow for more efficient and consistent estimation of the parameters. And it provides a sounder basis for inference about profound questions about the timing and inevitability of economic takeoff. It also explicitly introduces a notion of contingency in history, recognizing that many paths are plausible—not just the pristine solutions to a deterministic differential equation.

## 2 A deterministic model of long-term development

As a prelude to the stochastic model, this section presents a deterministic model of economic development, following Lee (1988) and Kremer (1993). The section examines the conditions under which the levels and growth rates of factors in the system are stable. And it analyzes the limiting behavior when the system diverges. This analysis will bring some perspective to the explosive dynamics in human history. And it will help show how the mathematical form of the stochastic model connects with Solovian growth theory.

### 2.1 A model

A single, global production process produces output $Y$ from inputs $y_0, \ldots, y_k$. The input $y_0$ is special: it is technology—non-rival, imperfectly excludable, highly persistent (Romer 1990). The $k$ remaining factors can include capital, labor, and natural resources. Production is Cobb-Douglas:

$$Y = \prod_{i=0}^{k} y_i^{\alpha_i}.$$

The exponents on the conventional factors, $\alpha_1, \ldots, \alpha_k$, sum to 1, giving constant returns to scale. However, when we take a factor such as resources as fixed, we will drop it from the model for simplicity, reducing the formal order of homogeneity. Within this structure, technology might be taken as factor-neutral, with $\alpha_0 = 1$, equating it with TFP. Or technology might augment labor alone, in which case the two would carry the same exponent.

As in the Solow-Swan model, the allocation of output is not explicitly grounded in optimizing behavior of agents. The equations of motion for the factors take a shared form, which captures three influences: reinvestment of output, modulation of this reinvestment by the level of technology, and exogenous depreciation or appreciation:

$$\dot{y}_i = s_i y_0^{\phi_i} Y + \delta_i y_i. \tag{6}$$

For depreciation, $\delta_i < 0$. The investment rates $s_i$ simultaneously specify the allocation of output and convert from its units to the units of the factors. The $y_0^{\phi_i}$ term allows the level of technology to influence reinvestment in each factor, isoelastically. $\phi_i$ can be interpreted as adjusting the quantity or the productivity of investment. The latter interpretation staves off violation of the constraint that total reinvestment cannot exceed output. This term generalizes a structure in Jones (1995) that makes technology modulate the productivity of investment in technology only.

An example motivates the general formulation. Indexing with letters instead of numbers, we set

$$Y = AK^{\alpha_K} P^{\alpha_P} H^{\alpha_H} R^{1-\alpha_K-\alpha_P-\alpha_H}, \tag{7}$$

in which the factors are technology ($A$, synonymous with $y_0$), excludable business investment capital ($K$), population ($P$), human capital ($H$), and natural resources ($R$). For now, $R$ is fixed at 1, and dropped. We equate population with labor force. As in Solow (1957), technology is factor-neutral.

Table 1 displays illustrative parameter choices for this model. I make three comments on the choices.

First, all factors are endogenous: for all $i$, $s_i > 0$. This structural symmetry makes sense in the long view,

since in the long run all inputs are affected by output. The symmetry also undermines many arguments for casting technology as augmenting a single factor such as labor (Uzawa 1961; Kennedy 1964; Drandakis and Phelps 1966; Acemoglu 2003).[5] Those arguments typically distill to the conclusion that for output growth to be stable, technology must augment a stable, exogenously growing input. Here there are no such. This is one reason technology is factor-neutral in (7).

Second, as in Lee (1988), the two components in the equation of motion for population can produce a Malthusian equilibrium. Holding $A$ fixed, investment of economic product in creation and sustenance of life ($s_P A^{\phi_P} Y$) can balance the predation of mortality ($\delta_P P$ with $\delta_P < 0$). However, that dynamic alone cannot explain the historically novel drop in worldwide fertility since midcentury.[6] Galor (2012) argues that the dominant causal channel for the fertility drop has run from improving technology in production to higher demand for human capital, to a parental investment shift from child quantity to child quality, as contemplated in Barro and Becker (1989). The $A^{\phi_P}$ factor, with $\phi_P < 0$, expresses a version of this effect, as does the $A^{\phi_H}$ factor in the human capital equation, with $\phi_H = -\phi_P$.[7]

The last comment is that the equation of motion for technology resembles the Rivera-Batiz and Romer (1991) "lab equipment" specification for innovation, which takes the output good as the input to the production of technological advance. This distinguishes it from another common form, in which a factor stock such as $P$ or $H$ is the input (Romer 1990). If returns to investment of the output good into innovation are increasing then $\phi_A > 0$. Or, each advance may make the next *harder* (Jones 1995). Substituting (7) into (6) and specializing the latter to technology,

$$\dot{A} = s_A A^{1+\phi_A} K^{\alpha_K} P^{\alpha_P} H^{\alpha_H} + \delta_A A. \tag{8}$$

In simulating growth history since 25,000 BCE, Jones (2001, p. 23) chooses an idea production function similar to $\dot{A} = A^{1+\phi_A} P^{\alpha_P}$ and takes, in the present notation, $1 + \phi_A = 0.5$. I therefore choose $\phi_A = -0.5$ for illustration.

Replicating the substitution in (8) for all factors produces a dynamical system in the four variables $A$, $K$, $P$, and $H$. The system can be restated in the abstractions of linear algebra. Returning to numerical indexes, define the column vector $\mathbf{y} = [y_\mathrm{i}]_{i=0,..,k}$ and the vectors $\mathbf{s}$, $\boldsymbol{\alpha}$, $\boldsymbol{\delta}$, and $\boldsymbol{\phi}$ analogously. Denote by $\circ$ the elementwise product of vectors. Define the exponential function of a vector as applying elementwise. More generally, define exponentiation of a column vector by a row vector or matrix via $\mathbf{v}^{\mathbf{U}} \coloneqq e^{\mathbf{U} \ln \mathbf{v}}$. Then the system is

$$\dot{\mathbf{y}} = \mathbf{s} \circ \mathbf{y}^{\mathbf{I}+\mathbf{B}} + \boldsymbol{\delta} \circ \mathbf{y}. \tag{9}$$

With the parameter choices in Table 1,

---

[5] Technology is commonly taken to augment labor alone. By the Uzawa Theorem, *if* an economy achieves constant, growth then technology can only augment exogenous, constantly growing factors. But here we will assume no factors are exogenous and as a result the system will not attain constant growth.

[6] Worldwide total fertility fell from 4.97 live births per woman in 1950–55 to 2.47 in 2015–20 (UN 2019, file FERT/4).

[7] However, in the "unified growth theory," it is the rate of change rather than the level of technology that influences education investment (Galor and Weil 2000, Lemma 1).

$$\mathbf{s} = \begin{bmatrix} 0.025 \\ 0.25 \\ 0.2 \\ 0.04 \end{bmatrix}, \boldsymbol{\alpha} = \begin{bmatrix} 1 \\ 0.3 \\ 0.3 \\ 0.3 \end{bmatrix}, \boldsymbol{\delta} = \begin{bmatrix} -0.001 \\ -0.03 \\ -0.02 \\ -0.02 \end{bmatrix}, \boldsymbol{\phi} = \begin{bmatrix} -0.5 \\ 0 \\ -0.1 \\ 0.1 \end{bmatrix}, \mathbf{I} + \mathbf{B} = \begin{bmatrix} 1+\phi_A & \alpha_K & \alpha_P & \alpha_H \\ 1+\phi_K & \alpha_K & \alpha_P & \alpha_H \\ 1+\phi_P & \alpha_K & \alpha_P & \alpha_H \\ 1+\phi_H & \alpha_K & \alpha_P & \alpha_H \end{bmatrix} = \begin{bmatrix} 0.5 & 0.3 & 0.3 & 0.3 \\ 1 & 0.3 & 0.3 & 0.3 \\ 0.9 & 0.3 & 0.3 & 0.3 \\ 1.1 & 0.3 & 0.3 & 0.3 \end{bmatrix}. \tag{10}$$

Output is

$$Y = \mathbf{y}^{\boldsymbol{\alpha}'}, \tag{11}$$

where the prime indicates transposition. We can also write (9) as

$$\dot{\ln}\,\mathbf{y} = \mathbf{s} \circ \mathbf{y}^{\mathbf{B}} + \boldsymbol{\delta}. \tag{12}$$

Formally, $\mathbf{B}$ is constructed as follows. Let $\boldsymbol{\iota}$ be the $(k+1)$-vector of 1's; $\mathbf{I}$ an identity matrix; and $[\![\boldsymbol{\phi}]\!]$ the square matrix whose entries are 0 except in the 0-indexed column, which holds $\boldsymbol{\phi}$. Then

$$\mathbf{B} = \boldsymbol{\iota}\boldsymbol{\alpha}' + [\![\boldsymbol{\phi}]\!] - \mathbf{I} \tag{13}$$

The concise statement (12) invites a generalization: $\mathbf{B}$ could be freed from the single-output assumption embedded in (13) and allowed any entries. For example, factors other than technology could modulate fertility. Several broad cases then emerge. If $\mathbf{B} = \mathbf{0}$, the model is purely exogenous. If some but not all rows of $\mathbf{B}$ are 0's, the model is partially endogenous: the factors corresponding to the zeroed rows are exogenous.[8] If $\mathbf{B}$ is irreducible—if the graph of influences implied by $\mathbf{B}$ is strongly connected (Meyer 2010, p. 671)—then the system is *fully endogenous*. Every factor affects the growth of every factor, directly or indirectly.

Even with the single-output restriction (13), the system evidently admits no general closed-form solution.[9] To illustrate potential dynamics, Figure 3 therefore depicts two simulations of (12) with the parameters in Table 1. Time is measured in years: the depreciation rates are annual. The two simulations differ only in their starting point, and only slightly. Since the process we seek to model begins in ancient times, it takes population as the initially plentiful factor: it starts at 1 while the other inputs start at a shared smaller value. In one scenario this value is 0.03117; in the other, 0.03107. With the lower value, depreciation rules the day (or epoch), creating a poverty trap. The higher starting value is chosen so that the system escapes the trap and then explodes around 1800 (at the risk of seeming pretension to realism).

The simulations confirm that a multivariate endogenous growth system can maintain something close to stasis for a long stretch, then explode. But it is evidently impossible to determine without simulation whether and when it will do so from given starting values. In the next subsections, we will therefore focus on the equilibrium and explosion dynamics of such systems, which are more tractable.

## 2.2 Equilibria

We will examine the conditions for existence and stability of equilibria in (12), first in the levels of inputs and output, then in their growth rates.

[8] Partial endogeneity differs from semi-endogeneity. For example, the Romer (1990) model is partially endogenous in that the labor and human capital stocks are fixed while capital and technology are endogenous; but it is not semi-endogenous in the sense of Jones (1995).

[9] If $\boldsymbol{\delta} = \mathbf{0}$, the system admits a one-parameter subset of solutions in closed form: $\mathbf{y} = \mathbf{m}/(t_c - t)^{\mathbf{k}}$, in which $\mathbf{k} \coloneqq \mathbf{B}^{-1}\boldsymbol{\iota}$, $\mathbf{m} \coloneqq (\mathbf{k}/\mathbf{s})^{\mathbf{B}^{-1}}$, and in which exponentiation and division of vectors takes place elementwise. $t_c$ is the sole free parameter, the time of joint explosion or collapse.

If one of the factors is exogenous, then it is not very interesting to ask whether the system can achieve stasis (zero growth) in all variables, for that easily attained outcome is the basis for the neoclassical model. So assume that $\mathbf{s} > \mathbf{0}$. And assume that the exponent matrix $\mathbf{B}$ is invertible. Then, setting $\dot{\ln}\,\mathbf{y}$ to $\mathbf{0}$ in (12) and solving, the solution to the first-order condition for stasis is

$$\mathbf{y}^* = (-\boldsymbol{\delta}/\mathbf{s})^{\mathbf{B}^{-1}}, \tag{14}$$

where division of vectors takes place elementwise.[10] By (11), output at this point is $Y^* = (-\boldsymbol{\delta}/\mathbf{s})^{\boldsymbol{\alpha}'\mathbf{B}^{-1}}$. Applying a standard method in the analysis of continuous-time dynamical systems, we check whether at this point of stasis the Jacobian of (12) is stable, i.e., whether all its eigenvalues have negative real part. If so, then the equilibrium $\mathbf{y}^*$ is asymptotically stable, meaning that the system converges to $\mathbf{y}^*$ if starting adequately close to it. The Jacobian works out to

$$\frac{\partial \dot{\ln}\,\mathbf{y}}{\partial \ln \mathbf{y}} = \mathbf{s} \circ \mathbf{y}^{\mathbf{B}} \circ \mathbf{B}, \tag{15}$$

where the second $\circ$ operator broadcasts across the columns of $\mathbf{B}$, applying to each in turn.[11] Substituting with (14) into (15),

$$\frac{\partial \dot{\ln}\,\mathbf{y}}{\partial \ln \mathbf{y}} = -\boldsymbol{\delta} \circ \mathbf{B}.$$

The conditions under which $-\boldsymbol{\delta} \circ \mathbf{B}$ is stable can be characterized in two ways, one simpler and sufficient, one more exact.

The simpler method focuses on the sign of the determinant $|\boldsymbol{\delta} \circ \mathbf{B}|$. If it is negative then the number of eigenvalues of $\boldsymbol{\delta} \circ \mathbf{B}$ with negative real part is odd and cannot be 0. Therefore the number of eigenvalues of $-\boldsymbol{\delta} \circ \mathbf{B}$ with *positive* real part is odd and non-zero, and $-\boldsymbol{\delta} \circ \mathbf{B}$ is unstable. Thus a sufficient condition for instability is $0 > |\boldsymbol{\delta} \circ \mathbf{B}| = |\text{diag}(-\boldsymbol{\delta})||-\mathbf{B}|$. When $\boldsymbol{\delta} < \mathbf{0}$—when exogenous influences only cause net depreciation—this condition distills to $-|-\mathbf{B}| > 0$. In the single-output model (13), it works out that $-|-\mathbf{B}| = \boldsymbol{\alpha}'\boldsymbol{\phi} + (\boldsymbol{\alpha}'\boldsymbol{\iota} - 1)(1 - \phi_A)$ (see appendix A.1). So a sufficient condition for *instability* in state is

$$\boldsymbol{\alpha}'\boldsymbol{\phi} \ + (1 - \phi_A)(\boldsymbol{\alpha}'\boldsymbol{\iota} - 1) > 0. \tag{16}$$

This condition is more likely satisfied when technological advance tends to raise productivity of reinvestment in factors ($\boldsymbol{\alpha}'\boldsymbol{\phi}$ is high) and when returns to returns to scale in all inputs, including technology ($\boldsymbol{\alpha}'\boldsymbol{\iota}$) are high. An intuitive case is when $\boldsymbol{\alpha}'_{(0)}\boldsymbol{\iota}_{(0)} = 1$ and $\boldsymbol{\alpha}'_{(0)}\boldsymbol{\phi}_{(0)} = 0$, where the $(0)$ subscripts indicate deletion of the 0-indexed entries for technology. That case embraces the Jones (1995) model, in which conventional factors experience constant returns to scale ($\boldsymbol{\alpha}'_{(0)}\boldsymbol{\iota}_{(0)} = 1$) and technology is the only input whose reinvestment productivity rate varies with the level of technology ($\boldsymbol{\phi}_{(0)} = \mathbf{0}$). Then, (16) merely demands $\alpha_A > 0$: if the marginal product of technology is positive, which it effectively is by definition, a state of constant output in the fully endogenous model is unstable.

[10] If an entry of $\mathbf{s}$ is 0, then the corresponding (exogenous) factor cannot experience zero growth unless its entry in $\boldsymbol{\delta}$ is also 0, in which case growth in (12) is always zero and the corresponding entry of $\mathbf{y}$ is not identified by the first-order condition.

[11] Equivalently, $\partial \dot{\ln}\,\mathbf{y}/\partial \ln \mathbf{y} = \text{diag}(\mathbf{s})\ \text{diag}(\mathbf{y}^{\mathbf{B}})\ \mathbf{B}$.

That intuitive case does not quite embrace the example simulated earlier, which has $\boldsymbol{\alpha}'_{(0)}\boldsymbol{\iota}_{(0)} = 0.9$ rather than 1, the difference owing to the dropping of the static factor, natural resources. (The simulated example does satisfy $\boldsymbol{\alpha}'_{(0)}\boldsymbol{\phi}_{(0)} = 0$ because $\phi_K = 0$, $\phi_P = -\phi_H$, and $\alpha_P = \alpha_H$.) But this opening for stability is narrow: (16) still holds as long as $\phi_A > -9$.[12] Even the pessimistic assessment of Bloom et al. (2020) of the current marginal productivity of R&D does not point to such a low value. Their methods using aggregate data for the U.S. in 1930–2015 suggest $\phi_A \approx -3$.[13]

The more precise statement of the conditions for stability develops the characteristic equation of $-\boldsymbol{\delta} \circ \mathbf{B}$ and reiterates that the solutions must have negative real part. Notice that we can rewrite (16), which originates in $|\boldsymbol{\delta} \circ \mathbf{B}| < 0$, in a certain way. We have

$$|\boldsymbol{\delta} \circ \mathbf{B}| < 0 \Longleftrightarrow \boldsymbol{\alpha}'\left(\boldsymbol{\iota} + \frac{\boldsymbol{\phi}}{1-\phi_A}\right) > 1. \tag{17}$$

(This assumes $\phi_A < 1$.) A parallel statement applies to the characteristic equation:

$$|-\boldsymbol{\delta} \circ \mathbf{B} - \lambda\mathbf{I}| = 0 \Longleftrightarrow \left(\frac{\boldsymbol{\alpha}}{\boldsymbol{\iota} + \frac{\lambda}{-\boldsymbol{\delta}}}\right)'\left(\boldsymbol{\iota} + \frac{\boldsymbol{\phi}}{1 + \frac{\lambda}{-\delta_A} - \phi_A}\right) = 1. \tag{18}$$

in which $\lambda$ is an eigenvalue and $\lambda/-\boldsymbol{\delta} \coloneqq \lambda\boldsymbol{\iota}/-\boldsymbol{\delta}$. (See appendix A.3, which also discusses some degenerate exceptions.) Plugging $\lambda = 0$ into the left of the second equation in (18) yields the corresponding expression in (17)—which we have just seen exceeds 1 under plausible parameter choices. By examination, decreasing the complex expression in (18) to achieve equality requires driving $\lambda$ in a positive direction; and a positive $\lambda$ indicates instability. In particular, a root $\lambda$ must shift more in the positive direction the more that technological advance stimulates further advance ($\phi_A$ is higher) or stimulates other factors (the other elements of $\boldsymbol{\phi}$ are higher). That is, positive stimulus to factors from technology destabilizes stasis even more.

In sum, in the fully endogenous model, constancy of output is unstable under reasonable parameter choices.

There might be greater hope for stable *growth* in an endogenous system. Defining $\mathbf{z} = \ln \dot{\mathbf{y}}$, we can derive a direct equation of motion for $\mathbf{z}$ by differentiating (12) with respect to time. It works out that

$$\dot{\mathbf{z}} = (\mathbf{z} - \boldsymbol{\delta}) \circ \mathbf{B}\mathbf{z}. \tag{19}$$

---

[12] In the example developed in section 2.1, $\mathbf{I} - \boldsymbol{\delta} \circ \mathbf{B}$ is positive, so the Perron-Frobenius theorem pertains. The positive eigenvalue of $-\boldsymbol{\delta} \circ \mathbf{B}$ is 1 less than the Perron root of $\mathbf{I} - \boldsymbol{\delta} \circ \mathbf{B}$ and the associated eigenvector is the (positive) Perron vector of $\mathbf{I} - \boldsymbol{\delta} \circ \mathbf{B}$. Rather balanced growth or shrinkage in all factors constitutes the most purely destabilizing direction away from stasis.

[13] Bloom et al. (2020, Table 7, row 1) estimates a parameter $\beta$ at 3.1, which corresponds to a value of $-2.1$ for the $\phi$ of Jones (1995) (Bloom et al. 2020, note 25). However, those papers conceive of labor as the input to the production of ideas, rather than the money-denominated reinvestment flow $s_A Y$ as here. Bloom et al. thus deflates that flow by an index of high-skill wages in order to arrive at a measure of research effort. To modify the Bloom et al. calculations to the "lab equipment" specification used here, in which the research input is $s_A Y$, we deflate instead by the consumer price index. The estimates become $\beta = 4.0$ and $\phi_A = -3.0$. See this paper's code archive.

This allows us to view the system as one in which growth rates are the state variables, and to analyze steady states in the same way as before.[14,15] The corresponding Jacobian is

$$\frac{\partial \dot{\mathbf{z}}}{\partial \mathbf{z}} = (\mathbf{z} - \boldsymbol{\delta}) \circ \mathbf{B} + \operatorname{diag}(\mathbf{B}\mathbf{z}). \tag{20}$$

Suppose $\mathbf{z}^*$ is a root of the right side of (19), a self-perpetuating vector of input growth rates. If all inputs receive some reinvestment, so that none is purely exogenous, then when $\mathbf{z} = \mathbf{z}^*$, the first factor on the right of (19) is entirely non-zero. Then, for constant growth ($\dot{\mathbf{z}} = \mathbf{0}$), the second term, $\mathbf{B}\mathbf{z}^*$, must be $\mathbf{0}$. If $\mathbf{B}$ has full rank then $\mathbf{z}^* = \mathbf{0}$—which returns us to the analysis of stasis just above. And in the single-output economy (13), for $\mathbf{B}$ to be rank-deficient, $|\mathbf{B}| = 0$, and equality rather than inequality must hold in (16), which was just cast as unrealistic.[16]

Turning that reasoning around, if none of the elements of $\boldsymbol{\delta}$ is positive (exogenous factors are net-depreciative) and if $\mathbf{B}$ has full rank, then constant, positive growth is in general impossible in the model (12). Typical behavior, rather, is for factor growth rates to converge to their respective depreciation rates or diverge to positive infinity.

The prospect for stable growth brightens in the *partially* endogenous model, especially if technology becomes exogenous. So suppose there is at least one exogenous factor in the model. Use *ex* and *en* subscripts to denote the parts of the various vectors and matrices corresponding to exogenous and endogenous factors. For the exogenous factors, $\mathbf{z}_{ex} = \boldsymbol{\delta}_{ex}$ and the corresponding rows of $\mathbf{B}$ may be taken as $\mathbf{0}$. As for the endogenous factors, since the entries of $\mathbf{z}_{en} - \boldsymbol{\delta}_{en}$ are non-zero, for (19) to be $\mathbf{0}$ still requires

$$\mathbf{0} = (\mathbf{B}\mathbf{z}^*)_{en} = \mathbf{B}_{en,en}\mathbf{z}^*_{en} + \mathbf{B}_{en,ex}\mathbf{z}^*_{ex} = \mathbf{B}_{en,en}\mathbf{z}^*_{en} + \mathbf{B}_{en,ex}\boldsymbol{\delta}_{ex}.$$

Then, assuming $\mathbf{B}_{en,en}$ is invertible, solving for $\mathbf{z}^*_{en}$ gives the steady-state growth rates for endogenous factors:

$$\mathbf{z}^*_{en} = -\mathbf{B}^{-1}_{en,en}\mathbf{B}_{en,ex}\boldsymbol{\delta}_{ex}.$$

The equilibrium output growth rate is

$$Z^* = \boldsymbol{\alpha}'_{en}\mathbf{z}^*_{en} + \boldsymbol{\alpha}'_{ex}\mathbf{z}^*_{ex} = \left(\boldsymbol{\alpha}'_{ex} - \boldsymbol{\alpha}'_{en}\mathbf{B}^{-1}_{en,en}\mathbf{B}_{en,ex}\right)\boldsymbol{\delta}_{ex}.$$

As in the neoclassical model, the growth rates of exogenous factors ($\boldsymbol{\delta}_{ex}$) operate as multipliers in the equilibrium output growth rate. To investigate stability here, we plug the formulas for $\mathbf{z}^*_{ex}$ and $\mathbf{z}^*_{en}$ into the Jacobian (20), and check its eigenvalues. We get:

$$\frac{\partial \dot{\mathbf{z}}}{\partial \mathbf{z}} = \begin{bmatrix} (\mathbf{z}^*_{en} - \boldsymbol{\delta}_{en}) \circ \mathbf{B}_{en,en} & (\mathbf{z}^*_{en} - \boldsymbol{\delta}_{en}) \circ \mathbf{B}_{en,ex} \\ \mathbf{0} & \mathbf{0} \end{bmatrix}.$$

The non-zero eigenvalues, corresponding to endogenous factors, are those of the upper-left block.

The earlier results on the stability of $-\boldsymbol{\delta} \circ \mathbf{B}$, the Jacobian at stasis, transfer by analogy to $(\mathbf{z}^*_{en} - \boldsymbol{\delta}_{en}) \circ \mathbf{B}_{en,en}$. Before we assumed that the first factor in $-\boldsymbol{\delta} \circ \mathbf{B}$, namely $-\boldsymbol{\delta}$, was positive. Now we have assumed the same for the first factor of $(\mathbf{z}^*_{en} - \boldsymbol{\delta}_{en}) \circ \mathbf{B}_{en,en}$. Replacing $-\boldsymbol{\delta}$ with $\mathbf{z}^*_{en} - \boldsymbol{\delta}_{en}$, (17) and (18) transmogrify to

$$\left|(\mathbf{z}^*_{en} - \boldsymbol{\delta}_{en}) \circ \mathbf{B}_{en,en}\right| < 0 \Rightarrow \boldsymbol{\alpha}'_{en}\left(\boldsymbol{\iota}_{en} + \frac{\boldsymbol{\phi}_{en}}{1 - \phi_A}\right) > 1, \tag{21}$$

$$\left|(\mathbf{z}^*_{en} - \boldsymbol{\delta}_{en}) \circ \mathbf{B}_{en,en} - \lambda \mathbf{I}_{en}\right| = 0 \Rightarrow \left(\frac{\boldsymbol{\alpha}_{en}}{\boldsymbol{\iota}_{en} + \frac{\lambda}{\mathbf{z}^*_{en} - \boldsymbol{\delta}_{en}}}\right)'\left(\boldsymbol{\iota}_{en} + \frac{\boldsymbol{\phi}_{en}}{1 + \frac{\lambda}{z^*_A - \delta_A} - \phi_A}\right) = 1. \tag{22}$$

---

[14] One cannot move so neatly to growth space under the more general constant-elasticity-of-substitution production.
[15] As long as $\mathbf{z} \geq \boldsymbol{\delta}$, it is an attainable growth state, in the sense that there is a vector of positive factor levels $\mathbf{y}$ at which the system grows at $\mathbf{z}$. In particular, solving for $\mathbf{y}$ in (12) gives $\mathbf{y} = ((\mathbf{z} - \boldsymbol{\delta})/\mathbf{s})^{\mathbf{B}^{-1}}$.
[16] This conclusion is an instance of Growiec (2007, eq. 4).

These two conditions—one simpler, one exact—speak to the stability of growth in the partially endogenous system. Indeed, they create room for stable, constant, positive growth. If we imagine that, $\boldsymbol{\alpha}'_{en}\boldsymbol{\iota}_{en}$, the endogenous contribution to production, falls, then $\boldsymbol{\alpha}'_{en}\boldsymbol{\phi}_{en}$ will tend too as well, reducing the likelihood that the sufficient condition for instability of equilibrium growth (21) will be satisfied. By the same reasoning as after (18), this will also make it more plausible for eigenvalue solutions in (22) to have negative real part. The scope for stability is especially great if technology is exogenous, for then $\boldsymbol{\phi}$ does not figure in the construction of $\mathbf{B}_{en,en}$, and we take $\phi_A = 0$ and $\boldsymbol{\phi}_{en} = \mathbf{0}$ in (21) and (22). Then (21) is simply $\boldsymbol{\alpha}'_{en}\boldsymbol{\iota}_{en} > 1$, which we assume is impossible, since technology is not among the endogenous inputs and the remaining inputs have at most constant returns to scale. More generally, the left side of the inequalities in (17) and (21) emerges as the *degree of endogenous scale effect*; when this exceeds 1, the system contains explosive propensity. Meanwhile, Appendix A.4 shows that in the partially endogenous case, it is precisely when (21) does *not* hold that the equilibrium growth rate $Z^*$ is, under reasonable assumptions, positive.

This exploration of stability in a mathematical family embracing the Solow-Swan model with Cobb-Douglas production illustrates a few points. First, within this larger family, instability is the rule, in both levels and growth rates. From this standpoint, two features of the Solow-Swan model nevertheless assure convergence toward constant, positive growth: the presence of exogenous factors, and the assumption that these factors grow at constant rate. In effect, constant growth is injected into the Solow-Swan model and rather than emerging from it (Jones 2003).

## 2.3 Disequilibria

As a fully endogenous system approaches a singularity in one variable, that development will drive all other variables to diverge at the same moment, or to collapse to the pure-depreciation path. For once one entry of $\mathbf{y}$ goes to zero or infinity, all entries of $\mathbf{y}^{\mathbf{B}} (= e^{\mathbf{B}\ln\mathbf{y}})$ in the equation of motion (12) must do so. Typically, the path of each diverging variable will be increasingly well approximated by a *univariate* model, in which depreciation loses relevance and the acceleration of growth becomes dominated by the largest (or only) positive eigenvalue of $\mathbf{B}$. And as shown in the discussion after (16), an endogenous system with the single-output structure (13) has a positive eigenvalue under a broad range of reasonable parameter values.

For a more precise statement and demonstration of these assertions, let $\{\lambda_i\}_{i=0,\dots,k}$ be the eigenvalues of $\mathbf{B}$, indexed from highest to lowest real part, and $\{\mathbf{v}_i\}$ a corresponding eigenvector basis. Assume $\lambda_0$ is real. This is the case, for example, in the single-output system under instability condition (16). Then, appendix A.2 shows, the eigenvalues of $\mathbf{B}$ are $-1$, with multiplicity $k - 1$, and

$$\lambda_\pm \coloneqq \frac{\boldsymbol{\alpha}'\boldsymbol{\iota} + \phi_0}{2} - 1 \pm \sqrt{\left(\frac{\boldsymbol{\alpha}'\boldsymbol{\iota} + \phi_0}{2}\right)^2 + \boldsymbol{\alpha}'(\boldsymbol{\phi} - \phi_0\boldsymbol{\iota})}. \tag{23}$$

Condition (16) works out to be equivalent to $\lambda_- < 0$ and $\lambda_+ > 0$, so under assertedly plausible parameter values that satisfy (16), $\lambda_+$ is the sole positive eigenvalue.

Write

$$\ln \mathbf{y} = \sum_i c_i \mathbf{v}_i \tag{24}$$

for some $\mathbf{y}$-dependent coefficients $c_i$. Substituting that into the equation of motion (12), assuming that the potentially superexponential growth term $\mathbf{y}^{\mathrm{B}}$ is large enough to justify dropping the depreciation term $\boldsymbol{\delta}$, and taking the logarithm of both sides,

$$\ln \ln \dot{\mathbf{y}} \approx \ln\left(\mathbf{s} \circ \left(e^{\sum c_i \mathbf{v}_i}\right)^{\mathbf{B}}\right) = \ln\left(\mathbf{s} \circ e^{\mathbf{B}\sum c_i \mathbf{v}_i}\right) = \ln \mathbf{s} + \sum \lambda_i c_i \mathbf{v}_i. \tag{25}$$

In the $\mathrm{c}_0 \to \infty$ limit, assuming $\mathbf{s} > \mathbf{0}$, the elementwise ratio of (25) to (24) is $\lambda_0 \boldsymbol{\iota}$. Thus as the eigenvector with the greatest eigenvalue comes to dominate the composition of $\ln \mathbf{y}$,

$$\ln \ln \dot{\mathbf{y}} \approx \lambda_0 \ln \mathbf{y}. \tag{26}$$

Exponentiating both sides of that approximation,

$$\ln \dot{\mathbf{y}} \approx \mathbf{y}^{\lambda_0} \tag{27}$$

This is a special case of the general system (12), with a scalar matrix replacing $\mathbf{B}$; and it is in this sense that as the system approaches a singularity, it approximates a collection of univariate systems with shared scaling factor $\lambda_0$. Moreover, by working with the explicit superexponential growth solution (11) it can be checked that if all entries of $\mathbf{y}$ are exploding according to (27), then $Y = \mathbf{y}^{\boldsymbol{\alpha}'}$ explodes according to

$$\ln \dot{Y} \approx Y^{\lambda_0/\boldsymbol{\alpha}'\boldsymbol{\iota}}. \tag{28}$$

Partly on the basis of the observation that a diverging multivariate system approaches a set of univariate systems with a shared scaling rate, we will turn in the next section to a univariate model for GWP. The matrix $\mathbf{B}$ will become a scalar $B$. If we take the univariate empirical model as a reduction of a multivariate theoretical model that is more realistic but too complex to identify, the discussion above motivates viewing the estimated $B$ as containing information about the largest eigenvalue of an otherwise unknown $\mathbf{B}$. Equation (23) suggests one way to interpret estimates of $B$. If we revert to the example in the previous subsection featuring zero average effect of technological advance on reinvestment in non-technology inputs ($\boldsymbol{\alpha}'_{(0)}\boldsymbol{\phi}_{(0)} = 0$), constant returns in those inputs ($\boldsymbol{\alpha}'_{(0)}\boldsymbol{\iota}_{(0)} = 1$), and technology as TFP ($\alpha_A = 1$), then by (23) the largest eigenvalue is

$$\lambda_0 = \lambda_+ = \frac{\phi_A}{2} + \sqrt{1 + \frac{\phi_A^2}{4}}.$$

Solving, $\phi_A = \lambda_0 - 1/\lambda_0$. If in light of (28) we interpret an estimate of $B$ as $\lambda_0/\boldsymbol{\alpha}'\boldsymbol{\iota} = \lambda_0/2$, we obtain

$$\phi_A \approx 2B - \frac{1}{2B}. \tag{29}$$

This approximation will link univariate estimation results to the returns elasticity for investment of output in TFP in a multifactor context.

# 3 Stochastic modeling

The models contemplated in the previous section are deterministic: parameter values and initial conditions exactly govern the path of the system for its lifetime. This section introduces stochasticity. The resulting model dynamically incorporates shocks while still allowing a superexponential component of growth and a component of constant appreciation or depreciation. This allows for different outcomes to result from the same starting point and produces fat-tailed distributions at each time point. Yet the distributions can be stated analytically, which facilitates fitting to data.

## 3.1 A univariate stochastic model

The stochastic model developed here is chosen to be univariate, for tractability. In addition, the previous section produced a broader rationale: as multivariate systems explode, they converge to collections of univariate systems. In a deterministic multivariate model, the timing of any takeoff depends on the starting values and exact dynamics in a way that is complex and hard to analyze. In a discrete-time variant, the dynamics might be chaotic: deterministic yet sensitive to parameters or starting values, with a fractal boundary between parameter regions leading to qualitatively different outcomes. A stochastic model can capture some of the character of an analytically intractable, determinate process with a more tractable, less determinate process.

We start with the univariate subcase of the multivariate system (9):

$$dy = (sy^{1+B} + \delta y)dt. \tag{30}$$

When $B, \delta \neq 0$, this equation is solved by a change of variables, $y := x^{-1/B}$. That produces

$$dx = (-\delta Bx - Bs)dt, \tag{31}$$

whose particular solution is

$$x = \left(x_0 - \frac{s}{-\delta}\right)e^{-\delta Bt} + \frac{s}{-\delta}, \tag{32}$$

where $x_0$ is the value at $t = 0$. The solution for $y$ follows directly. Notice that if there is depreciation ($\delta < 0$) but also explosive propensity ($B > 0$), $x$'s distance from $s/(-\delta)$ grows exponentially—whether upward or downward depending on which side of that value $x$ starts on. If downward, then $x$ reaches 0 in finite time. It follows that $y = x^{-1/B}$ decays over infinite time or explodes in finite time depending on whether $y$ starts above or below $(-\delta/s)^{1/B}$. In Figure 3, the multivariate system exhibited a similar bifurcation.

The stochastic model adds to (30) a random term with a particular form:

$$dY_t = (sY_t^{1+B} + \delta Y_t)dt + \sigma\sqrt{Y_t Y_t^{1+B}}\,dW_t. \tag{33}$$

Following typographic convention in the literature on stochastic differential equations (SDEs), I have replaced the deterministic $y$ with the random $Y_t$. The two familiar terms in the multiplier on $dt$ constitute the drift coefficient. The novel $dW_t$ represents the progression of a Wiener process, whose sample paths are continuous random walks with fractal complexity, and whose cumulative variance at time $t$ equals $t$. For reasons soon to be stated, the multiplier on $dW_t$, the diffusion coefficient, is proportional to the geometric mean of the two drift components. Together, these definitions make $Y_t$ a random variable whose distribution at each time $t$ is determined by $Y_0$ as well as $s, B, \delta$, and $\sigma$. As a shorthand, I will call (33) the "superexponential" diffusion even though it is also capable of exponential and subexponential evolution if $B \leq 0$.

The mathematical construct of the Wiener process, represented by $dW_t$, is the heart of the stochastic differential equation. It is what allows the data generating process across a microsecond, if compounded enough, to equal that across a millennium. It addresses the methodological concern emphasized in section 1.

Stochasticity induces a distinction between distribution and instance. We can imagine an infinite number of "rollouts" of world history all beginning at the same GWP level in 10,000 BCE, and all evolving according to (33) with the same parameters. In one, the wheel is invented a thousand years early; in another, it never is. Yet the *distribution* for GWP at each moment, conditional on a fixed starting point, evolves deterministically.

The diffusion coefficient in (33), $\sigma\sqrt{Y_t Y_t^{1+B}}$, governs how the variance of the stochastic component grows with $Y_t$; it too is chosen for tractability. For with this specification, assuming $B \neq 0$, the SDE remains solvable via the change of variables used to solve (30):

$$Y_t \coloneqq X_t^{-1/B}. \tag{34}$$

To see this, define, as the inverse of that transformation, $f(Y_t) \coloneqq Y_t^{-B}$. Itô's stochastic calculus dictates how increments of the randomly evolving variable $X_t$ depend on those of $Y_t$ (Oksendal 2014):

$$dX_t = f'(Y_t)dY_t + \frac{1}{2}f''(Y_t)dY_t^2, \tag{35}$$

where $dY_t^2$ represents an infinitesimal increment of the quadratic variation of $Y_t$.[17] Since in the Itô calculus $dW_t \cdot dW_t = dt$ and $dt \cdot dt = dt \cdot dW_t = 0$, substituting (33) into (35) results in

$$dX_t = \left(-\delta B X_t - Bs + \frac{\sigma^2}{2}B(B+1)\right)dt + \sigma B\sqrt{X_t}dW_t. \tag{36}$$

Collecting the constant terms with

$$a \coloneqq \frac{\sigma^2 B^2}{2}, b \coloneqq -B\delta, c \coloneqq -Bs + \frac{\sigma^2 B(B+1)}{2} \tag{37}$$

gives

$$dX_t = (bX_t + c)dt + \sqrt{2aX_t}dW_t. \tag{38}$$

This SDE stochastically extends (31). It corresponds to the Feller (1951b) diffusion, and is applied in finance as the Cox-Ingersoll-Ross (CIR; 1985) model, though CIR impose $b < 0$ and $c > 0$. The distribution of $X_t|X_0$, or *solution*, admits an analytical form, something most SDEs do not.[18]

Cox (1996 [1975]) first proposed extending a CIR-type model via a power transform like (34).[19]

---

[17] We use the Itô interpretation of stochastic differential equations, according to which each infinitesimal innovation $dW_t$ is independent of all developments up to time $t$.
[18] The solutions are solutions of the Kolmogorov forward/Feynman-Kac equation discussed in appendix B.
[19] As CIR (1976, note 6) observes, Feller (1967, pp. 325–26) earlier pointed out the general principle that one stochastic process with a known generator can be constructed from another via a monotone transformation.

However, Cox then imposes $s = 0$, in the above parameterization. This produces the Constant Elasticity of Variance (CEV) model in his context and a purely exogenous growth system in ours. Cox further requires $-2 \leq B < 0$. The diffusion (33) embraces the CEV as a special case, along with one-dimensional Brownian motion, the Bessel process, the squared Bessel, and, in a limit, geometric Brownian motion. Figure 15 in appendix B.3 shows how these and other common models are connected through parameter restrictions and the power transform.

As a model for very long-term economic series, the superexponential diffusion has virtues and drawbacks. The analytical formulas for the solutions obviate the need for Monte Carlo simulation during maximum likelihood estimation (on the complexities of which, see Hurn, Jeisman, and Lindsay 2007). In addition, its connection to the Feller/CIR assures existence and uniqueness of solutions in a context in which general theory does not assure such (Cox and Ross 1976, note 6). An SDE capable of superexponential growth will typically contain a super-linear drift component such as $Y_t^2 dt$ (appearing in (33) when $B = 1$). This term is not Lipschitz continuous over the positive reals: there is no global upper bound on the magnitude of its slope with respect to $Y_t$. Yet Lipschitz continuity is assumed in standard proofs of existence and uniqueness of SDE solutions (e.g., Oksendal 2014, Theorem 5.2.1).[20] It is therefore not certain that a *general* SDE $dY_t = b(Y_t)dt + \sigma(Y_t)dW_t$, with $b(\cdot)$ super-linear, has (unique) solutions.

The price of that assurance in this special "superexponential" model (33) is the fixing of the functional form for the diffusion coefficient in relation to the drift coefficient, as the parameter $B$ enters both. Any departure from that restriction would make it impossible to reach the Feller/CIR in (36). This inflexibility may cause model fits to misestimate the stochasticity in economic history.

Another drawback lies in the univariate character of the model. For millennia, income per person enjoyed little clear trend. The obvious explanation for this stability, the Malthusian model, requires two interacting factors as in Lee (1988). A univariate model may fail to reproduce both the long stretch of relative stasis and the recent centuries of takeoff.

## 3.2 Transition densities and sample paths under the Feller/CIR diffusion

Feller (1951b) first studied the solutions to (38)—in particular how they are shaped by conditions on the behavior of sample paths at the $X_t = 0$ boundary. Despite that work and the popularization of the Feller diffusion through CIR (1985), I have found no unified treatment of the solutions and the corresponding boundary conditions. There are two standard solutions.But like Clark Kent and Superman, they never appear in the same place at the same time, which gives our knowledge of them a folkloric quality. Here I will only state those two solutions, leaving their demonstration to appendix B. These statements then transfer to $Y_t$, the potentially superexponential variable, via the power transform.

We first define two families of probability densities, which will form the basis of distributions for $X_t | X_0$

[20] In fact, (38) is not Lipschitz either, since $\sqrt{2aX}$ has unbounded slope as $X \to 0$. Nevertheless, theory assures existence and uniqueness of solutions (Yamada and Watanabe 1971). Alternatively, one can observe that for $B = 1$, the terms of (33) are locally Lipschitz, which assures existence and uniqueness until an explosion time (Van Handel 2007, Theorem 5.6.2). Again, the result for the Feller/CIR variable $X_t$ bootstraps to $Y_t$.

at each time $t$. Let $f_\Gamma(z;\alpha) = e^{-z}z^{\alpha-1}/\Gamma(\alpha)$ be the standard gamma density. Then the noncentral $\chi^2$ density is

$$f_{\chi^2}(x;\lambda,\nu) \coloneqq \sum_{m=0}^{\infty} f_\Gamma(\lambda;m+1)f_\Gamma(x;m+\nu+1) \tag{39}$$

And what I call the Feller density is

$$f_{-\chi^2}(x;\lambda,\nu) \coloneqq \sum_{m=0}^{\infty} f_\Gamma(\lambda;m-\nu+1)f_\Gamma(x;m+1). \tag{40}$$

$\lambda$ is a location parameter for the distributions and $\nu$, a shape parameter. Write $f_{\pm\chi^2}$ for $f_{\chi^2}$ and $f_{-\chi^2}$ as a pair. Figure 4 plots the two for $\lambda = 1$ and $\nu = -3.0, -2.5, \ldots, +3.0$. For each value of $\nu$, $f_{\chi^2}$ and $f_{-\chi^2}$ are the same color in the figure; $f_{\chi^2}$ is dotted while $f_{-\chi^2}$ is solid. When $\nu$ is an integer, the two coincide. Otherwise, the two fork toward the left; in some cases $f_{\chi^2}$ goes to infinity.

To fashion these densities into time-indexed diffusion solutions, we set their inputs to depend on the Feller/CIR parameters $a$, $b$, and $c$, the initial value $X_0$, and $t$. Specifically, as motivated in Appendix B, we plug into (39) and (40) with a series of definitions:

$$\begin{gathered} Z_t \coloneqq \frac{e^{-bt}}{a}X_t \\ \tilde{t} \coloneqq \int_0^t e^{-bs}ds = \frac{1}{b}[1-e^{-bt}] \\ \tilde{Z}_{\tilde{t}} \coloneqq Z_t,\ \ \tilde{c} \coloneqq c/a \\ x \coloneqq \tilde{Z}_{\tilde{t}}/\tilde{t},\ \ \lambda \coloneqq \tilde{Z}_0/\tilde{t},\ \ \nu \coloneqq \tilde{c}-1 \end{gathered} \tag{41}$$

After multiplying by the Jacobians of the changes in variables from $X_t$ to $Z_t$ to $x$, the densities for $X_t|X_0$ are

$$f^*_{\pm\chi^2}(X_t;X_0,t,a,b,c) \coloneqq \frac{e^{-bt}}{a\tilde{t}}f_{\pm\chi^2}(x;\lambda,\nu). \tag{42}$$

To translate these solutions for the Feller/CIR into ones for the superexponential, we transform lastly with $Y_t = X_t^{-1/B}$ while multiplying once more by a corresponding Jacobian:

$$f^{*B}_{\pm\chi^2}(Y_t;Y_0,t,s,B,\delta,\sigma) \coloneqq \frac{e^{-bt}|B|Y_t^{-B-1}}{a\tilde{t}}f_{\pm\chi^2}(x;\lambda,\nu). \tag{43}$$

But for the moment, we focus on $f^*_{\pm\chi^2}$, not $f^{*B}_{\pm\chi^2}$. The two solutions $f^*_{\pm\chi^2}$ differ in the behavior of sample paths at the $X_t = 0$ boundary. $f^*_{\chi^2}$ is realized by instantaneous reflection there and $f^*_{-\chi^2}$ by absorption. These behaviors, as well as others intermediate between them, are components of the stochastic modeling vocabulary. Absorption at zero might represent permanent civilizational collapse while reflection could represent recovery.

The reflecting and absorbing solutions each lose physical or economic plausibility for certain ranges of the constant drift factor $c$ (equivalently, of $\nu = c/a - 1$). For at $X_t = 0$, the drift term $cdt$ in (38) becomes the only non-zero component, the only source of motion. The term can only drive reflecting (upward) motion if $c > 0$. So the reflecting solution $f^*_{\chi^2}$ requires $c \geq 0$. It turns out that when $0 < c < a$, the modesty of the upward drift causes the density to approach infinity near the reflecting boundary. In contrast, when $c > a$, the density approaches 0 near the boundary. The stronger upward drift makes the boundary fully escapable if $X_0 = 0$ and unattainable if $X_0 > 0$. Meanwhile, the absorbing solution $f^*_{-\chi^2}$ is compatible with negative drift ($c \leq 0$) or modest positive drift $0 < c < a$. Because of the absorption, probability mass accumulates at the boundary, in the amount of $1 - F_\Gamma(\lambda; -\nu)$, where $F_\Gamma(\cdot;\cdot)$ is the cumulative standard gamma distribution. This mass point is not expressed in (42).

Figure 5 depicts $f^*_{\pm\chi^2}$, the absorbing and reflecting solutions for the Feller/CIR diffusion, for selected parameters. In all plots, $X_0 = a = 1$; and $b = -1$, bestowing a component of exponential decay. From top to bottom, $\nu = -1.5, -0.5, +0.5$, meaning $c = -0.5, +0.5, +1.5$. In each plot, time runs along the horizontal axis. Density values are shown by color, yellow indicating high and dark purple, low. The figures also show how the mean, median, and mode evolve.[21] The figures omit the implausible cases (as just discussed) of $f^*_{-\chi^2}$ with $\nu > 0$ and $f^*_{\chi^2}$ with $\nu < -1$.

Only when $0 < c < a$ ($-1 < \nu < 0$) are $f^*_{\pm\chi^2}$ distinct and yet both plausible. Feller (1951b) determines that within this parameter range, the Feller/CIR diffusion admits infinitely many plausible solutions, but does not find explicit statements of them. Members of the solution set differ in the boundary behavior of corresponding sample paths. Ito and McKean (1965) develops the theoretical tools for understanding these possibilities; but historically, the tools have not been applied to the Feller/CIR.[22]

Theory points to two additional dimensions of boundary behavior when $0 < c < a$, which are potentially relevant for modeling the human trajectory. First, the $X_t = 0$ boundary may be *sticky* or *slowly reflecting*. On a given sample path, each sojourn at a sticky boundary is instantaneous. Yet the path may return with infinitesimal immediacy and infinite frequency, at least within a finite spell, so that total time spent at the boundary is positive. Like a fat Cantor set, the set of times at which the path is at zero has positive Lebesgue measure yet is nowhere dense (Engelbert and Peskir 2014). Peskir (forthcoming) develops sticky solutions for the squared Bessel process; Peskir and Roodman (forthcoming) extends these results to the Feller/CIR.

The second aspect of boundary behavior that becomes compatible with valid solutions when drift is modestly positive ($0 < c < a$) is perhaps easier to visualize, if less intuitively named: elasticity. While at the boundary, a path has a fixed, positive probability per unit time of being *killed*. A killed process does not stay at zero for eternity. Rather, it halts. That distinction becomes meaningful in conjunction with stickiness. Without the added possibility of killing, a particle will inevitably escape a sticky boundary, just as every atom in a radioactive sample will eventually decay. With killing, the path might halt before escaping. An elastic boundary combines stochastic killing with stickiness.

Transferring the various Feller/CIR boundary behaviors to the superexponential diffusion introduces a conceptual complication. If $X_t$ reflects (absorbs) at 0 then $Y_t = X_t^{-1/B}$ with $B > 0$ reflects (absorbs) at $+\infty$. This notion is best interpreted probabilistically: when the boundary $+\infty$ is reflecting, the probability that a sample path will dwell permanently there, i.e., explode, is zero; at all times, probability mass is diffused across $(0, \infty)$. When $+\infty$ is absorbing, the probability of explosion is positive.

---

[21] Means are computed analytically, according to formulas in section B.9.2. Medians and modes are computed numerically, starting from the analytical distribution formulas.

[22] Molchanov (1967, remark 1) proposes forming the infinite family of convex linear combinations of $f^*_{\chi^2}$ and $f^*_{-\chi^2}$, apparently with fixed coefficients. However, these novel solutions are not in general Markovian. The rate of escape of probability mass from the boundary does not remain in fixed proportion to the amount of probability present there.

## 3.3 Monte Carlo testing

To test the superexponential diffusion model as the basis for a maximum likelihood (ML) estimator, and to compare it to NLS as applied in Kremer (1993), I run a Monte Carlo simulation. First, I fit the model to a GWP series using data and methods reported in the following sections (estimates from column 4 of Table 3, below). Using the resulting parameter estimates, along with the starting GWP value of $1.6 billion for 10,000 BCE, I generate 10,000 sample paths according to a finite-difference (Euler-Maruyama) approximation of (33) with a time step of 0.1 years. I run each path until it reaches 25,000 years or $100 trillion.[23] I sample each to mimic the observation spacing in the GWP series introduced in the next section, with the same number of observations, 36, and the same relative spacings. For example, if a path reaches $100 trillion in 6,000 years, about half the time of the actual series, then the final observations come every five years instead of every ten. To each path so sampled, I apply NLS for (5) and ML for (33).[24]

The simulation shows that ML estimation of the diffusion model is more efficient. NLS converges 98.8% of the time and ML 97.1%. When ML converges, NLS has a much higher dispersion: in estimating the scale effect $B$, NLS and ML have standard deviations of 0.136 and 0.05, respectively. ML is essentially unbiased when it converges (bias is $-0.0008$, two-tailed $p = 0.13$) while the NLS has a bias of $+0.019$ (two-tailed $p = 0.000$). See Figure 6.

Theory foreordains this result: ML is efficient when correctly specified. Yet the exercise helps quantify the added precision from ML. And it is meaningful in that while the NLS model is internally inconsistent, as noted in section 1, the DGP most consistent with NLS in spirit is the superexponential diffusion.

# 4 Data

The empirics below will focus on GWP observed over the very long term, while paying modest attention to global population, GWP per capita, and frontier GDP/capita. The univariate stochastic model defined in the previous section is appropriate for GWP in that it posits that a fraction of society's product is reinvested in the sources of its productivity ($sY_t^{1+B}$) and also allows depreciation ($\delta Y_t$). The model is somewhat less apt for population since it would cast population as the sole source of its own "reinvestment," thus omitting the crucial role of economic variables. The model is also less apt for a ratio such as GWP/capita. For while it is standard in models that make population exogenous to mathematically abstract from it by taking variables on a per-capita basis

[23] Both estimators apply here tend to diverge for sample paths that, unlike the historical series, are dominated by decay. So I discard paths that end lower than they start. I generate 1,222 paths and retain 1,000.
[24] Since the NLS estimating equation $\ln \dot{y}_{t_i} = sy_{t_{i-1}}^B + \delta + \epsilon_i$ is linear but for exponentiation by $B$, it is fit by analytically concentrating out the other parameters, then iteratively maximizing the profile likelihood for $B$. The NLS estimates are the starting points for the ML fits. As in Kremer (1993) the dependent variable in NLS is the compound annual growth rate. To address heteroskedasticity, observations are weighted by the time spans between them.

(Sala-i-Martin and Barro 2004, p. 28), such a factoring becomes impossible when population is endogenous. Then, GWP/capita is an inherently multifactor concept: GWP/capita per se is not invested in GWP/capita. Withal, the graphs of population and GWP/capita also look like hockey sticks, so it is interesting how well the superexponential diffusion fits them.

The GWP and population series gathered here are extended back a million years—with caution. Szathmáry (2015) perceives seven major transitions in the history of life. Most occurred when competing units—genes, prokaryotes, individual animals—ganged up. The last of the seven was the development of natural language 40,000–50,000 years ago. Arguably it was then that technology gained its modern, alchemical character. Through language, humans could share ideas more quickly and flexibly than any organism before. In light of this structural break, it may be best to apply the models developed above only to data from the last few myriads of years.

Many authors provide long-term population series (Carr-Saunders 1936; Huxley 1950, Woytinksy and Woytinksy 1953, Putnam 1953, Bennett 1954; Deevey 1960; Cipolla 1962; Clark 1967; Durand 1967, 1977; McEvedy and Jones 1978; Biraben 1979; Blaxter 1986; Maddison 2001). The estimates are not independent. Rather, a consensus has evolved as authors have copied or adjusted earlier figures or extended them back, incorporating information from sources such as ancient censuses and studies of population densities supportable with given forms of agriculture.[25] As a result, available global population figures rarely disagree on overlap by more than 30% (Durand 1977, Table 5; Cohen 1995, Appendix 2). Among these sources, McEvedy and Jones (1978) provides the most observations before the Common Era. Maddison (2001, p. 230) generally prefers their figures and Kremer (1993) relies on them for 10,000 BCE–1900 CE.[26,27] I also use them.

Estimates of population before 10,000 BCE are especially uncertain—population not only of *Homo sapiens*, but the genus *Homo*. Deevey (1960, p. 196) posits a human population of 0.125 million in 1 million BCE, 1.0 million in 300,000 BCE, and 3.34 million in 25,000 BCE. Kremer (1993) incorporates these numbers. But, calling them "forty year old speculations," Hanson (2000) prefers the estimates of 20,000 in 2 million BCE (Hawks et al. 2007) and 0.5 million in the period 0.5–1.0 million BCE (Weiss 1984). However, these newer values may not be more reliable. Hawks et al. (2007) infer from genetic evidence the occurrence of a population bottleneck 2 million years ago; but that speaks to the size of a breeding population from which modern people are descended, as distinct from the total population then extant. Meanwhile, the Weiss (1984) calculation is in the manner of Deevey and appears to contain an order-of-magnitude error.[28] I follow Kremer (1993) in using the Deevey

---

[25] On the intellectual history, see Caldwell and Schindlmayr (2002).
[26] However, for Asia in 1 CE, Maddison (2001, Appendix B) tends to take higher estimates from other sources.
[27] A recent synthesis of these estimates appears in Klein Goldewijk et al. (2017), which cites for pre-modern values McEvedy and Jones (1978), Maddison (2001), and Livi-Bacci (2007), the last of which appears to copy Biraben (1979). While the methodology of the synthesis is not documented, there is no suggestion that it injects new information. The main contribution appears to be interpolation of estimates with comprehensive geographic resolution and even temporal spacing. Such imputations are presumably valuable as inputs to climate modeling. But they can harm when the imputed variable is the *object* of modeling, by adding spurious information content.
[28] Weiss (1984, p. 641) assumes an inhabited area of 13.3 million square miles, 34.5 million km$^2$, and a population

figures since these have not been obviously improved upon.

Estimates of GWP and GWP/capita are dominated by the lifework of Angus Maddison. Maddison (2001, 2003) calculates GWP for 1, 1000, 1500, 1600, and 1700, 1820, 1870, 1913, and more recent years. Only De Long (1998) has extended the Maddison estimates back; starting from the Maddison (1995) estimates for 1820–1992, De Long develops a series starting in 1 million BCE.

I take the Maddison (2010) global estimates as the spine for GWP/capita series. In so doing, I again eschew some previous tactics for extending series to ancient dates—in this case, those of De Long (1998). To extend the Maddison (1995) GWP series before 1820, De Long makes a Malthusian argument: when the binding limit on population is humans' productive capacity, a change in GWP/capita should be followed by a change of the same sign in population. De Long observes that GWP/capita and population growth are indeed strongly, positively correlated "from the early nineteenth century until roughly World War II," with a slope of $1,165 per percentage point of population growth. De Long applies this association to population growth rates from Kremer to estimate GWP/capita back to 1 million BCE. But it may not be that meaningful to extrapolate the relationship between population growth and GWP/capita from an era of divergence from Malthusian equilibrium to an era dominated by it.[29]

The other De Long tactic discarded here takes on board the argument of Nordhaus (1997) that standard measures of output growth grossly underestimate, because they do not fully capture the gains in quality and diversity of goods and services. As a rough adjustment, De Long injects an extra quadrupling into GWP growth between 1800 and 2000. Yet seemingly the reasoning applies before 1800 too: in many places, product quality and diversity increased between 800 and 1800. Since I cannot quantify these gains, I avoid such adjustments.

I anchor a GWP/capita series in Maddison's values for 1, 1000, and 1500–2010, while taking a subsistence level of $400/year for 10,000 BCE and earlier. Before 1500, I interpolate log GWP/capita to dates for which population estimates are available by assuming that log GWP/capita grew in proportion with log population.[30] This makes some sense given the long-term, positive association between population and income. But it may

---

density of 0.28/km$^2$, which yields 9.65 million people rather than the stated 0.5 million. In correspondence, Kenneth Weiss was quick to assume that there is an error in the paper. For 300,000 BCE, Deevey (1960) has 0.012/km$^2$ and an inhabited area of 85 million km$^2$, for a total of 1 million.

[29] The conceptual basis for the extrapolation is further muddied in the De Long (1998) implementation by a mix of time period lengths: e.g., the observation of GWP/capita in 1800 is associated with average population growth in the next 50 years while that in 1960 is matched to the next five. In the event, the De Long (1998) extrapolations did not perform especially well in predicting the pre-1820 estimates that Maddison subsequently published. For example, where Maddison (2010) perceives a decline between 1 and 1000 CE from $467 to $453 and then a rise by 1500 to $566, the De Long (1998) method produces the opposite: a substantial rise in those first 1000 years, from $404 to $494, then a slight climb in the next 500, to $512. In addition, reverse-engineering reveals a few undocumented and debatable choices in De Long (1998). The regression that is for the basis for extrapolation is run on a data series that itself contains imputations, e.g., for 1800, 1850, 1875, 1920, 1925, and 1940. And before serving as the independent variable in the extrapolation, pre-1800 population growth rates are smoothed via a three-observation moving average, without adjustment for the uneven spacing of observations. For example, the smoothed, growth rate for 1 CE is the simple average of the annualized growth rates for 200 BCE–1 CE, 1–14 CE, and 14–200 CE.

[30] I also incorporate updates for Western Europe from the Maddison Project (Bolt and van Zanden 2014), which in turn draws on new scholarship on the evolution of material standards of living in that region. But the Maddison Project updates do not estimate global totals, so Maddison (2010) remains the foundation.

understate short-term Malthusian effects. For example, average income in Europe is thought to have risen after the Black Death. Yet in the series constructed here, since GWP/capita is interpolated in positive association with population, GWP/capita falls between 1300 and 1400, from $551 to $541. In the empirics, I check for sensitivity to switching to De Long's figures.

Growth in GWP/capita can be factored into frontier growth and catch-up growth. Frontier growth is important because in can be the only source of permanent overall GWP/capita growth. I proxy for the frontier with France, a choice driven by a combination of data availability and France's proximity to the frontier in modern and Roman times. The Maddison Project (Bolt et al. 2018) estimates income in the Gallic region back to 1 CE. Coverage of France is strong starting in 1280, thanks to the analysis of Ridolfi (2016) of wages and prices in primary sources such as records for work on the Chartres Cathedral. I extend this series back by copying GWP/capita values for 5000 and 10,000 BCE. I extend it forward from 2016 with growth rates from the IMF (2020, series NGDPRPPPPCPCH).

Table 2 displays the data for population, GWP/capita, GWP (their product), and GDP/capita for France.

As explained in footnote 27, population estimates in the HYDE climate modeling database are not used here. However, one element of that data set is taken up: an indicator of the uncertainty of estimates. Klein Goldewijk et al. (2017, p. 8) expresses the uncertainty as ±1% for the year 2000 and thereafter, ±5% for 1900, ±25% for 1700, ±75% for 1 CE, and ±100% 10,000 BCE. Values in between are linearly interpolated. The authors do not ascribe credence values to these ranges. I incorporate this uncertainty indicator, call it $h_t$, as observation weights $1/(1+2h_t^2)$ in the regressions.[31] This weights ancient observations—already deemphasized by their sparsity—a third as much as the modern ones, since $h_t = 1$ for former and $\sim 0.01$ for the latter.[32]

One potential elaboration would be to study data for continental regions such as Eurasia and Africa. But it is not pursued here. Fits to regional data before significant global integration, say in 1500, could provide independent tests of the model. Unfortunately, the available data do not appear to support regionalization of the time series approach. The series for Eurasia dominates the world total throughout, so restricting to the region does not affect results much. The series for the Americas before contact with Europe is too short and uncertain

[31] I also set $h_t = 1$ for $t <$10,000 BCE.

[32] A rough theory for the weighting runs as follows. Since we estimate with dynamic maximum likelihood, the principled question is how measurement error affects the log likelihood for each observation $Y_{t_i}$ conditional on the previous one, $Y_{t_{i-1}}$. If the process were pure Brownian motion, with $b = c = 0$, and if there were no measurement error, then the distribution $Y_{t_i}|Y_{t_{i-1}}$ would be normal, with variance $2at$, with $t \coloneqq t_i - t_{i-1}$. The log likelihood would be $-0.5 \cdot (\ln 2at\tau + (Y_{t_i} - Y_{t_{i-1}})^2/2at)$, where $\tau = 2\pi$ is the circle constant. But if $Y_{t_i}$ and $Y_{t_{i-1}}$ are measured with normally distributed errors of variance $h^2 Y_{t_i}^2$ and $h^2 Y_{t_{i-1}}^2$, and if these errors are independent, then $Y_{t_i} - Y_{t_{i-1}}$ is measured with normal error of variance of approximately $h^2 Y_{t_i}^2 + h^2 Y_{t_{i-1}}^2 \approx 2h^2 Y_{t_i}^2$. If this error is in turn independent of the process, then it adds to the variance of measured $Y_{t_i}$ while preserving normality. The likelihood becomes $-0.5 \times [\ln(2at + 2h^2 Y_{t_i}^2)\tau + (Y_{t_i} - Y_{t_{i-1}})^2/(2at + 2h^2 Y_{t_i}^2)]$. Focusing on the second term, which contains the squared change, incorporating measurement error as described multiplies the term by $1/(1 + 2h^2 Y_{t_i}^2/2at)$. If we take $Y_{t_i}^2$ as a scale proxy for the rising variance $2at$, then we can take the ratio as roughly $1/(1+2h^2)$.

to support formal modeling.[33] The same goes for Africa, if distinguished from Eurasia.[34]

# 5 Estimation

Using ML, I fit the superexponential diffusion to the data. The modeling is dynamic: the log likelihood for each observation is computed with (43), conditioning on the previous observation. The reflecting and absorbing models ($f_{\pm\chi}^{*B}$) are estimated, and the better fit reported. Sticky models turn out not to be relevant because $\nu$ is estimated to be well outside [0,1]. Observations are weighted by the HYDE-based weights described above. In addition, to test the hypothesis of exponential growth, I fit the CEV. Recall that this imposes $s = 0$ in (33), the marker of constant, exogenous growth.

## 5.1 Estimates for GWP

I model four versions of the GWP series: with and without pre–10,000 BCE observations, with and without annual data after 1950. The latter variation bears explaining. If the diffusion model is correct, it requires no adjustment for variation in observation spacing. For example, fitting the model to a series with annual rather than decennial data after 1950 would not spuriously overweight the postwar era. The likelihood assigned to each observation would properly quantify its information content. That said, an assumption that may become suspect at high sampling frequency is that the series is Markovian. It is possible, for example, that since 1950, the series contains a trend from which GWP cyclically deviates and returns—in other words, that the true Markov process is hidden.[35] Modeling the annual data available starting in 1950 might mix some short-term, trend-reverting behavior into what is before then a sparser series. The negative association between level and subsequent change would downward-bias estimates of the long-term propensity for growth. Two parameters figure in the relevant term in the growth equation, $sY_t^{1+B}$; since this phenomenon would occur only in the modern, high observations, the downward bias might load mainly onto $B$. This motivates estimation with only decennial data after 1950 (including the 2019 observation).

Table 3 reports results for the four samples. In estimation, the primary parameters are not those in the SDE (33) for the variable $Y_t$—not $s$, $B$, $\delta$, and $\sigma$. Instead, the estimation parameters are more closely linked to the Feller/CIR SDE (38) for the related variable $X_t$. The three are $\ln a$, $b$, and $\nu = c/a - 1$. The final estimation parameter is an exponent, $\gamma$, which equals $-1/B$.[36] The superexponential SDE parameters are then derived and their standard errors computed with the delta method; all are also in Table 3.[37]

---

[33] McEvedy and Jones (1978) provide estimates only for 10,000, 9000, 6000, 5000 BCE and 1, 1000, and 1500 CE.
[34] Diamond (1997, ch. 10) argues that the predominately north-south orientation of the African land mass created natural ecological barriers to dissemination of innovations, which long isolated sub-Saharan Africa from Eurasia.
[35] Regarding GDP/capita in frontier economies, a stream of literature starting with Nelson and Plosser (1982) favors the unit-root view. Other research challenges it.
[36] Usually, the power series representations of the densities $f_{\pm\chi^2}$ in (39) and (40) converge fast enough, requiring computation of at most the few million terms around the peak term. For extremely large values of $\lambda x$, estimation is instead based on Hankel's asymptotic expansion of the modified Bessel function of the first kind (NIST, eq. 10.40.1) or, if $\nu \gg x\lambda$, the large-order approximation (NIST, eq. 10.41.1).
[37] The translation inverts the definitions (37): $s = \gamma^2 a(1 + \gamma\nu)$, $\delta = b\gamma$, $\sigma = \gamma\sqrt{2a}$. And $B = -1/\gamma$.

Results for the full series are in column 1. They put the depreciation rate of the productive potential of humanity at $-3.95 \times 10^{-5}$/year (standard error $1.15 \times 10^{-5}$). Such a low rate is implausible if read as applying to physical or human capital, but can make sense if the essential productive stock in the long run is ideas. The investment rate $s$ is also small, if with expected sign, at $1.49 \times 10^{-4}$ (s.e. $2.81 \times 10^{-5}$). Perhaps for most of history, lived near subsistence, our species invested slowly in knowledge. Or perhaps the *productivity* of such investment was low. Meanwhile, the scale effect $B$ is estimated at 0.518 (s.e. 0.0275). According to equation (29), the implied returns elasticity for investment of output in TFP is about 0.07, suggesting constant returns.

These results flow from the absorbing-barrier model, whose fit yields a higher likelihood than the reflecting model. Since the "barrier" here is in the first instance the zero boundary for the Feller/CIR variable $X_t$, and GWP is modeled as $Y_t = X_t^{-1/B}$, and since $B$ is estimated to be positive, $Y_t$ experiences an absorbing barrier at $+\infty$. In the mathematical universe of the fitted model, there is positive probability of explosion.

Column 1 of Table 3 includes results that interrogate this propensity for explosion. Given an initial GWP level $Y_0$, the probability that a sample path will *not* eventually explode is $F_\Gamma(Y_0^{-B}b/a\,; -\nu)$. (Appendix B.9.1 develops this and related formulas.) Setting $Y_0$ to the GWP value for 1 million BCE, \$0.05 billion, this probability is 0.978. Conditioning instead on the 2019 GWP of \$73.6 trillion, the probability of no eventual explosion is just $5 \times 10^{-139}$. Giving non-exploding sample paths an explosion date of $\infty$, the median wait till explosion is $-\left(\ln(1 - bY_0/aF_\Gamma^{-1}(0.5; -\nu))\right)/b$. In this case, starting from 2019's GWP, the median explosion date is 2060 (s.e. ~9 years). Strange as that result is, a likelihood ratio test rejects the exponentially growing CEV model as an alternative ($p = 0.000$).

The results are perhaps better conveyed graphically. As the estimates hold that $\delta < 0$ and $B > 0$, they correspond to the case discussed after (32), in which the starting value determines whether the explosive propensity overcomes the downdraft of depreciation. But in the stochastic model, both outcomes can eventuate from any starting point. Figure 7 illustrates this richness by plotting 99 rollouts of the full-sample model fit. The empirical GWP series is in red and the simulation rollouts in greys. One rollout attains sustained growth and explodes within a quarter million years. The rest never escape decay. This simulation accords with the analytically derived 0.978 chance of no eventual explosion.

On its face, this result says that the attainment of civilization—indeed, the survival of our genus to this point—was improbable. This tension between model and reality admits at least two explanations. One is that the model is roughly correct. If so, then history's defiance of this prediction of extinction illustrates the anthropic principle (Carter 1974). Conditioning on the fact that such a paper as this has been written, the retrospective odds of takeoff are high. The other explanation is that the model is substantially incomplete—which surely it is. The inability of the univariate model to capture the Malthusian equilibration between income per person and population appears to render it unrealistically unstable: notice *all* the sample paths diverge from observed history. And it may ask too much to fix the same parameter values before and after the evolution of language.

To better convey the model fit, Figure 8 changes the presentation in several ways. Instead of 99 rollouts from the starting point, 10,000 are run. Another 10,000 are run from the 2019 observation. For each bundle, the median as a function of time is depicted with a black line, while the 5th, 10th, etc., quantiles are marked with changes in shading. The time axis is now logarithmic in years till the median explosion date of 2060. Finally, to convey uncertainty in the model fit along with the *modeled* stochasticity, each rollout incorporates a different set of parameters, drawn randomly from a joint distribution. For this purpose, the estimates of the primary parameters, $\ln a$ , $b$, $\nu$, and $\gamma$ are assumed to be distributed multivariate normal, with the covariance produced by the ML estimator. In the figure, the distributions emerging from the initial and final observations clash.

Retaining only decennial observations after 1950 changes the results statistically but not qualitatively (Table 3, column 2). The estimate of the overall scale effect $B$ rises from 0.515 to 0.629 (s.e. 0.0407). This suggests that trend reversion in the annual data does downward-bias the estimates of the scale effect. The median explosion year moves from 2060 to an eye-popping 2041.

Next Deevey (1960)'s three pre–10,000 BCE observations of GWP are dropped (columns 3 and 4 of Table 3). The scale effect $B$ falls to 0.429 (s.e. 0.0416) with annual data after 1950, and to 0.552 (s.e. 0.0492) with decennial data. Under the latter estimates, the probability of no eventual explosion from the starting GWP value of \$1.6 billion in 10,000 BCE is a mere $1.63 \times 10^{-10}$, quite opposite the first estimate of 0.978. Conditioning on 2019 GWP, the results now put the median explosion at 2047 (s.e. 8 years).

I mildly prefer this last specification since it skirts the structural break of the birth of language and perhaps reduces misspecification by taking decennial observations after 1950. The scale effect estimate for GWP, 0.55, again implies approximately constant returns to investment of output into TFP (0.20, s.e. 0.18). Sample paths under the fit mimic history: see Figure 9 and Figure 10. In the diffusion plot in Figure 10, observed GWP largely stays between the 40th and 60th quantiles of the simulations. And the diffusion emerging from the last observation coheres with that emerging from the first.

## 5.2 Robustness and goodness of fit

While it is clear that GWP growth has on average risen with GWP, much of the data buttressing that generalization is uncertain. And errors in the ancient observations might distort the estimation of the probability and timing of explosion. To test for such sensitivity, I rerun the preferred regression after multiplying all GWP observations by $e^{-h_t}$, where $h_t$ is the HYDE uncertainty indicator. Recall that $h_t$ ranges from 1 in the earliest observations to .01 in the most modern. I do the same after multiplying by $e^{h_t}$. The results are not sensitive to these changes. The first one lowers the estimate of $B$ from 0.552 to 0.492 and defers the median explosion year from 2047 to 2050. The second yields 0.616 and 2043.

I also estimate against the De Long (1998) GWP series for 10,000 BCE–2000 CE in order to check for sensitivity to the method of interpolating GWP/capita. This change matters more, but not qualitatively. With the preferred series halted in 2000 for comparability, switching to the De Long data lowers the estimate of $B$ from 0.595 to 0.517. The median explosion, conditioning on the 2000 GWP value, shifts from 2031 to 2022.

A more cutting concern is about goodness of fit. The superexponential growth component of the model captures the first-order pattern of acceleration. The stochastic component captures how GWP oscillated around this longer term trend—but not fully, it turns out. I check goodness of fit as follows. If the best-fit model is correct, then the quantiles of the observations within the distributions predicted for them under the best fit should be i.i.d. uniform. I check the uniformity of these quantiles with the Kolmogorov-Smirnov (KS) test and check their independence by examining their serial correlation. As shown along the bottom of Table 3, the regressions with annual data after 1950 fail both tests. Those with decennial data after 1950 do pass the KS test, but perhaps only because shrinking the sample reduces power.

For the preferred regression (Table 3, col. 4), Figure 11 plots the quantiles that are the subject of these tests. While the distribution of the quantiles may be formally indistinguishable from uniform, the series indeed contains episodic patterns that violate the Markovian diffusion model. The data points for 1820, 1870, and 1913, the zenith of the industrial revolution, ascend steadily. Then, starting in 1960, the quantiles descend monotonically, indicating that GWP growth has failed to accelerate as much as predicted. While the univariate diffusion model captures important aspects of GWP history, it leaves much out.

Since Figure 11 is built from a fit of data through 2019, it effectively asks whether GWP history is surprising conditional on it having happened. That is appropriate for testing goodness of fit. With modifications, the figure can avoid this circularity and better address whether certain historical developments broke with the past. Figure 12 is like Figure 11 except that the quantile for each observation is computed after fitting the absorbing-boundary model only to *previous* observations. Also, like the diffusion plots in Figure 8 and Figure 10, the distributions are simulated while incorporating parameter uncertainty (with 10,000 rollouts and 10,000 time steps). Convergence often fails for the earliest observations, evidently because of small samples, but stabilizes around 1500. Figure 12 accentuates the impression left by Figure 11. Observations during the first century of the industrial revolution are high while those since 1960 are low.

## 5.3 Estimates for series other than GWP

As argued at the start of section 4, the univariate stochastic model is more suited to GWP than population and income per capita. Nevertheless, it is interesting to test it against the other variables. Estimates from the preferred sample—starting in 10,000 BCE, decennial observations after 1950—are in Table 4. Three new variables are modeled: world population; GWP/capita; and GDP/capita in the frontier economy of France. Diffusion and quantile plots constructed the same way as Figure 10 and Figure 11 are gathered in Figure 13.

In all cases, the likelihood is again maximized by assuming an absorbing barrier at $X_t = 0$ and taking $B > 0$, which maps that barrier to +∞ for $Y_t$. In all cases, therefore, the model perceives a positive probability of explosion. The results for population broadly resemble those for GWP, which is perhaps expected in light of the dominance of population growth in GWP growth for most of history. Compare Table 4's column 1 for GWP (which is copied from Table 3, column 4) with the next column. The estimates for $s$, $B$, and $\delta$ hardly differ statistically. As a result, the estimate of $B$ is about half the 1.03 attained via NLS in Kremer (1993, Table VI, col. 1). The stochasticity coefficient $\sigma$ is larger than for GWP, perhaps to accommodate the surprise of the global fertility decline. This surprise presumably also explains the deferment of the median explosion year to 2175 (s.e. 47.6).

The fits to GWP/capita and frontier GDP/capita are worse—see columns 3 and 4 of Table 4 and the bottom two-thirds of Figure 13. The superexponential tendency is now measured as greater, presumably because of the suddenness of the acceleration of these variables a few centuries ago. The exponent $B$ is estimated at a high 1.699 (s.e. 0.335) for GWP/capita and 0.945 (s.e. 0.262) for frontier GDP/capita. This helps explain why the median rollout under the best fits for both GWP/capita and frontier GDP/capita explodes by 6600 BCE.

## 6 On the meaning of singularities

The results just presented are at once unsurprising and implausible: unsurprising because the acceleration in the expansion of the human system over the very long term is recognized; implausible because of the conflict with the recent experience of relatively steady global growth, not to mention the laws of physics. What should we make of the conflicts?

It is easy to dismiss the outsider-model's implications for historical base rates or base distributions for growth changes—to conclude that a good model for the past tells us little about the future. After all, the progression of human affairs is complex and could follow different patterns in different eras. Yet to declare a complete structural break is to risk, in the wording of Occam's razor, multiplying entities without necessity. Aiming to harvest the most insight from the tension between outside and inside views, I will start from the claim that the outside view's superexponential dynamic remains relevant, then yield ground.

Johansen and Sornette (2001) observe that singularities in model are a sign that it is being stretched beyond the realm of state space for which it is appropriate. Beyond that realm, some factor once neglected no longer can be, like the fact that the speed of light is the same in all inertial reference frames. "Singularities are always mathematical idealisations of natural phenomena: they are not present in reality but foreshadow an important transition or change of regime." (Johansen and Sornette 2001, p. 479).

In GWP growth, what might be the factor once neglected that no longer can be? I see three candidates. First, the superexponential singularity is formally an artifact of passing to the infinitesimal limit, the very analytical step that makes possible the ordinary and stochastic calculi. In real economies, technological advance and reinvestment of output of capital do not occur as a continuous sequence of infinitesimal impulses, each instantaneously propagated. This is why finite-difference simulations of the abstract economies defined here do not in fact reach infinite output but terminate upon numerical overflow. With an adequate system for representing large numbers, the simulations could be continued for an arbitrarily long time. The infinitesimal approximation may hold well enough at a hypothetical growth rate of 1% per year, but perhaps not at 1% per hour.

The upshot of this concession, however, seems only to be that while GWP will not go to infinity, it will still get stupendously big. Does that suffice for realism? Conceivably. In the last century or so, human beings have built machines that solve challenges evolution took millions of years to solve: locomotion on land, flight, vision, calculation, communication. We have not made machines as good as us at formulating and carrying out effective plans in complex environments, which arguably is the Holy Grail of AI research. But we might yet. This would open major new production possibilities (Hanson 2001; Yudkowsky 2008; Bostrom 2014). Even more radically, if AI is doing the economic modeling a century from now, it may count the welfare of artificial minds in GWP. Their number would presumably dwarf the human population. As absurd as this scenario may sound, an AI revolution could be seen as a stage in the unfolding that began with talkative, toolmaking apes.

While it is impossible to dismiss such fantastic possibilities, we should not let them distract from other shortcomings in the models developed here. A second factor that becomes dangerous to ignore as the models diverge is the consumption side, that is, the allocation of output among consumption and investment in various inputs. The neoclassically inspired model in section 2 articulates a Cobb-Douglas production process, which implies a microtheory of optimizing agents on the production side. But while it gives flexibility to investment rates by modulating them isoelastically with the technology level, this formulation is fixed-coefficient in spirit. It is not grounded in a theory of optimizing agents that adjust to, and even foresee, radically changing circumstances. Fixed coefficients may do limited harm in a model converging to a steady state. They become more problematic in a system that diverges.

The third neglected factor in this paper's model links most directly to physics. The model does not recognize that the output of the terrestrial economy is effectively capped by the flow of negative entropy from the sun and the earth's interior (Georgescu-Roegen 1971; Daly 1977). Conventional modeling obscures the link between economic activity and entropy by taking production to extract flows of output from stocks of inputs while leaving the inputs unchanged, which is thermodynamically all but impossible. Depreciation is modeled as decoupled from how much the stocks are used in production, which still implies that production per se has no thermodynamic consequence. The limit imposed by nature is not theoretically inviolable since the locus of valuation of output is the mind of the consumer, and humans might have a taste for entropy; but that possibility seems academic.

In the multivariate model in section 2.1, the finitude of natural resources does in fact damp economic growth. The factor $R$ is fixed at one and assigned a Cobb-Douglas exponent of 0.1, so that the remaining non-technology inputs have diminishing returns to scale, at 0.9. But that does not prevent explosion. Might the relationship between the artificial and the natural be captured more realistically within this mathematical framework? The best that can be done is to introduce a combination of exogenous appreciation in natural resources and endogenous disinvestment in them, the entropic corrosion from economic activity. A simulation appears in Figure 14. The model and initial values are the same as the exploding case in Figure 3 except that the reinvestment coefficient $s_R$ and appreciation rate $\delta_R$ are changed from 0 and 0 to $-0.01$ and 0.001. Like population, the stock of resources $R$ is taken as initially plentiful, at 1. From there, the slow increase in $R$ (in green) hastens the explosion by about 1000 years. But the approach of explosion initiates a plummet in $R$, which quickly brings output ($Y$, in red) to zero. Making the new factor suffer endogenous harm deflects the explosive impulse downward.

The scenario is, one hopes, unrealistic, not least because of the second neglected factor, the response of agents to changing circumstances. Yet the augmented model suffices to demonstrate that a fully endogenous, accelerating-growth model need not generate infinities. Thus the presence of infinities in certain simplifications of such models—ones neglecting natural resource dynamics—is not a logical basis for dismissal of the whole class.

Sadly, natural resource depletion is not the only plausible route to Armageddon (Bostrom and Ćirković

2008; Ord 2020). Some categories of global catastrophic risk are ancient, such as pandemics, while others are made possible by modernity, including nuclear winter and, conceivably, an AI-sparked collapse. This may be the true message of the infinities: not that society will literally explode or implode, but that the human project is intrinsically unstable. When more broadly contemplating the human past and prospect, the traditional focus in growth theory on the steady state seems narrow.

## 7 Conclusion

Solow and Swan built the neoclassical model in the 1950s to explain the relative stability of frontier-economy development over previous decades. Coming after the Depression and during the Cold War, the model spoke to concerns of the day. It offered hope that Western economies could experience steadily rising prosperity. The models generated this prediction through a kind of humility, leaving determinants of major inputs outside the theory, and tending to endow the inputs with constant growth as a placeholder. When studying the very long term, removing those asymmetric restrictions seems more realistic as a matter of principle and evidence. In the very long run, surely all the conventional factors—population, capital, etc.—are endogenous to output and to each other. And fully endogenous models diverge, thereby easily explaining the acceleration of growth in the historical record. Perhaps this framework, along with the stochastic data generating that gives it econometric rigor, resonates more with concerns of our day: pandemics, the propensity of the human system to attack its ecological foundations, the benefits and dangers of artificial intelligence.

A throughline of this paper, motivated by the outside-view's interest in base rates, has been the search for a mathematical representation of history that balances parsimony and realism, if emphasizing parsimony. A four-parameter, univariate model that integrates scale effects, depreciation, and dynamic stochasticity can recognizably approximate GWP history since 10,000 BCE. The scale effect is estimated at 0.55 for GWP. This suggests a returns elasticity for investment in TFP of about zero over the very long term. Thus, while it may be that ideas have gotten harder to find in the last 50 years, there is not a clear trend over the longer sweep of history.

A model emphasizing parsimony deserves critique for realism; and a model with a coherent statistical foundation *supports* such critique, for it can quantify its own failings. The outsider model developed here struggles to reproduce the combination of a million years of near-stasis and a few millennia of explosion. Even when restricted to the more explosive phase, the model is surprised by the industrial revolution and the relatively slow, steady growth of recent decades—phenomena that more richly specified models can better explain, such as the "unified growth theory" of Galor and Weil (2000).

By the same token, the natural inside view of the growth prospect in the current century—that population growth will slow or halt and per-capita GWP growth will converge to a modest frontier-economy rate—deserves critique for neglecting the base rates revealing in the longer-term history. This paper's investigation suggests that a naïve inside view underestimates the propensity for instability in the human system.

## References

Acemoglu, Daron. 2003. "Labor- and Capital-Augmenting Technical Change." *Journal of the European Economic Association* 1 (1): 1–37. DOI: 10.1162/154247603322256756.

Aghion, Philippe, and Peter Howitt. 1992. "A Model of Growth through Creative Destruction." *Econometrica* 60 (2): 323–51. DOI: 10.3386/w3223.

Arrow, Kenneth J. 1962. "The Economic Implications of Learning by Doing." *Review of Economic Studies* 29 (3): 155–73. DOI: 10.2307/2295952.

Barro, Robert J., and Gary S. Becker. 1989. "Fertility Choice in a Model of Economic Growth." *Econometrica* 57 (2): 481–501. DOI: 10.2307/1912563.

Barro, Robert J., and Jong-Wha Lee. 2000. "International Data on Educational Attainment Updates and Implications." Working Paper Series. National Bureau of Economic Research. DOI: 10.3386/w7911.

Barro, Robert J., and Xavier Sala-i-Martin. 2004. *Economic Growth*. MIT Press.

Bartlett, Maurice S. 1952. "Approximate Confidence Intervals. II. More than One Unknown Parameter." *Biometrika* 40(3/4): 306–17. DOI: 10.2307/2333349.

Becker, Gary S., Kevin M. Murphy, and Robert Tamura. 1990. "Human Capital, Fertility, and Economic Growth." *Journal of Political Economy* 98 (5): S12–37. DOI: 10.1086/261723.

Bennett, Merrill Kelly. 1954. *The World's Food: A Study of the Interrelations of World Populations, National Diets, and Food Potentials*. Harper & Brothers.

Biraben, Jean-Noël. 1979. "Essai Sur L'évolution du Nombre des Hommes." *Population* 34 (1): 13–25. DOI: 10.2307/1531855.

Blaxter, Kenneth Lyon. 1986. *People, Food and Resources*. Cambridge University Press.

Bloom, Nicholas, Charles I. Jones, John Van Reenen, and Michael Webb. 2020. "Are Ideas Getting Harder to Find?" *American Economic Review*. DOI: 10.1257/aer.20180338.

Bolt, Jutta, and Jan Luiten van Zanden. 2014. "The Maddison Project: Collaborative Research on Historical National Accounts." *Economic History Review* 38 (March). DOI: 10.1111/1468-0289.12032.

Bolt, Jutta, Robert Inklaar, Herman de Jong, and Jan Luiten van Zanden. 2018. "Rebasing 'Maddison': New Income Comparisons and the Shape of Long-Run Economic Development." rug.nl/ggdc/html_publications/memorandum/gd174.pdf.

Boserup, Ester. 1965. *The Conditions of Agricultural Growth*. Aldine Publishing Company.

Bostrom, Nick. 2014. *Superintelligence*. Oxford University Press.

Bostrom, Nick, and Milan M. Ćirković, 2008. "Introduction," in Nick Bostrom and Milan M. Ćirković, eds., *Global Catastrophic Risks*. Oxford University Press.

Caldwell, John C., and Thomas Schindlmayr. 2002. "Historical Population Estimates: Unraveling the Consensus." *Population and Development Review* 28 (2): 183–204. DOI: 10.1111/j.1728-4457.2002.00183.x.

Carr-Saunders, Alexander Morris. 1936. *World Population: Past Growth and Present Trends*. Clarendon Press.

Carter, Brandon. 1974. "Large Number Coincidences and the Anthropic Principle in Cosmology." *Symposium -*

*International Astronomical Union*. DOI: 10.1017/s0074180900235638.

Cipolla, Carlo. 1962. *The Economic History of World Population*. 5th ed. Pelican.

Clark, Colin. 1977. *Population Growth and Land Use*. 2nd ed. St. Martin's Press.

Cohen, Joel E. 1995. *How Many People Can the Earth Support?* W. W. Norton.

Cox, John C. 1996. "The Constant Elasticity of Variance Option Pricing Model." *Journal of Portfolio Management* 23 (5): 15–17. DOI: 10.3905/jpm.1996.015.

Cox, John C., Jonathan E. Ingersoll, and Stephen A. Ross. 1985. "An Intertemporal General Equilibrium Model of Asset Prices." *Econometrica* 53 (2): 363–84. DOI: 10.2307/1911241.

Cox, John C., and Stephen A. Ross. 1976. "The Valuation of Options for Alternative Stochastic Processes." *Journal of Financial Economics* 3 (1): 145–66. DOI: 10.1016/0304-405X(76)90023-4.

Crafts, Nicholas. 2018. "The Productivity Slowdown: Is It the 'New Normal'?" *Oxford Review of Economic Policy* 34 (3): 443–60. DOI: 10.1093/oxrep/gry001.

Daly, Herman. 1977. *Steady-State Economics*. W.H. Freeman.

Deevey, Edward S., Jr. 1960. "The Human Population." *Scientific American* 203 (September):195–204. DOI: 10.1038/scientificamerican0960-194.

De Long, J. Bradford. 1998 "Estimates of World GDP, One Million B.C.–Present." delong.typepad.com/print/20061012_LRWGDP.pdf.

Diamond, Jared. 1997. *Guns, Germs, and Steel*. W.W. Norton.

Doepke, Matthias. 2004. "Accounting for Fertility Decline during the Transition to Growth." *Journal of Economic Growth* 9 (3): 347–83. DOI: 10.1023/B:JOEG.0000038935.84627.e4.

Dolgonosov, Boris M. 2016. "Knowledge Production and World Population Dynamics." *Technological Forecasting and Social Change* 103: 127–41. DOI: 10.1016/j.techfore.2015.10.023.

Drandakis, E. M., and E. S. Phelps. 1966. "A Model of Induced Invention, Growth and Distribution." *Economic Journal* 76 (304): 823–40. DOI: 10.2307/2229086.

Durand, John D. 1967. "The Modern Expansion of World Population." *Proceedings of the American Philosophical Society* 111 (3): 136–59. jstor.org/stable/985711.

Durand, John D. 1977. "Historical Estimates of World Population: An Evaluation." *Population and Development Review* 3 (3): 253–96. DOI: 10.2307/1971891.

Engelbert, Hans-Jürgen, and Goran Peskir. 2014. "Stochastic Differential Equations for Sticky Brownian Motion." *Stochastics* 86 (6): 993–1021. DOI: 10.1080/17442508.2014.899600.

Feller, William. 1951a. "Diffusion Processes in Genetics." In *Proceedings of the Second Berkeley Symposium on Mathematical Statistics and Probability*. The Regents of the University of California. projecteuclid.org/euclid.bsmsp/1200500231.

Feller, William. 1951b. "Two Singular Diffusion Problems." *Annals of Mathematics* 54 (1): 173–82. DOI: 10.2307/1969318.

Fernández-Villaverde, Jesús. 2001. "Was Malthus Right? Economic Growth and Population Dynamics."

sas.upenn.edu/~jesusfv/pennversion.pdf.

Flyvberg, Bent. 2006. "From Nobel Prize to Project: Management: Getting Risks Right." *Project Management Journal* 37 (3): 5–15. DOI: 10.1177/875697280603700302.

Galor, Oded. 2012. "The Demographic Transition: Causes and Consequences." *Cliometrica* 6 (1): 1–28. DOI: 10.1007/s11698-011-0062-7.

Galor, Oded, and Omer Moav. 2000. "Ability-Biased Technological Transition, Wage Inequality, and Economic Growth." *Quarterly Journal of Economics* 115 (2): 469–97. DOI: 10.1162/003355300554827.

Galor, Oded, and David N. Weil. 2000. "Population, Technology, and Growth: From Malthusian Stagnation to the Demographic Transition and Beyond." *American Economic Review* 90 (4): 806–28. DOI: 10.1257/aer.90.4.806.

Georgescu-Roegen, Nicholas. 1971. *The Entropy Law and the Economic Process*. Harvard University Press.

Godwin, William. 1820. *Of Population: An Enquiry Concerning the Power of Increase in the Numbers of Mankind, Being an Answer to Mr. Malthus's Essay on That Subject*. Longman. books.google.com/?id=5HzQAAdx6DsC.

Göing-Jaeschke, Anja, and Marc Yor. 2003. "A Survey and Some Generalizations of Bessel Processes." *Bernoulli* 9 (2): 313–49. DOI: 10.3150/bj/1068128980.

Goodfriend, Marvin, and John McDermott. 1995. "Early Development." *American Economic Review* 85 (1): 116–33. jstor.org/stable/2117999.

Green, John. 1960. "Growth Models, Capital and Stability." *Economic Journal* 70 (277): 57–73. DOI: 10.2307/2227482.

Grossman, Gene M., and Elhanan Helpman. 1991. "Quality Ladders in the Theory of Growth." *Review of Economic Studies* 58 (1): 43–61. DOI: 10.2307/2298044.

Growiec, Jakub. 2007. "Beyond the Linearity Critique: The Knife-Edge Assumption of Steady-State Growth." *Economic Theory* 31 (3): 489–99. DOI: 10.1007/s00199-006-0107-5.

Hansen, Gary D., and Edward C. Prescott. 2002. "Malthus to Solow." *American Economic Review* 92 (4): 1205–17. DOI: 10.1257/00028280260344731.

Hanson, Robin. 2000a. "Long-Term Growth as a Sequence of Exponential Modes." Revised December 2000. citeseerx.ist.psu.edu/viewdoc/summary?doi=10.1.1.70.7068.

Hanson, Robin. 2001. "Economic growth given machine intelligence." University of California, Berkeley. mason.gmu.edu/~rhanson/aigrow.pdf.

Hawks, John, Keith Hunley, Sang-Hee Lee, and Milford Wolpoff. 2000. "Population Bottlenecks and Pleistocene Human Evolution." *Molecular Biology and Evolution* 17 (1): 2–22. DOI: 10.1093/oxfordjournals.molbev.a026233.

Hazan, Moshe, and Binyamin Berdugo. 2002. "Child Labour, Fertility, and Economic Growth." *Economic Journal* 112 (482): 810–28. DOI: 10.1111/1468-0297.00066.

Hurn, Aubrey, Jospeh Jeisman, and Kenneth Lindsay. 2007. "Seeing the Wood for the Trees: A Critical Evaluation of Methods to Estimate the Parameters of Stochastic Differential Equations." *Journal of Financial*

*Econometrics* 5 (3): 390–455. DOI: 10.1093/jjfinec/nbm009.

Huxley, Julian. 1950. "Population and Human Destiny." *Harper's Magazine*. September.

International Monetary Fund (IMF). 2020. World Economic Outlook Database, April 2020. imf.org/external/pubs/ft/weo/2020/01/weodata/index.aspx.

Itô, Kiyosi, and Henry P. McKean. 1965. *Diffusion Processes and their Sample Paths*. Springer-Verlag.

Jeanblanc, Moniqe, Marc Yor, and Marc Chesney. 2009. *Mathematical Methods for Financial Markets*. Springer.

Johansen, Anders, and Didier Sornette. 2001. "Finite-Time Singularity in the Dynamics of the World Population, Economic and Financial Indices." *Physica A: Statistical Mechanics and Its Applications* 294 (3): 465–502. DOI: 10.1016/S0378-4371(01)00105-4.

Jones, Charles I. 1995. "R&D–Based Models of Economic Growth." *Journal of Political Economy* 103 (4): 759–84. DOI: 10.1086/262002.

Jones, Charles I. 2001. "Was an Industrial Revolution Inevitable? Economic Growth Over the Very Long Run." *Advances in Macroeconomics* 1 (2): 470. DOI: 10.2202/1534-6013.1028.

Jones, Charles I. 2003. "Population and Ideas: A Theory of Endogenous Growth," in Philippe Aghion et al., eds., *Knowledge, Information, and Expectations in Modern Macroeconomics: In Honor of Edmunds S. Phelps.* Princeton University Press.

Jones, Charles I., and Paul M. Romer. 2010. "The New Kaldor Facts: Ideas, Institutions, Population, and Human Capital." *American Economic Journal: Macroeconomics* 2 (1): 224–45. DOI: 10.2307/25760291.

Kahneman, Daniel, and Dan Lovallo. 1993. "Timid Choices and Bold Forecasts: A Cognitive Perspective on Risk Taking." *Management Science* 39 (1): 17–31. DOI: 10.1287/mnsc.39.1.17.

Kahneman, Daniel, and Amos Tversky. 1979. "Intuitive prediction: Biases and Corrective Procedures." In Spyros G. Makridakis and Steven C. Wheelwright, eds. *Studies in the Management Sciences: Forecasting* 12. North Holland.

Kapitza, Sergei P. 1996. "The Phenomenological Theory of World Population Growth." *Physics-Uspekhi* 39 (1): 57. DOI: 10.1070/PU1996v039n01ABEH000127.

Kaulakys, Bronislovas, Vygintas Gontis, and Miglius Alaburda. 2005. "Point Process Model of 1/f Noise versus a Sum of Lorentzians." *Physical Review E* 71 (5 Pt 1): 051105. DOI: 10.1103/PhysRevE.71.051105.

Kennedy, Charles. 1964. "Induced Bias in Innovation and the Theory of Distribution." *Economic Journal* 74 (295): 541–47. DOI: 10.2307/2228295.

Kimmel, Marek, and David E. Axelrod. 2015. *Branching Processes in Biology*. Springer.

Klein Goldewijk, Kees, Arthur Beusen, Jonathan Doelman, and Elke Stehfest. 2017. "Anthropogenic Land Use Estimates for the Holocene – HYDE 3.2." *Earth System Science Data* 9: 927–53. DOI: 10.5194/essd-9-927-2017.

Kögel, Tomas, and Alexia Prskawetz. 2001. "Agricultural Productivity Growth and Escape from the Malthusian Trap." *Journal of Economic Growth* 6 (4): 337–57. DOI: 10.1023/A:1012742531003.

Korotayev, Andrey. 2007. "Compact Mathematical Models of World-System Development: How They Can Help

Us to Clarify Our Understanding of Globalization Processes." In *Globalization as Evolutionary Process*, 153–80. Routledge. DOI: 10.4324/9780203937297.

Kremer, Michael. 1993. "Population Growth and Technological Change: One Million B.C. to 1990." *Quarterly Journal of Economics* 108 (3): 681–716. DOI: 10.2307/2118405.

Kuznets, Simon. 1960. "Population Change and Aggregate Output." In George B. Roberts, Chairman, Universities-National Bureau Committee for Economic Research, *Demographic and Economic Change in Developed Countries*. Columbia University Press. nber.org/chapters/c2392.

Lagerlöf, Nils-Petter. 2003a. "From Malthus to Modern Growth: Can Epidemics Explain the Three Regimes?" *International Economic Review* 44 (2): 755–77. DOI: 10.1111/1468-2354.t01-1-00088.

Lagerlöf, Nils-Petter. 2003b. "Mortality and Early Growth in England, France and Sweden." *Scandinavian Journal of Economics* 105 (3): 419–40. DOI: 10.1111/1467-9442.t01-2-00006.

Laitner, John. 2000. "Structural Change and Economic Growth." *Review of Economic Studies* 67 (3): 545–61. DOI: 10.1111/1467-937X.00143.

Lee, Ronald D. 1988. "Induced Population Growth and Induced Technological Progress: Their Interaction in the Accelerating Stage." *Mathematical Population Studies* 1 (3): 265–88, 317. DOI: 10.1080/08898488809525278.

Livi-Bacci, Massimo. 2007. *A Concise History of World Population*. 4th ed. Wiley-Blackwell.

Lucas, Robert E. Jr. 2000. Some Macroeconomics for the 21st Century. *Journal of Economics Perspectives* 14 (1): 159–68. DOI: 10.1257/jep.14.1.159.

Maddison, Angus. 1995. *Monitoring the World Economy*. Paris: OECD Development Centre.

Maddison, Angus. 2001. *The World Economy: A Millennial Perspective*. Paris: OECD Development Centre.

Maddison, Angus. 2003. *The World Economy: Historical Statistics*. Paris: OECD Development Centre.

Maddison, Angus. 2010. "Statistics on World Population, GDP and Per Capita GDP, 1–2008 AD." ggdc.net/maddison/oriindex.htm.

Mankiw, N. Gregory, David Romer, and David N. Weil. 1992. "A Contribution to the Empirics of Economic Growth." *Quarterly Journal of Economics* 107 (2): 407–37. DOI: 10.2307/2118477.

McEvedy, Colin, and Richard Jones. 1978. *Atlas of World Population History*. arabgeographers.net/up/uploads/14299936761.pdf.

Meyer, Carl D. 2010. *Matrix Analysis and Applied Linear Algebra*. Society for Industrial and Applied Mathematics.

Meyer, François. 1947. *L'accélération Évolutive: Essai sur le Rythme Évolutif et son Interprétation Quantique*. Librairie des Sciences et des Arts.

Meyer, François. 1974. *La Surchauffe de la Croissance: Essai sur la Dynamique de L'évolution*. Fayard.

Molchanov, Stanislav. A. 1967. "Martin Boundaries for Invariant Markov Processes on a Solvable Group." *Theory of Probability and Its Applications* 12 (2): 310–14. DOI: 10.1137/1112036.

Nelson, Charles R., and Charles R. Plosser. 1982. "Trends and Random Walks in Macroeconomic Time Series: Some Evidence and Implications." *Journal of Monetary Economics* 10 (2): 139–62. DOI: 10.1016/0304-

3932(82)90012-5.

Nordhaus, William D. 1997. "Traditional Productivity Estimates Are Asleep at the (Technological) Switch." *Economic Journal* 107 (444): 1548–59. DOI: 10.1111/j.1468-0297.1997.tb00065.x.

Nuño, Galo, and Benjamin Moll. 2018. "Social Optima in Economies with Heterogeneous Agents." *Review of Economic Dynamics* 28 (April): 150–80. DOI: 10.1016/j.red.2017.08.003.

Oksendal, Bernt. 2014. *Stochastic Differential Equations: An Introduction with Applications*. 6th ed. Springer.

Ord, Toby. 2020. *The Precipice: Existential Risk and the Future of Humanity*. Hachette Books.

Peskir, Goran. Forthcoming. "Sticky Bessel Diffusions." University of Manchester.

Peskir, Goran, and David Roodman. Forthcoming. "Sticky Feller Diffusions." University of Manchester.

Putnam, Palmer C. 1953. *Energy in the Future*. Van Nostrand.

Ridolfi, Leonardo. 2017. "The French Economy in the Longue Durée. A Study on Real Wages, Working Days and Economic Performance from Louis IX to the Revolution (1250-1789)." IMT school for Advanced Studies Lucca. DOI: 10.6092/imtlucca/e-theses/211.

Rivera-Batiz, Luis A., and Paul M. Romer. 1991. "Economic Integration and Endogenous Growth." *Quarterly Journal of Economics* 106 (2): 531–55. DOI: 10.2307/2937946.

Romer, Paul M. 1986. "Increasing Returns and Long-Run Growth." *Journal of Political Economy* 94 (5): 1002–37. DOI: 10.1086/261420.

Romer, Paul M. 1990. "Endogenous Technological Change." *Journal of Political Economy* 98 (5, Part 2): S71–102. DOI: 10.1086/261725.

Simon, Julian Lincoln. 1977. *The Economics of Population Growth*. Princeton University Press.

Simon, Julian Lincoln. 1986. *Theory of Population and Economic Growth*. Basil Blackwell.

Solow, Robert M. 1956. "A Contribution to the Theory of Economic Growth." *Quarterly Journal of Economics*. DOI: 10.2307/1884513.

Solow, Robert M. 1957. "Technical Change and the Aggregate Production Function." *Review of Economics and Statistics* 39 (3): 312–20. DOI: 10.2307/1926047.

Solow, Robert. 2000. *Growth Theory: An Exposition*. Oxford University Press.

Swan, Trevor W. 1956. "Economic Growth and Capital Accumulation." *Economic Record* 32 (2): 334–61. DOI: 10.1111/j.1475-4932.1956.tb00434.x.

Szathmáry, Eörs. 2015. "Toward Major Evolutionary Transitions Theory 2.0." *Proceedings of the National Academy of Sciences* 112 (33): 10104–11. DOI: 10.1073/pnas.1421398112.

Tamura, Robert. 2002. "Human Capital and the Switch from Agriculture to Industry." *Journal of Economic Dynamics & Control* 27 (2): 207–42. DOI: 10.1016/S0165-1889(01)00032-X.

Tetlock, Philip, and Dan Gardner. 2015. *Superforecasting: The Art and Science of Prediction*. Crown Publishers.

United Nations (UN). 2019. *World Population Prospects*. Accessed December 2, 2019. population.un.org/wpp/Download/Standard/Population.

Uzawa, Hirofumi. 1961. "Neutral Inventions and the Stability of Growth Equilibrium." *Review of Economic Studies*

28 (2): 117–24. DOI: 10.2307/2295709.

van Handel, Ramon. 2007. “Stochastic Calculus, Filtering, and Stochastic Control.” Lecture notes. web.math.princeton.edu/~rvan/acm217/ACM217.pdf

Varfolomeyev, S. D., and K. G. Gurevich. 2001. “The Hyperexponential Growth of the Human Population on a Macrohistorical Scale.” *Journal of Theoretical Biology* 212 (3): 367–72. DOI: 10.1006/jtbi.2001.2384.

Von Foerster, Heinz, Patricia M. Mora, and Lawrence W. Amiot. 1960. “Doomsday: Friday, 13 November, A.D. 2026.” *Science* 132 (3436): 1291–95. DOI: 10.1126/science.132.3436.1291.

Weiss, Kenneth. M. 1984. “On the Number of Members of the Genus Homo Who Have Ever Lived, and Some Evolutionary Implications.” *Human Biology* 56 (4): 637–49. jstor.org/stable/41463610.

World Bank. 2019. *World Development Indicators*. databank.worldbank.org/reports.aspx?source=world-development-indicators.

Woytinsky, Wladimir. S., and Emma S. Woytinsky. 1953. *World Population and Production: Trends and Outlook*. Twentieth Century Fund.

Yamada, Toshio, and Shinzo Watanabe. 1971. “On the Uniqueness of Solutions of Stochastic Differential Equations.” *Journal of Mathematics of Kyoto University* 11 (1): 155–67. DOI: 10.1215/kjm/1250523691.

Yudkowsky, Eliezer. 2008. “Artificial Intelligence as a Positive and Negative Factor in Global Risk.” In Nick Bostrom and Milan M. Ćirković, eds., *Global Catastrophic Risks*. Oxford University Press. intelligence.org/files/AIPosNegFactor.pdf.

**Table 1. Parameter values in simulated Cobb-Douglas economy**

| Input | Parameter | Value | Notes and sources |
|---|---|---|---|
| Technology ($A$) | $\alpha_A$ | 1 | Factor-neutrality |
| | $s_A$ | 0.025 | ~ global R&D/GDP (World Bank 2019, series GB.XPD.RSDV.GD.ZS) |
| | $\phi_A$ | −0.5 | Corresponds to $\phi = 0.5$ in Jones (2001, p.23) |
| | $\delta_A$ | −0.001 | Small but not 0 |
| Capital ($K$) | $\alpha_K$ | 0.3 | Mankiw, Romer, and Weil (1992, Table II, col. 1) |
| | $s_K$ | 0.25 | ~ global gross capital formation/GDP (World Bank 2019, series NE.GDI.TOTL.ZS) |
| | $\phi_K$ | 0 | |
| | $\delta_K$ | −0.03 | Mankiw, Romer, and Weil (1992, note 6) |
| Population ($P$) | $\alpha_P$ | 0.3 | Chosen so $\alpha_P + \alpha_H = 0.6$, close to typical value for labor share in GDP |
| | $s_P$ | 0.2 | ~ global health spending/GDP (World Bank 2019, series SH.XPD.CHEX.GD.ZS) after doubling to add nutrition, etc. |
| | $\phi_P$ | −0.1 | Sign from Galor (2012) finding that technological advance shifts investment from child quantity to quality; magnitude arbitrary |
| | $\delta_P$ | −0.02 | Corresponds to life expectancy of 50 years |
| Human capital ($H$) | $\alpha_H$ | 0.3 | Mankiw, Romer, and Weil (1992, Table II, col. 1) |
| | $s_H$ | 0.04 | ~ global education spending/GNI (World Bank 2019, series NY.ADJ.AEDU.GN.ZS) |
| | $\phi_H$ | 0.1 | $= -\phi_P$ |
| | $\delta_H$ | −0.02 | $= \delta_P$; Barro and Lee (2000, eq. 5) also equates population and human capital attrition |
| Natural resources ($R$) | $\alpha_R$ | 0.1 | |
| | $s_R$ | 0 | Held fixed |
| | $\phi_R$ | 0 | |
| | $\delta_R$ | 0 | |

Note: $\alpha$ parameters are exponents in Cobb-Douglas production of output. $s$ parameters are investment rates of output into inputs. $\phi$ parameters are elasticities of this investment to technology. $\delta$ parameters are exogenous appreciation/depreciation rates.

**Table 2. Preferred estimates of population and gross world product, 1 million BCE–2019 CE**

| Year | Population (million) | GWP/ capita (1990 $) | GWP (billion 1990 $) | GDP/ capita, France (2011 $) |
|---|---|---|---|---|
| 1000000 BCE | 0.125 | 400 | 0.05 | |
| 300000 BCE | 1.0 | 400 | 0.40 | |
| 25000 BCE | 3.34 | 400 | 1.34 | |
| 10000 BCE | 4 | 400 | 1.60 | 658 |
| 5000 BCE | 5 | 404 | 2.02 | 664 |
| 4000 BCE | 7 | 409 | 2.87 | |
| 3000 BCE | 14 | 421 | 5.90 | |
| 2000 BCE | 27 | 433 | 11.7 | |
| 1000 BCE | 50 | 444 | 22.2 | |
| 500 BCE | 100 | 457 | 45.7 | |
| 200 BCE | 150 | 465 | 69.7 | |
| 1 CE | 168 | 467 | 78.4 | 1,050 |
| 200 CE | 190 | 463 | 88.0 | |
| 400 CE | 190 | 463 | 88.0 | |
| 500 CE | 190 | 463 | 88.0 | |
| 600 CE | 200 | 462 | 92.3 | |
| 700 CE | 210 | 460 | 96.6 | |
| 800 CE | 220 | 459 | 101 | |
| 900 CE | 240 | 456 | 109 | |
| 1000 CE | 265 | 453 | 120 | |
| 1100 CE | 320 | 512 | 164 | |
| 1200 CE | 360 | 551 | 198 | |
| 1300 CE | 360 | 551 | 198 | 1,469 |
| 1400 CE | 350 | 541 | 190 | 1,853 |
| 1500 CE | 438 | 625 | 274 | 1,748 |
| 1600 CE | 556 | 629 | 350 | 1,661 |
| 1700 CE | 603 | 658 | 397 | 1,748 |
| 1820 CE | 1,042 | 712 | 741 | 1,867 |
| 1870 CE | 1,276 | 884 | 1,128 | 3,086 |
| 1900 CE | 1,563 | | | |
| 1913 CE | 1,793 | 1,543 | 2,767 | 5,733 |
| 1920 CE | 1,863 | | | |
| 1940 CE | 2,299 | 2,181 | 5,013 | 6,650 |
| 1950 CE | 2,528 | 2,104 | 5,318 | 8,531 |
| 1951 CE | 2,572 | 2,191 | 5,635 | 8,984 |
| 1952 CE | 2,618 | 2,250 | 5,891 | 9,154 |
| 1953 CE | 2,666 | 2,320 | 6,185 | 9,351 |
| 1954 CE | 2,717 | 2,353 | 6,393 | 9,731 |
| 1955 CE | 2,769 | 2,457 | 6,804 | 10,198 |
| 1956 CE | 2,823 | 2,524 | 7,125 | 10,608 |
| 1957 CE | 2,880 | 2,567 | 7,394 | 11,124 |
| 1958 CE | 2,939 | 2,596 | 7,631 | 11,277 |
| 1959 CE | 2,996 | 2,665 | 7,984 | 11,481 |
| 1960 CE | 3,042 | 2,764 | 8,407 | 12,170 |
| 1961 CE | 3,082 | 2,821 | 8,695 | 12,698 |
| 1962 CE | 3,136 | 2,902 | 9,101 | 13,271 |
| 1963 CE | 3,201 | 2,965 | 9,492 | 13,757 |
| 1964 CE | 3,266 | 3,118 | 10,184 | 14,509 |
| 1965 CE | 3,333 | 3,218 | 10,725 | 15,078 |
| 1966 CE | 3,402 | 3,326 | 11,315 | 15,701 |
| 1967 CE | 3,471 | 3,381 | 11,738 | 16,298 |
| 1968 CE | 3,543 | 3,494 | 12,379 | 16,890 |
| 1969 CE | 3,616 | 3,613 | 13,063 | 17,908 |
| 1970 CE | 3,691 | 3,725 | 13,751 | 18,771 |
| 1971 CE | 3,770 | 3,797 | 14,315 | 19,486 |
| 1972 CE | 3,846 | 3,901 | 15,004 | 20,175 |
| 1973 CE | 3,923 | 4,081 | 16,009 | 21,097 |
| 1974 CE | 3,998 | 4,097 | 16,378 | 21,571 |
| 1975 CE | 4,071 | 4,086 | 16,634 | 21,316 |
| 1976 CE | 4,141 | 4,213 | 17,448 | 22,154 |
| 1977 CE | 4,214 | 4,308 | 18,154 | 22,888 |
| 1978 CE | 4,286 | 4,422 | 18,953 | 23,427 |
| 1979 CE | 4,363 | 4,500 | 19,632 | 24,075 |
| 1980 CE | 4,440 | 4,511 | 20,026 | 24,292 |
| 1981 CE | 4,515 | 4,523 | 20,419 | 24,413 |
| 1982 CE | 4,587 | 4,500 | 20,644 | 24,893 |
| 1983 CE | 4,676 | 4,539 | 21,226 | 25,080 |
| 1984 CE | 4,757 | 4,668 | 22,201 | 25,305 |
| 1985 CE | 4,838 | 4,748 | 22,967 | 25,549 |
| 1986 CE | 4,921 | 4,832 | 23,779 | 26,047 |
| 1987 CE | 5,007 | 4,932 | 24,692 | 26,581 |
| 1988 CE | 5,093 | 5,056 | 25,753 | 27,621 |
| 1989 CE | 5,181 | 5,133 | 26,592 | 28,460 |
| 1990 CE | 5,269 | 5,149 | 27,133 | 29,031 |
| 1991 CE | 5,352 | 5,141 | 27,517 | 29,158 |
| 1992 CE | 5,436 | 5,163 | 28,065 | 29,449 |
| 1993 CE | 5,518 | 5,201 | 28,702 | 29,113 |
| 1994 CE | 5,599 | 5,305 | 29,705 | 29,656 |
| 1995 CE | 5,682 | 5,443 | 30,927 | 30,135 |
| 1996 CE | 5,762 | 5,547 | 31,962 | 30,415 |
| 1997 CE | 5,842 | 5,688 | 33,229 | 30,987 |
| 1998 CE | 5,921 | 5,718 | 33,855 | 31,941 |
| 1999 CE | 6,000 | 5,850 | 35,099 | 32,872 |
| 2000 CE | 6,077 | 6,057 | 36,806 | 33,967 |
| 2001 CE | 6,155 | 6,161 | 37,918 | 34,434 |
| 2002 CE | 6,232 | 6,303 | 39,281 | 34,618 |
| 2003 CE | 6,308 | 6,526 | 41,167 | 34,707 |
| 2004 CE | 6,374 | 6,782 | 43,228 | 35,465 |
| 2005 CE | 6,463 | 7,001 | 45,249 | 35,817 |
| 2006 CE | 6,544 | 7,276 | 47,610 | 36,439 |
| 2007 CE | 6,625 | 7,504 | 49,711 | 37,068 |
| 2008 CE | 6,707 | 7,626 | 51,148 | 36,928 |
| 2009 CE | 6,790 | 7,478 | 50,775 | 35,642 |
| 2010 CE | 6,873 | 7,814 | 53,704 | 36,141 |
| 2011 CE | 6,956 | 8,051 | 56,003 | 36,691 |
| 2012 CE | 7,040 | 8,234 | 57,969 | 36,571 |
| 2013 CE | 7,124 | 8,422 | 59,994 | 36,632 |
| 2014 CE | 7,207 | 8,623 | 62,147 | 36,527 |
| 2015 CE | 7,291 | 8,821 | 64,310 | 36,827 |
| 2016 CE | 7,374 | 9,021 | 66,519 | 37,124 |
| 2017 CE | 7,457 | 9,266 | 69,093 | 37,863 |
| 2018 CE | 7,539 | 9,493 | 71,566 | 38,458 |
| 2019 CE | 7,620 | 9,663 | 73,640 | 38,911 |

Notes: Population figures: for before 10,000 BCE from Deevey (1960); for 10,000 BCE–1400 CE from McEvedy and Jones (1978); for 1500–2010 from Maddison (2010); and for 2011–18 from UN (2019, file POP/1-1). GWP/capita figures: through 10,000 BCE set to $400, the subsistence estimate of Maddison (2001, p. 260); for 1 CE, 1000 CE, and 1500–2010 from Maddison (2010), incorporating revisions for Western Europe in Bolt and Van Zanden (2014); for other years before 1500 geometrically interpolated with respect to population; and for 2011–18 extrapolated forward using growth rates from IMF (2020, series NGDP_RPCH). Money values in Geary-Kharnis dollars of 1990. Figures for France from Bolt et al. (2018), except that those for 10,000 BCE and 5000 BCE are the GWP/capita figures converted from dollars of 1990 to dollars of 2011 and those after 2016 extrapolated forward using growth rates from IMF (2020, series NGDPRPPPPCPCH).

**Table 3. Diffusion fits to GWP**

| | 1 million BCE–2019 CE | | 10,000 BCE–2019 CE | |
|---|---|---|---|---|
| | All observations | Only decennial data after 1950 | All observations | Only decennial data after 1950 |
| Primary estimation parameters | | | | |
| $\ln a$ | −13.45 | −12.62 | −13.33 | −12.66 |
| | (0.186) | (0.253) | (0.229) | (0.281) |
| $b$ | $2.05 \times 10^{-5}$ | $6.49 \times 10^{-6}$ | $1.66 \times 10^{-4}$ | $1.86 \times 10^{-5}$ |
| | $(5.19 \times 10^{-6})$ | $(3.50 \times 10^{-6})$ | $(7.19 \times 10^{-5})$ | $(6.87 \times 10^{-5})$ |
| $\nu$ | −51.75 | −12.31 | −93.80 | −23.78 |
| | (9.520) | (4.019) | (16.38) | (7.439) |
| $\gamma$ | −1.930 | −1.588 | −2.329 | −1.813 |
| | (0.103) | (0.103) | (0.226) | (0.162) |
| Boundary type | Absorbing | Absorbing | Absorbing | Absorbing |
| Accessible boundary location (determined by sign of $\gamma$) | $\infty$ | $\infty$ | $\infty$ | $\infty$ |
| Derived superexponential diffusion parameters | | | | |
| $s$ | $1.49 \times 10^{-4}$ | $7.28 \times 10^{-5}$ | $3.62 \times 10^{-4}$ | $1.47 \times 10^{-4}$ |
| | $(2.81 \times 10^{-5})$ | $(2.20 \times 10^{-5})$ | $(1.20 \times 10^{-4})$ | $(5.83 \times 10^{-5})$ |
| $B$ | 0.518 | 0.630 | 0.429 | 0.552 |
| | (0.0275) | (0.0407) | (0.0416) | (0.0492) |
| $\delta$ | $-3.95 \times 10^{-5}$ | $-1.03 \times 10^{-5}$ | $-3.86 \times 10^{-4}$ | $-3.37 \times 10^{-5}$ |
| | $(1.15 \times 10^{-5})$ | $(6.00 \times 10^{-6})$ | $(1.98 \times 10^{-4})$ | $(1.27 \times 10^{-4})$ |
| $\sigma$ | $3.27 \times 10^{-3}$ | $4.08 \times 10^{-3}$ | $4.19 \times 10^{-3}$ | $4.57 \times 10^{-3}$ |
| | $(4.35 \times 10^{-4})$ | $(6.61 \times 10^{-4})$ | $(8.35 \times 10^{-4})$ | $(9.21 \times 10^{-4})$ |
| $\phi_A$ (returns elasticity for TFP) | 0.0708 | 0.465 | −0.306 | 0.197 |
| | (0.106) | (0.133) | (0.196) | (0.179) |
| (Unstable) steady state (billion $) | 0.0771 | 0.0448 | 1.158 | 0.0690 |
| | (0.0246) | (0.0270) | (0.688) | (0.425) |
| P[no eventual explosion \| initial GWP] | 0.978 | 0.611 | 0.146 | $1.63 \times 10^{-10}$ |
| | (0.0718) | (0.506) | (0.554) | $(1.06 \times 10^{-8})$ |
| Median predicted explosion year \| initial GWP | $\infty$ | $\infty$ | 3620 | 1527 |
| | | | (8413) | (3263) |
| P[no eventual explosion \| final GWP] | $5.00 \times 10^{-139}$ | $7.32 \times 10^{-44}$ | $3.79 \times 10^{-154}$ | $8.36 \times 10^{-70}$ |
| | $(3.14 \times 10^{-137})$ | $(2.36 \times 10^{-42})$ | $(2.43 \times 10^{-152})$ | $(4.02 \times 10^{-68})$ |
| Median predicted explosion year \| final GWP | 2060 | 2041 | 2073 | 2047 |
| | (9.117) | (6.559) | (14.09) | (8.122) |
| Likelihood ratio test: CEV model (exponential growth) | | | | |
| $\chi^2(1)$ | 39.50 | 62.92 | 57.93 | 61.72 |
| $p$ | 0.000 | 0.000 | 0.000 | 0.000 |
| Goodness of fit: quantiles of observations in predicted distributions i.i.d. uniform | | | | |
| Kolmogorov-Smirnov ($p$) | 0.000 | 0.488 | 0.005 | 0.489 |
| No serial correlation ($p$) | 0.000 | 0.003 | 0.000 | 0.007 |
| Observations | 100 | 38 | 97 | 35 |

Notes: In columns 2 and 4, post-1950 observations are decennial rather than annual, except that the 2019 observation is retained. Results are maximum likelihood estimates using a diffusion model corresponding to the stochastic differential equation $dX_t = (bX_t + c)dt + \sqrt{2aX}dW_t$ with $Y_t = X_t^\gamma$ representing GWP and $\nu \coloneqq c/a - 1$. Estimates of the superexponential diffusion parameters $s, B, \delta$, and $\sigma$ are then derived according to footnote 37. $\phi_A = 2B - 1/2B$ is the estimated scale effect in spending on innovation. "Steady state" is the level of zero drift, $(-\delta/s)^{1/B}$. Formulas for the probability of no eventual explosion and median wait till explosion are in the text. Likelihood ratio test is for the restriction to the constant-elasticity-of-variance model, which has $s = 0$, for exponential growth. Kolmogorov-Smirnov test is for the hypothesis that the quantiles of observations in predicted distributions are uniformly distributed. Standard errors in parentheses; those for derived quantities computed with the delta method. Observations weighted for precision as described in section 4.

**Table 4. Diffusion fits to GWP, population, GWP/capita, and France GDP/capita series, 10,000 BCE–2019, decennial observations after 1950**

| | GWP (billion $) | Population (million people) | GWP/capita ($) | France GDP/capita ($) |
|---|---|---|---|---|
| Primary estimation parameters | | | | |
| $\ln a$ | −12.66 | −13.69 | −19.32 | −15.24 |
| | (0.281) | (0.308) | (1.910) | (1.739) |
| $b$ | $1.86 \times 10^{-5}$ | $2.09 \times 10^{-5}$ | $2.91 \times 10^{-4}$ | $2.82 \times 10^{-4}$ |
| | $(6.87 \times 10^{-5})$ | $(5.91 \times 10^{-5})$ | $(4.13 \times 10^{-4})$ | $(3.73 \times 10^{-4})$ |
| $\nu$ | −23.78 | −39.30 | −4.810 | −4.520 |
| | (7.439) | (12.06) | (3.967) | (3.572) |
| $\gamma$ | −1.813 | −1.793 | −0.589 | −1.058 |
| | (0.162) | (0.218) | (0.116) | (0.293) |
| Boundary type | Absorbing | Absorbing | Absorbing | Absorbing |
| Accessible boundary location (determined by sign of $\gamma$) | ∞ | ∞ | ∞ | ∞ |
| Derived superexponential diffusion parameters | | | | |
| $s$ | $1.47 \times 10^{-4}$ | $8.32 \times 10^{-5}$ | $1.30 \times 10^{-8}$ | $1.42 \times 10^{-6}$ |
| | $(5.83 \times 10^{-5})$ | $(3.98 \times 10^{-5})$ | $(3.56 \times 10^{-8})$ | $(3.28 \times 10^{-6})$ |
| $B$ | 0.552 | 0.558 | 1.699 | 0.945 |
| | (0.0492) | (0.0679) | (0.335) | (0.262) |
| $\delta$ | $-3.37 \times 10^{-5}$ | $-3.74 \times 10^{-5}$ | $-1.71 \times 10^{-4}$ | $-2.99 \times 10^{-4}$ |
| | $(1.27 \times 10^{-4})$ | $(1.09 \times 10^{-4})$ | $(2.71 \times 10^{-4})$ | $(4.32 \times 10^{-4})$ |
| $\sigma$ | $4.57 \times 10^{-3}$ | $2.69 \times 10^{-3}$ | $5.31 \times 10^{-5}$ | $7.34 \times 10^{-4}$ |
| | $(9.21 \times 10^{-4})$ | $(6.77 \times 10^{-4})$ | $(6.13 \times 10^{-5})$ | $(8.38 \times 10^{-4})$ |
| (Unstable) steady state (billion $) | 0.0690 | 0.238 | 266.3 | 287.6 |
| | (0.425) | (1.067) | (133.0) | (288.2) |
| P[no eventual explosion \| initial GWP] | $1.63 \times 10^{-10}$ | $1.51 \times 10^{-14}$ | 0.160 | 0.172 |
| | $(1.06 \times 10^{-8})$ | $(2.43 \times 10^{-13})$ | (0.310) | (0.334) |
| Median predicted explosion year \| initial GWP | 1527 | 1831 | 6822 BCE | 6674 BCE |
| | (3263) | (2695) | (1725) | (2005) |
| P[no eventual explosion \| final GWP] | $8.36 \times 10^{-70}$ | $6.62 \times 10^{-83}$ | $6.75 \times 10^{-12}$ | $3.31 \times 10^{-8}$ |
| | $(4.02 \times 10^{-68})$ | $(3.75 \times 10^{-81})$ | $(9.13 \times 10^{-11})$ | $(3.31 \times 10^{-7})$ |
| Median predicted explosion year \| final GWP | 2047 | 2175 | 2028 | 2065 |
| | (8.122) | (47.60) | (5.044) | (43.11) |
| Likelihood-ratio test: CEV model (exponential growth) | | | | |
| $\chi^2(1)$ | 61.72 | 41.64 | 44.02 | 5.780 |
| $p$ | 0.000 | 0.000 | 0.000 | 0.016 |
| Goodness of fit: quantiles of observations in predicted distributions i.i.d. uniform | | | | |
| Kolmogorov-Smirnov ($p$) | 0.489 | 0.795 | 0.019 | 0.426 |
| No serial correlation ($p$) | 0.007 | 0.000 | 0.050 | 0.0467 |
| Observations | 35 | 37 | 35 | 21 |

Notes: See notes to Table 3.

**Figure 1. Gross world product, 10,000 BCE–2019 (billion $ of 1990)**

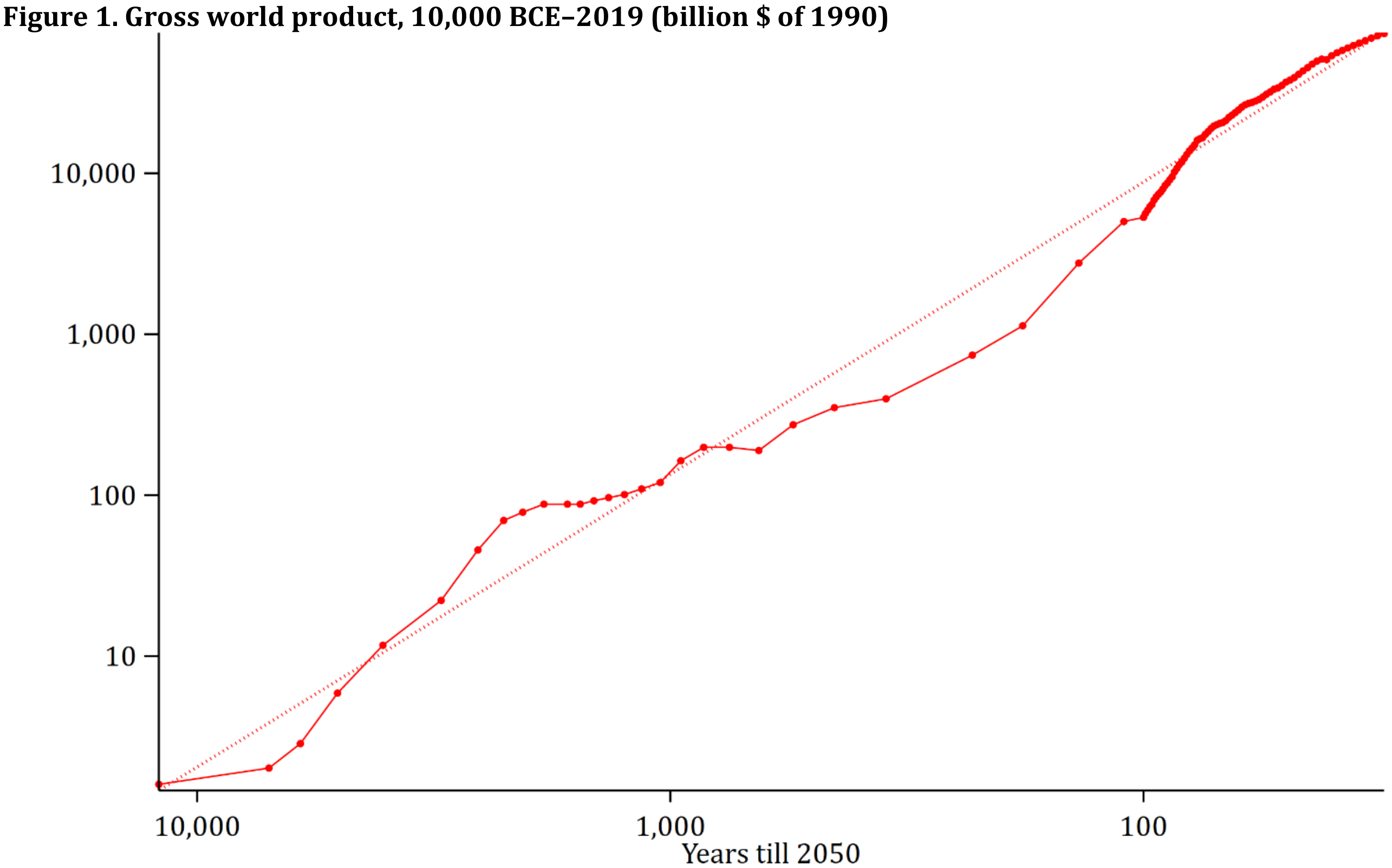

Note: Section 4 documents the construction of this data series.

**Figure 2. U.S. gross domestic product per capita, 1820–2016 ($ of 2011)**

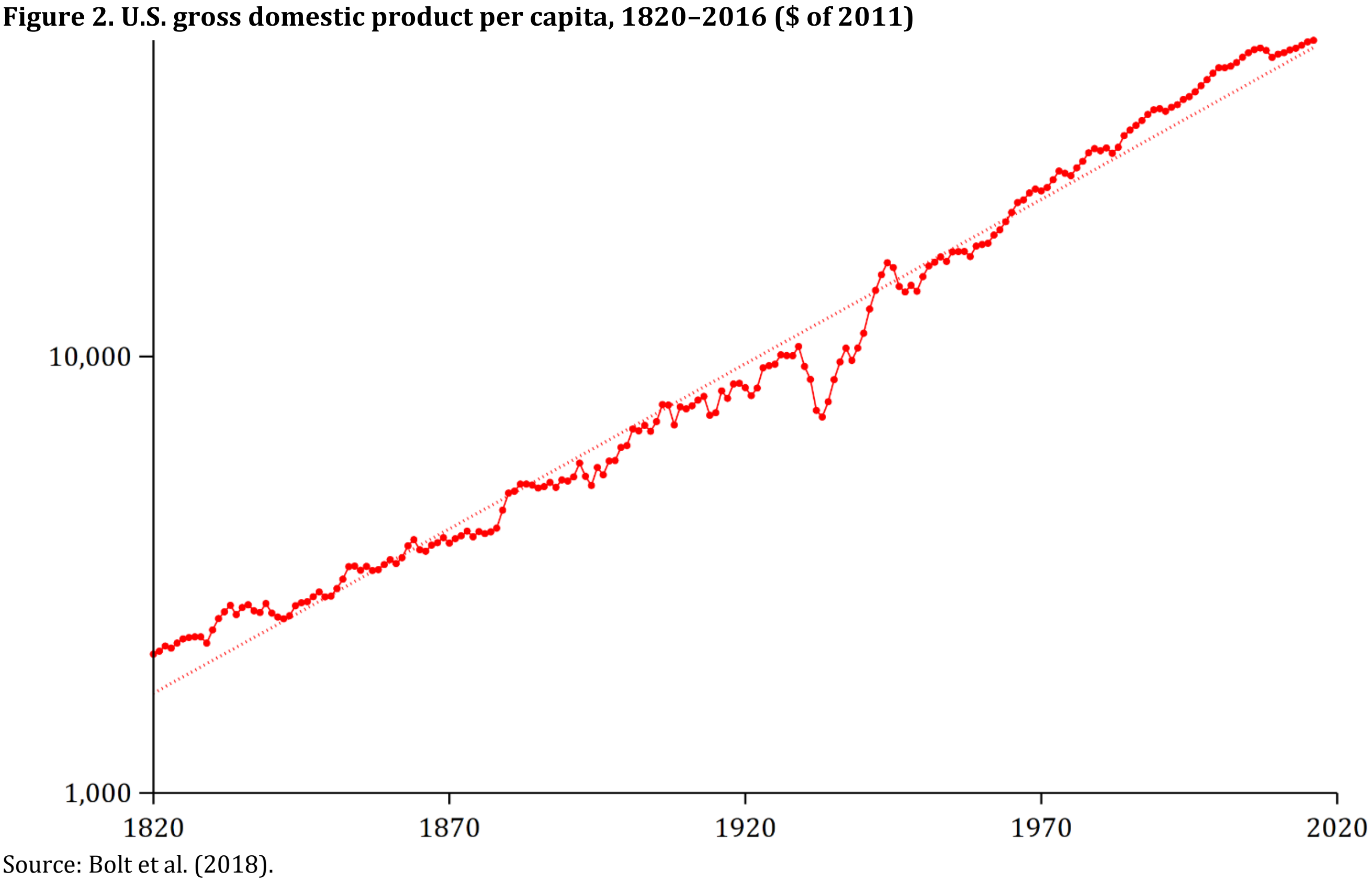

Source: Bolt et al. (2018).

**Figure 3. Factor stocks and output in simulated, fully endogenous Cobb-Douglas economy, two scenarios with slightly different starting points**

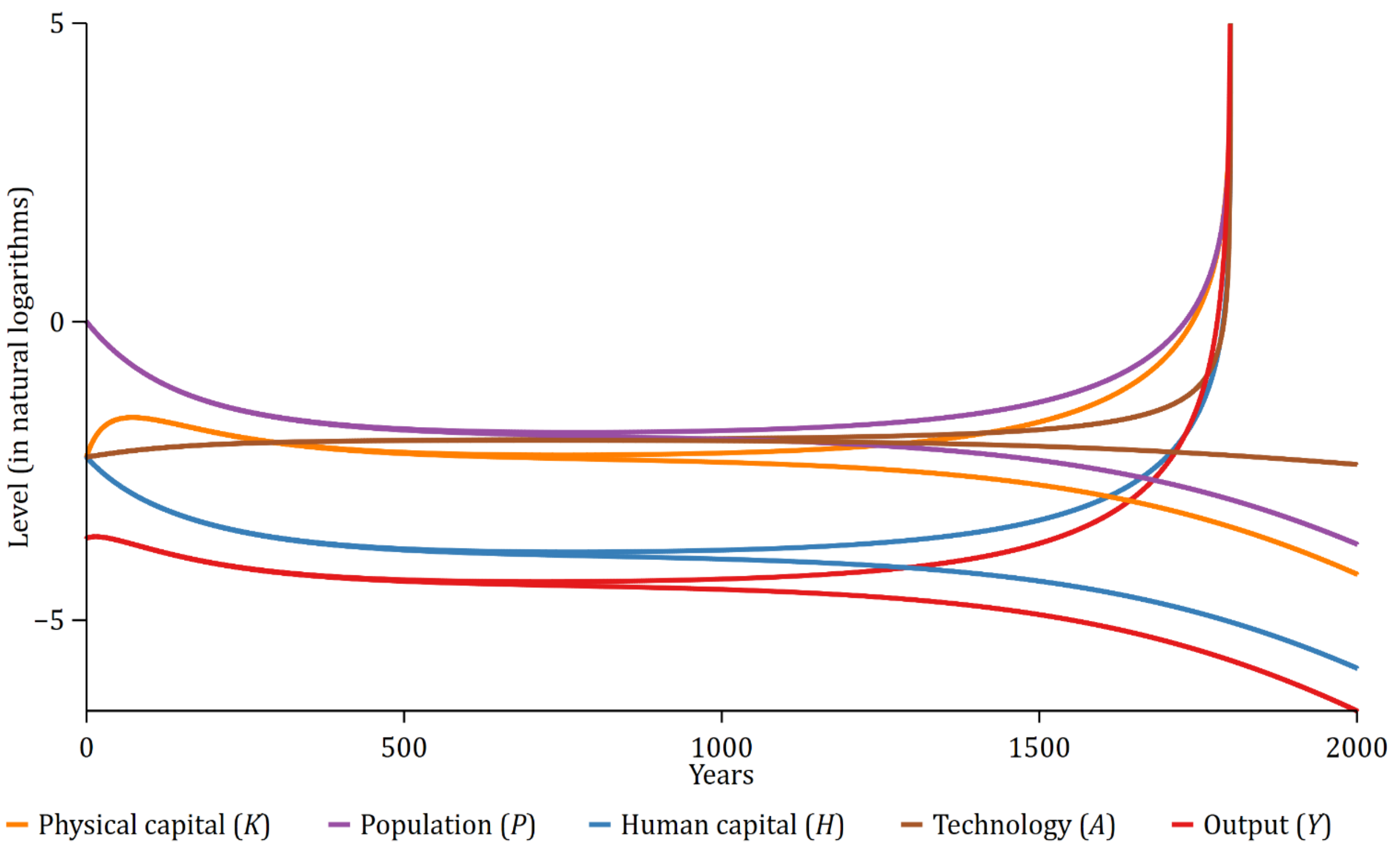

Note: Plots are based on simulations of the economy in (9) and (10). Time increment in simulations is $2 \times 10^{-5}$. Initial population, $P_0$, is 1 in both scenarios. In the exploding one, $A_0 = K_0 = H_0 = 0.03117$. In the other, $A_0 = K_0 = H_0 = 0.03107$.

**Figure 4. Plots of $f_{\pm\chi^2}(x;\lambda,\nu)$ for $\lambda = 1, \nu = -3.0, -2.5, \ldots, +3.0$**

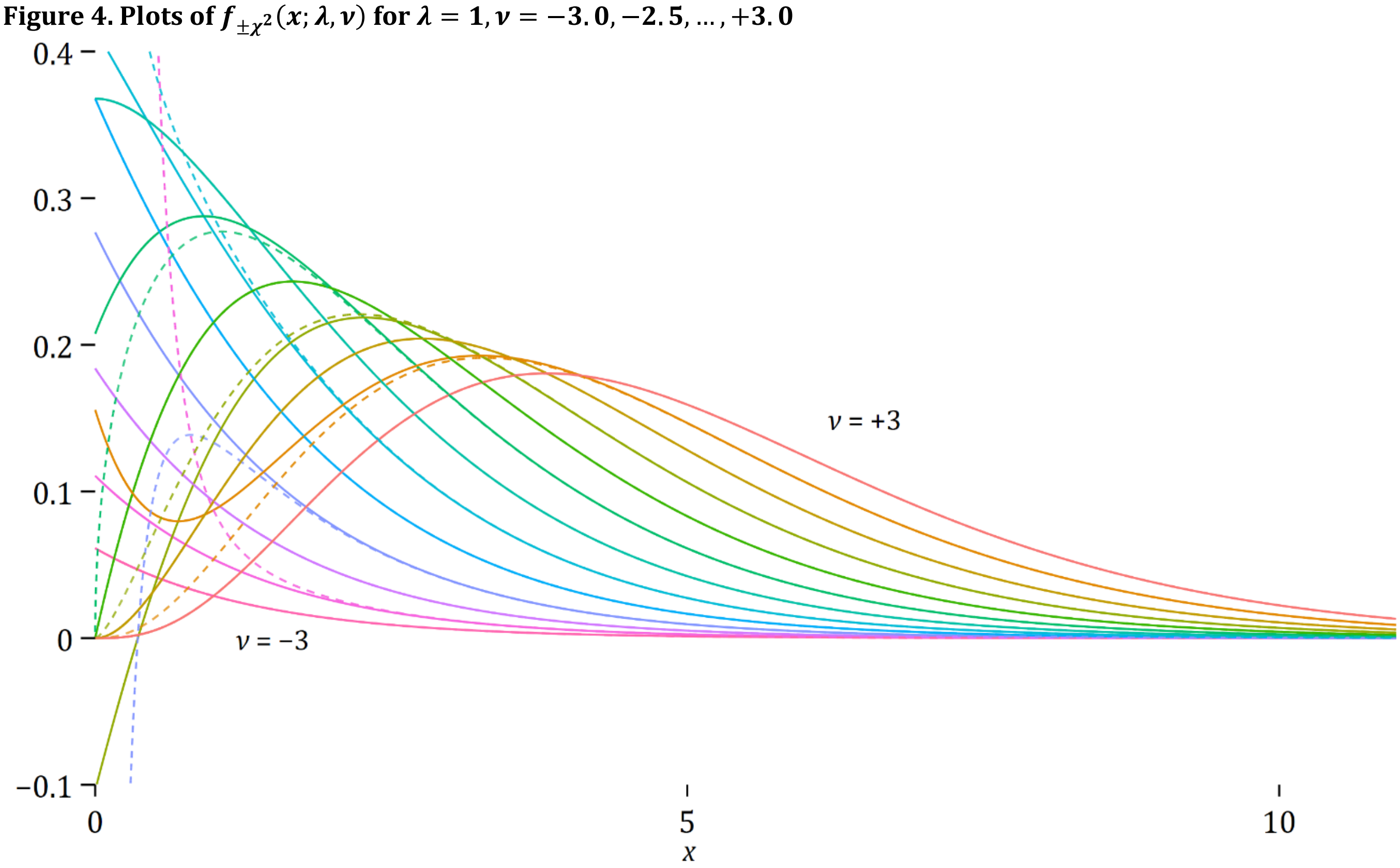

Notes: For each value of $\nu$, $f_{\chi^2}$ and $f_{-\chi^2}$ have the same color. When $\nu$ is an integer, the two coincide. Otherwise, $f_{\chi^2}$ is marked by short dashes and $f_{-\chi^2}$ by solid lines. Point mass accumulations at $x = 0$ under $f_{-\chi^2}$ are not depicted. $f_{\chi^2}$ is a valid probability distribution for $\nu \geq -1$, as is $f_{-\chi^2}$ for $\nu \leq 0$.

**Figure 5. Feller/CIR diffusion $f^*_{\pm\chi^2}(X_t; X_0, t, a, b, \nu)$ for $\nu = -1.5, -0.5, +0.5$, with $X_0 = a = 1, b = -1$**

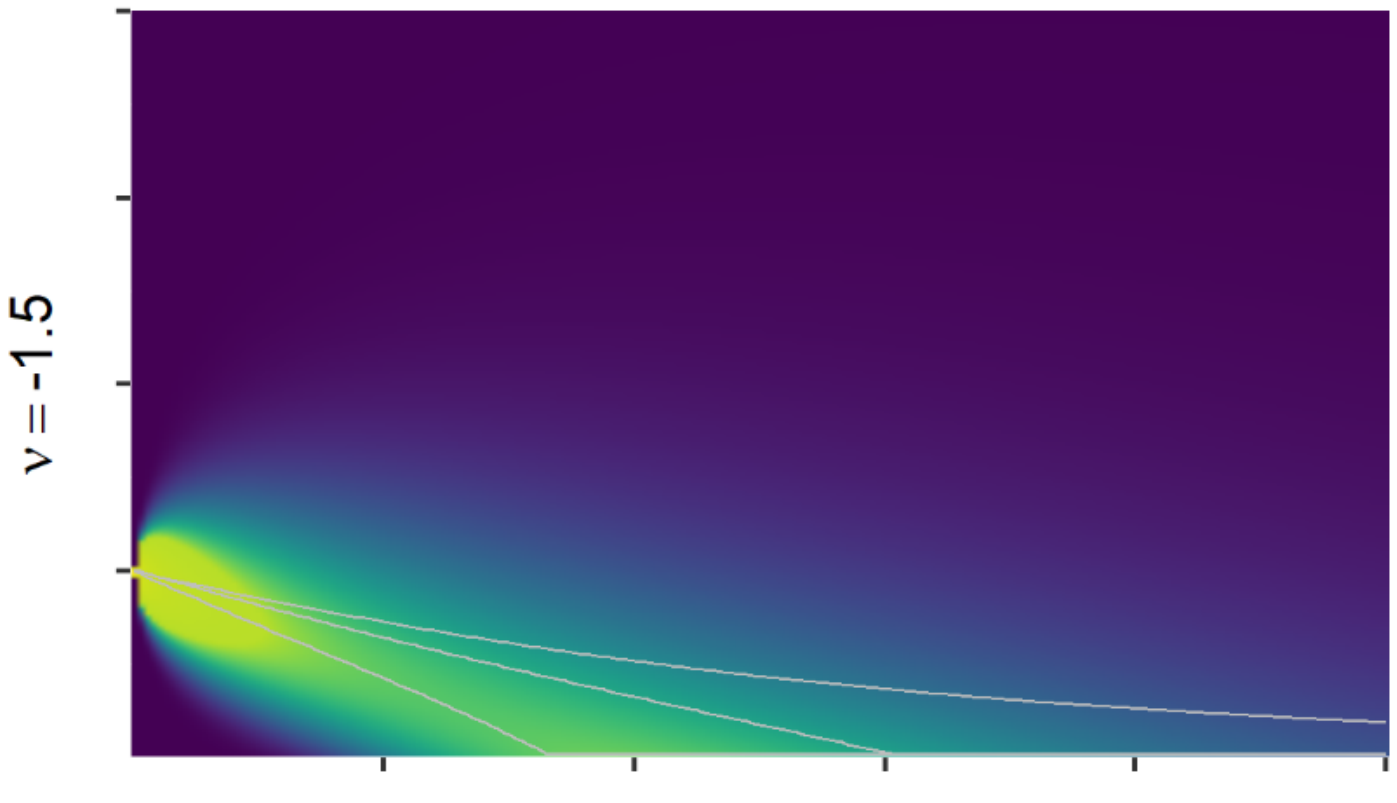

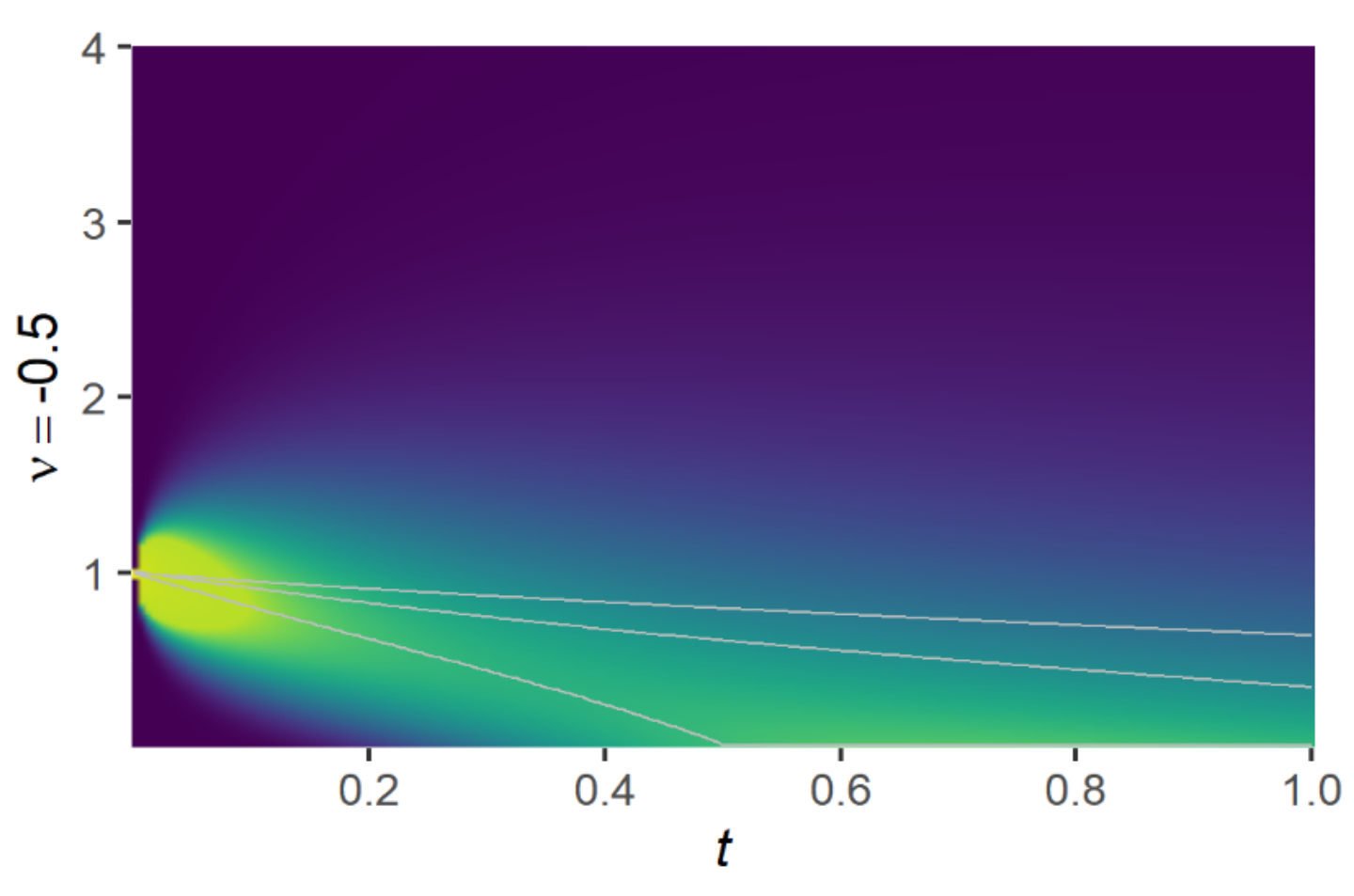

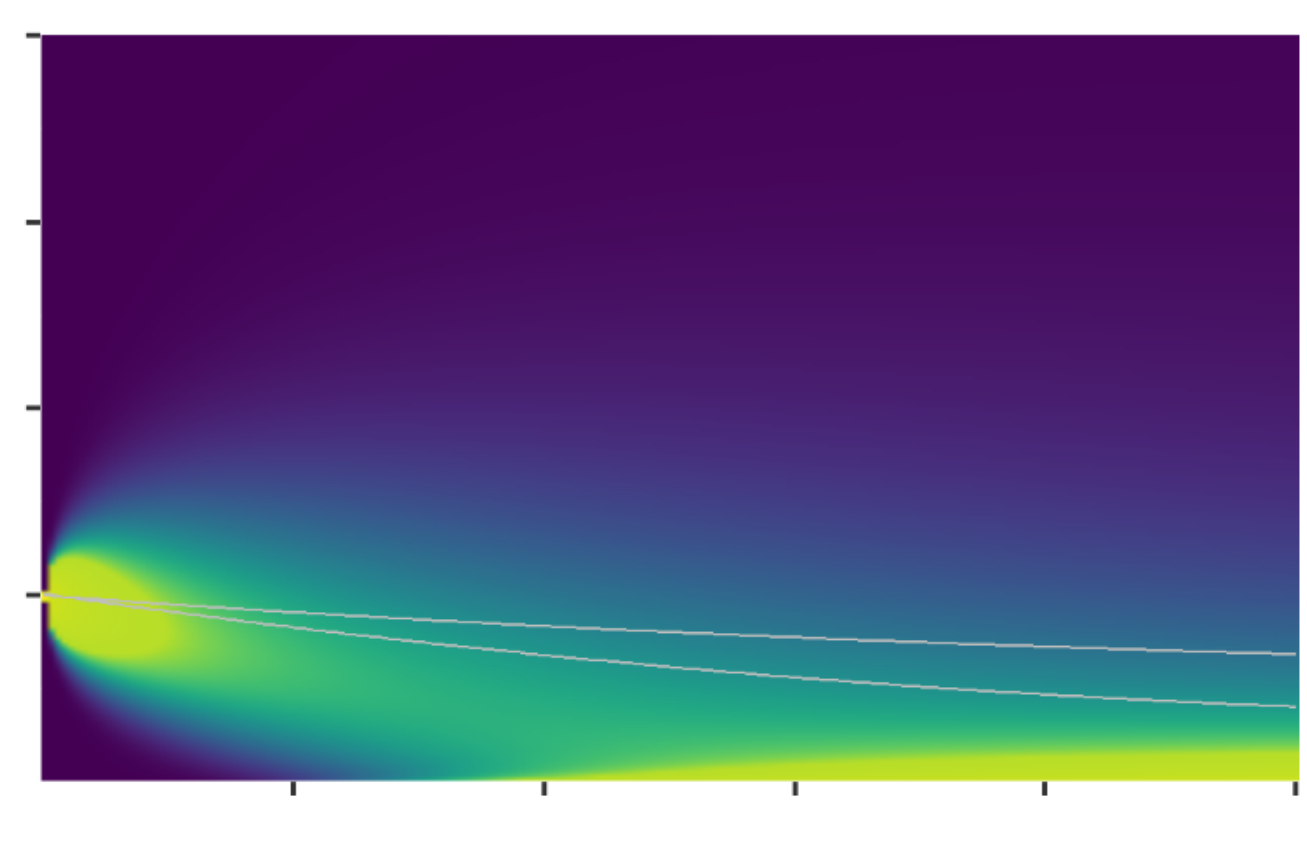

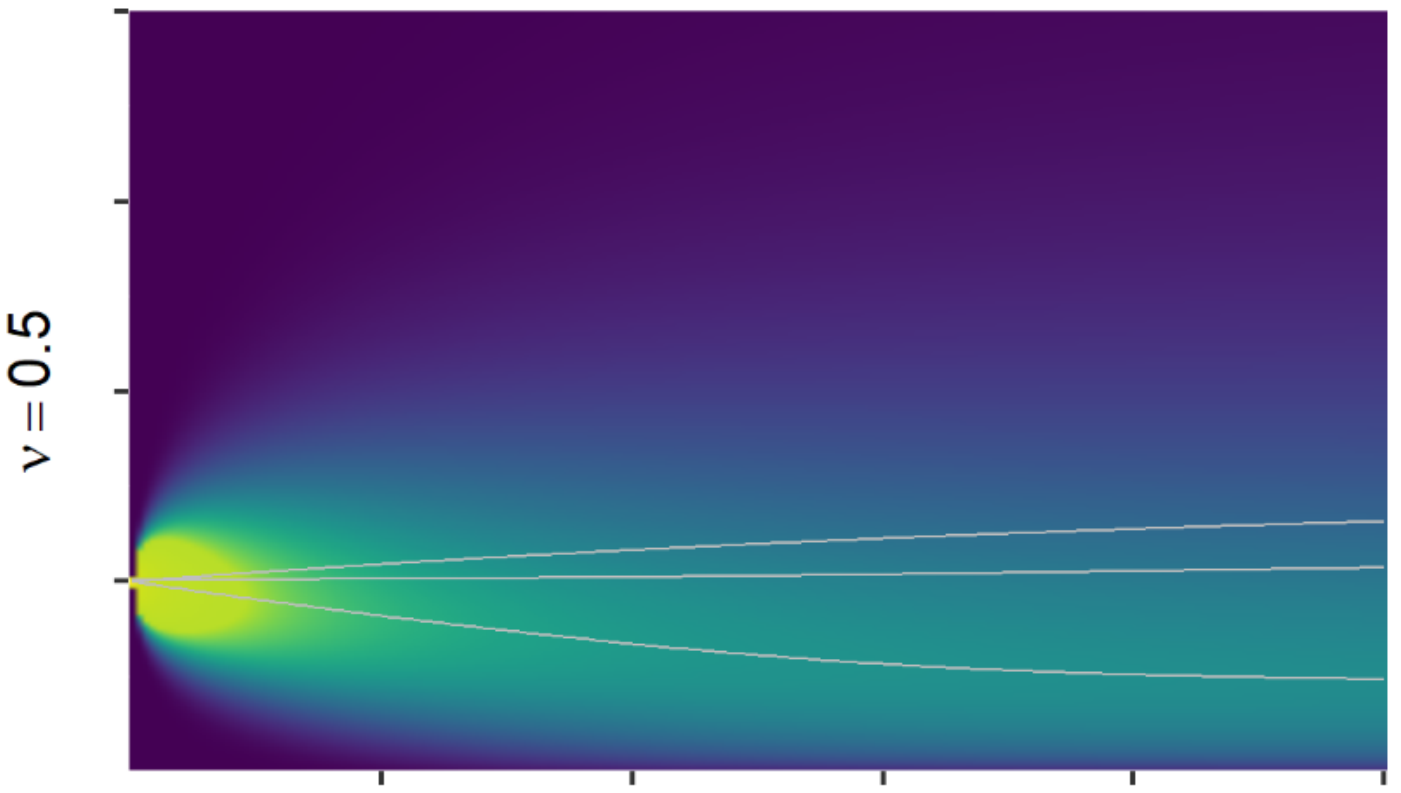

Notes: The absorbing- and reflecting-boundary solutions are $f^*_{-\chi^2}$ and $f^*_{\chi^2}$. In each plot, the horizontal coordinate is time, $t$, and the vertical is $X_t$. Dark purple indicates the lowest densities and yellow the highest. White lines show the evolution of, from upper to lower, the mean, median, and mode. But the mode is not defined and not plotted for the diffusion with unbounded density approaching zero. Plots are omitted for cases that do not yield proper diffusions. Point accumulations at $X_t = 0$ boundary, when it is absorbing, are not depicted because their densities are infinite.

**Figure 6. Distribution of nonlinear least squares and maximum likelihood estimates of $B$ in 10,000 Monte Carlo simulations**

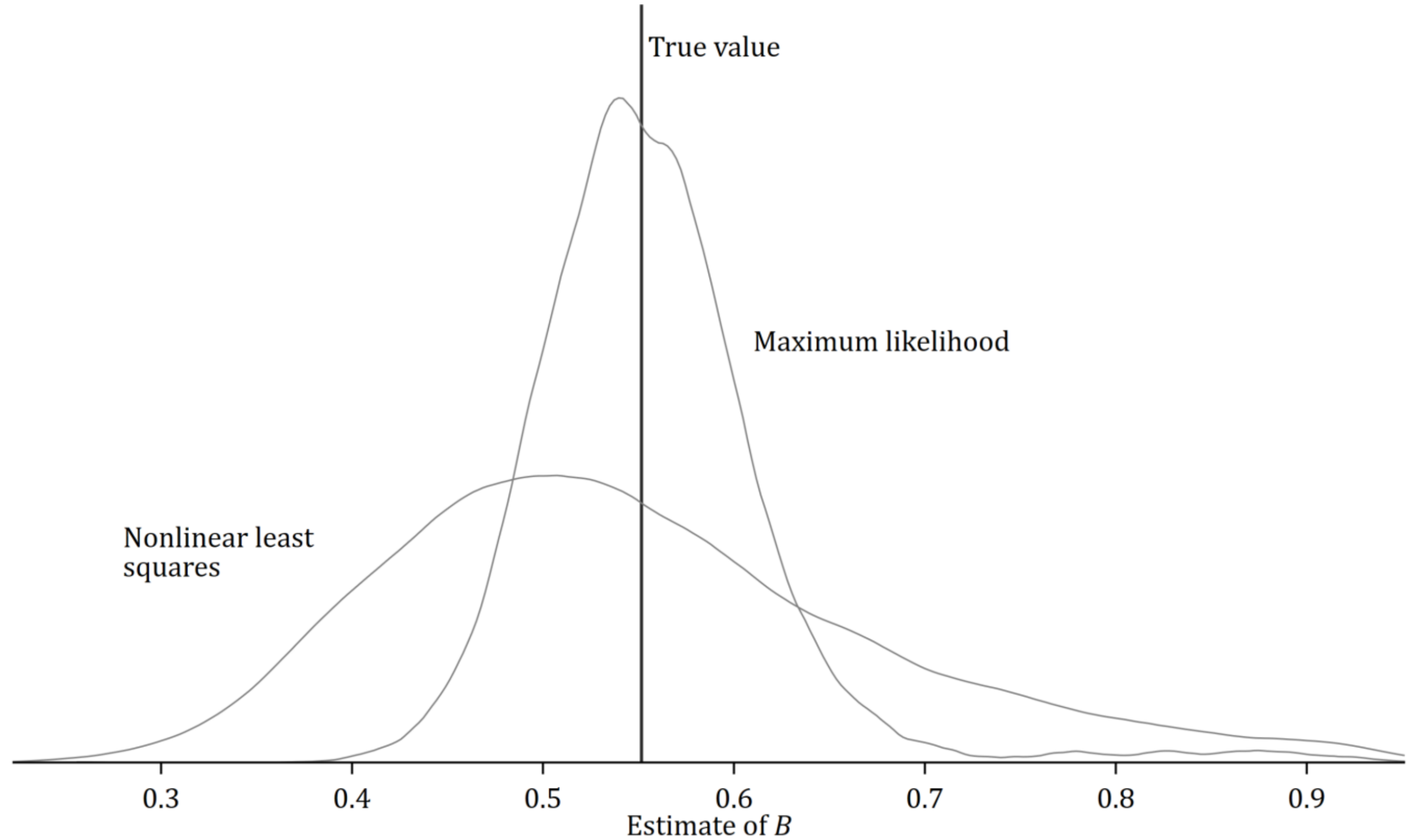

Notes: The figure shows the distribution of estimates of $B$ in (30) from nonlinear least squares as well as maximum likelihood fitting of the diffusion model. The true value of $B$ in the simulated data sets is 0.552. Each data set is built from a sample path generated with the Euler-Maruyama method for the process (33) with parameter values as estimated in Table 3, column 4; and with a starting value of $1.6 billion, the GWP level for 10,000 BCE. Each path terminates when it reaches $100 trillion or 25,000 years. Paths are sampled to mimic the empirical series in number and relative spacing of observations.

**Figure 7. 99 rollouts from the diffusion model fit to GWP, 1 million BCE–2019 CE**

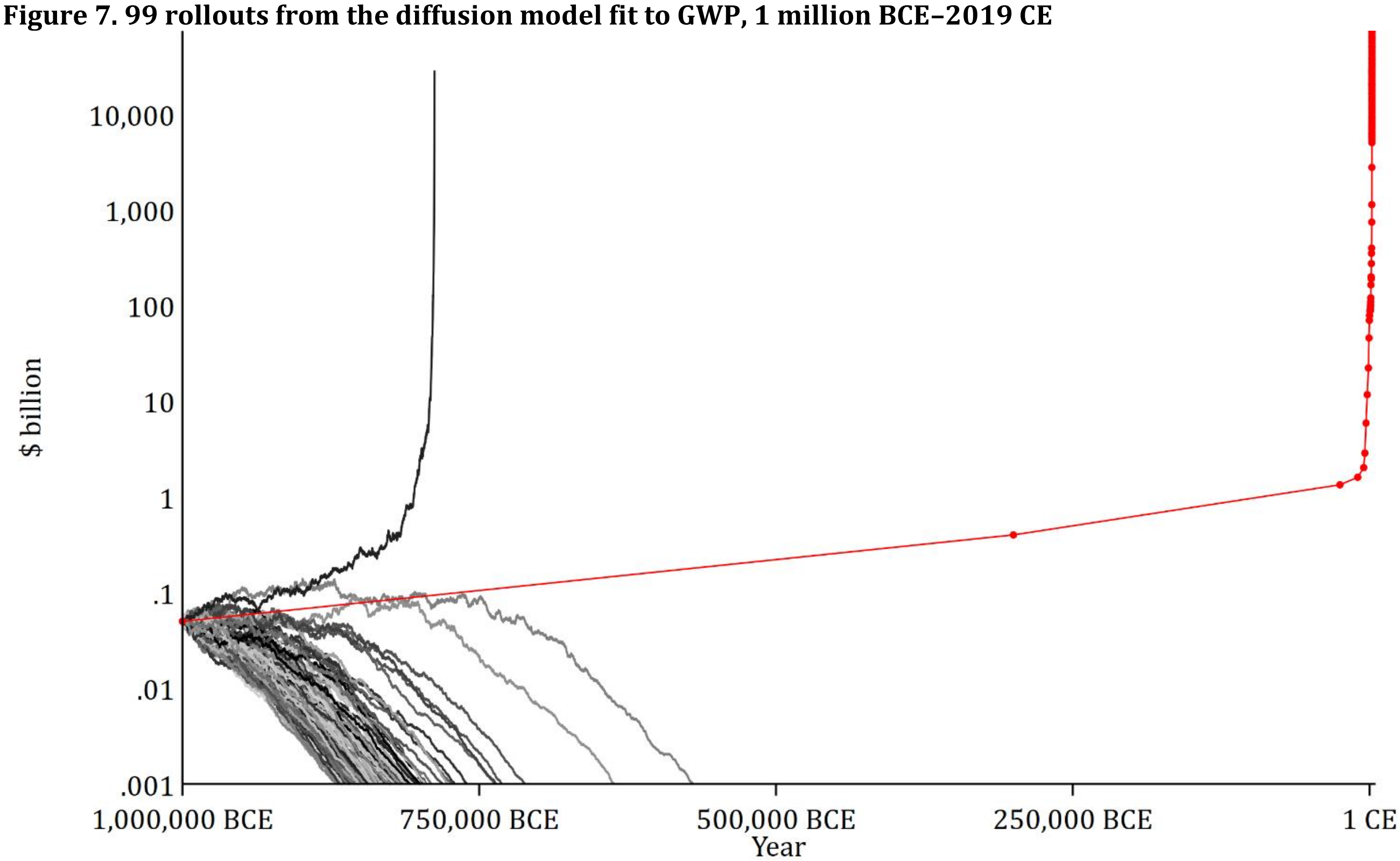

Notes: Historical series in red. Simulations conducted with the Euler-Maruyama method with 100,000 time steps.

**Figure 8. Distribution of 10,000 rollouts from the diffusion model for GWP, starting from both initial and final GWP values, incorporating modeled stochasticity and estimation uncertainty, 1 million BCE–2019**

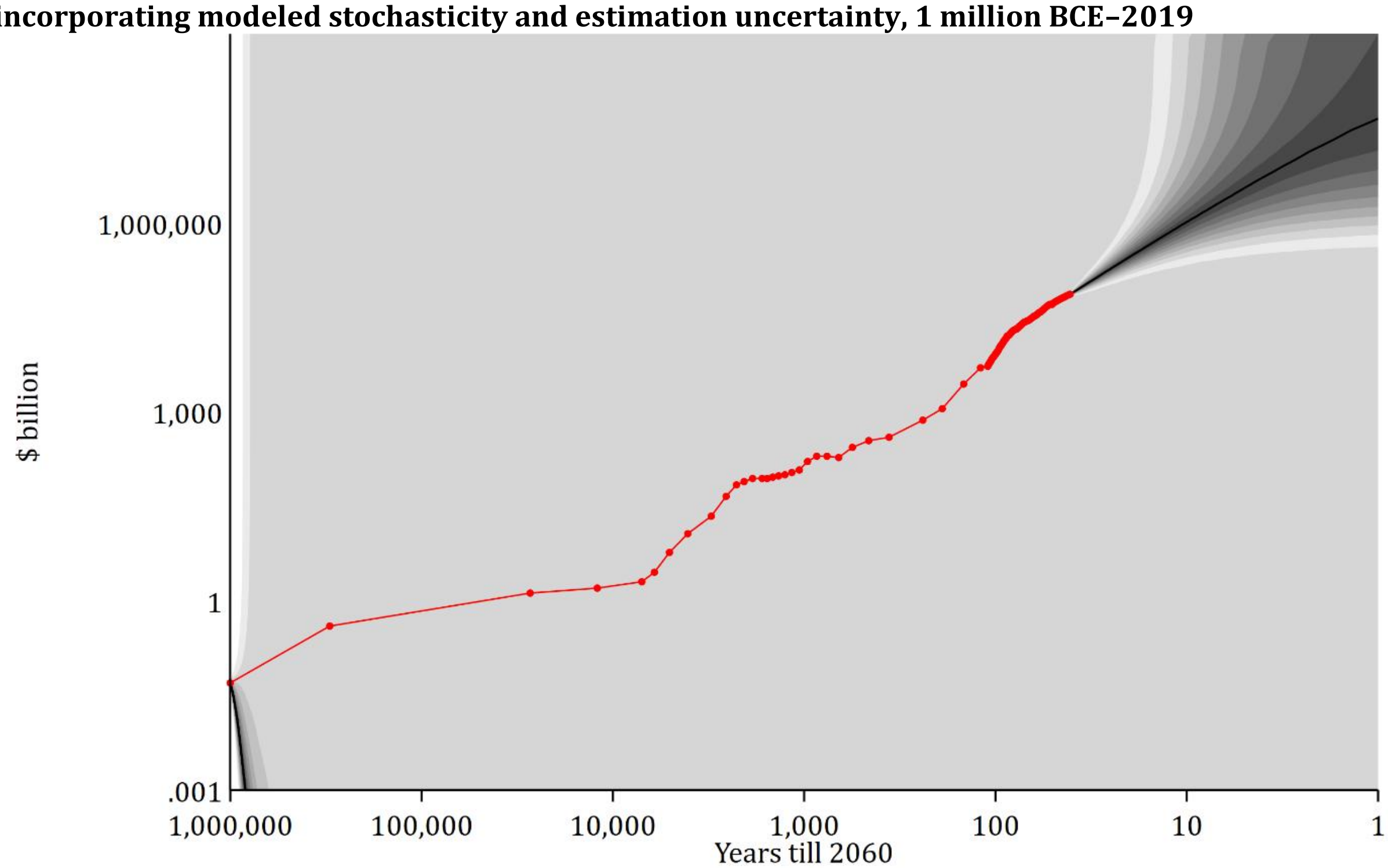

Notes: Grey bands indicate 5th, 10th, etc., quantiles of the distribution of 10,000 simulations, each starting from the initial or final values of the series. See also notes to previous figure.

**Figure 9. 99 rollouts from the diffusion model for GWP for 10,000 BCE–2019 CE, with decennial observations after 1950**

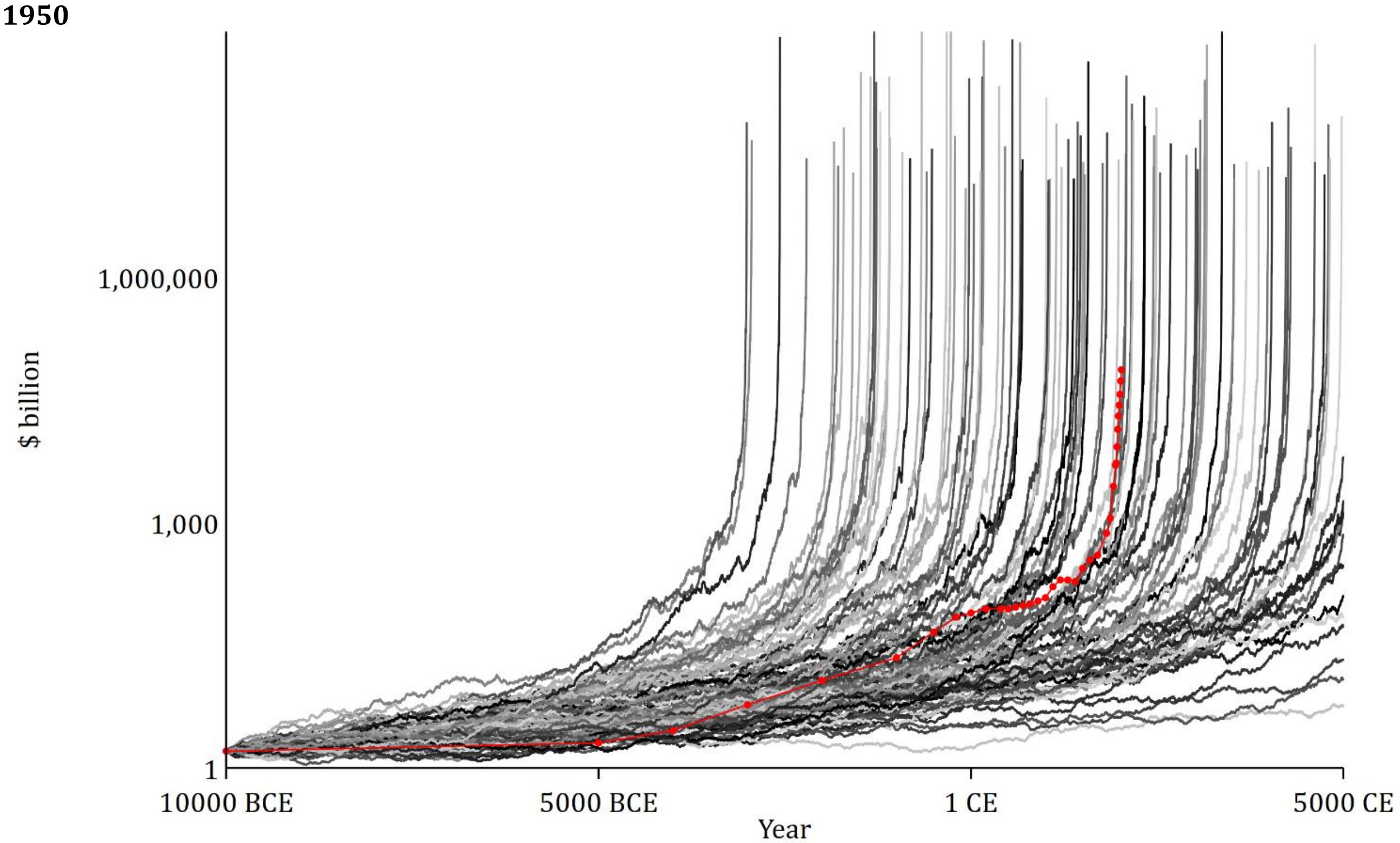

Notes: See Figure 7.

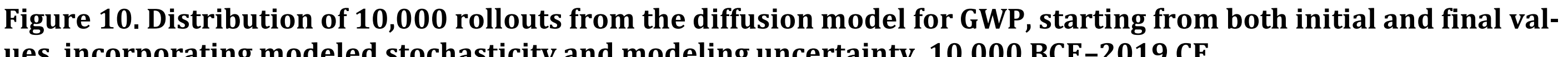

**Figure 10. Distribution of 10,000 rollouts from the diffusion model for GWP, starting from both initial and final values, incorporating modeled stochasticity and modeling uncertainty, 10,000 BCE–2019 CE**

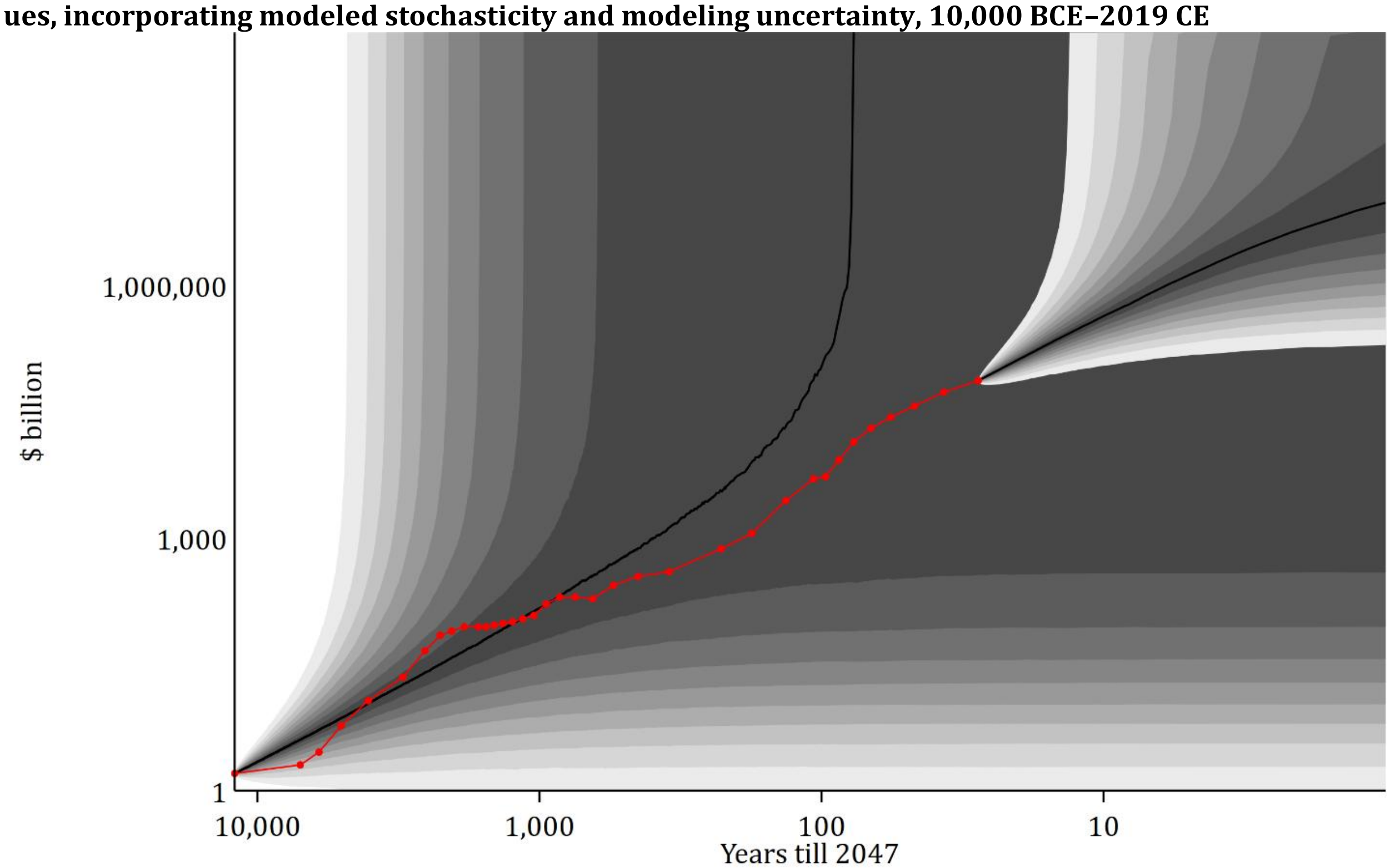

Notes: See Figure 8.

**Figure 11. Quantiles of GWP observations in distribution conditioned on previous observation, using parameters from preferred model fit (Table 3, col. 4)**

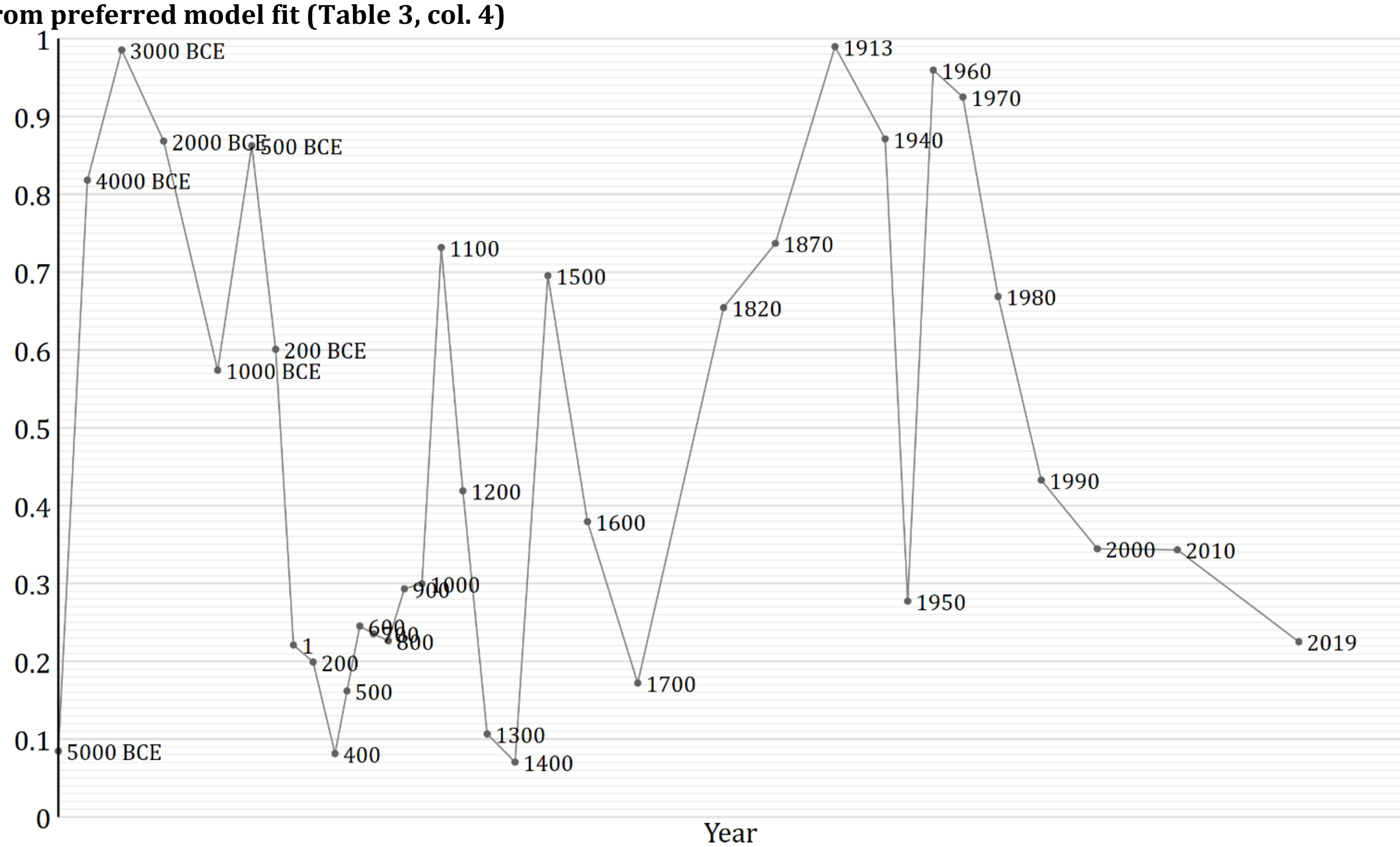

**Figure 12. Quantiles of GWP observations in distribution conditioned on previous observation, using parameters from diffusion model fit to previous observations only, incorporating modeled stochasticity as well as modeling uncertainty**

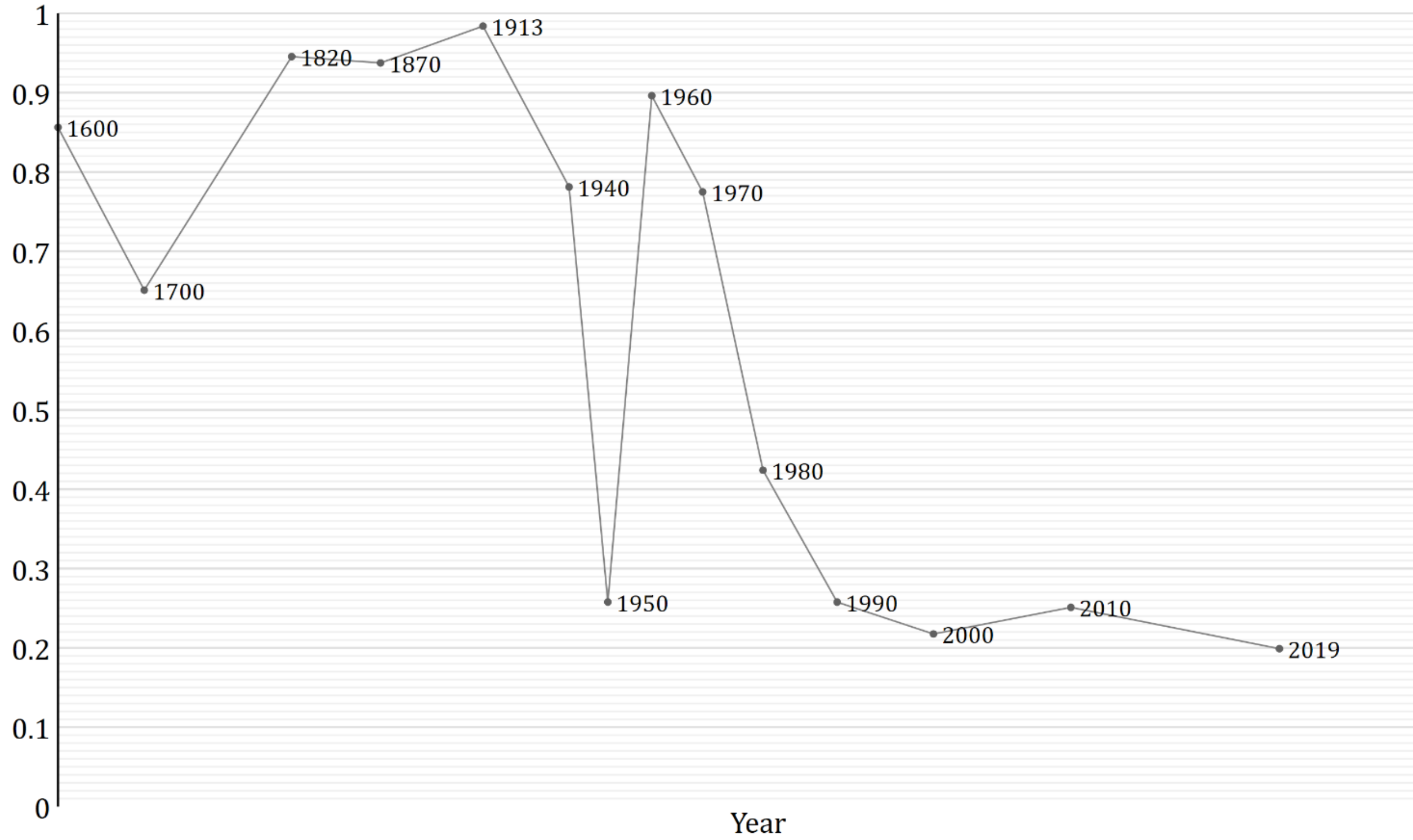

**Figure 13. Diffusion and quantile plots for diffusion fits to world population, GWP/capita, and GDP/capita in France, 10,000 BCE–2019 CE**

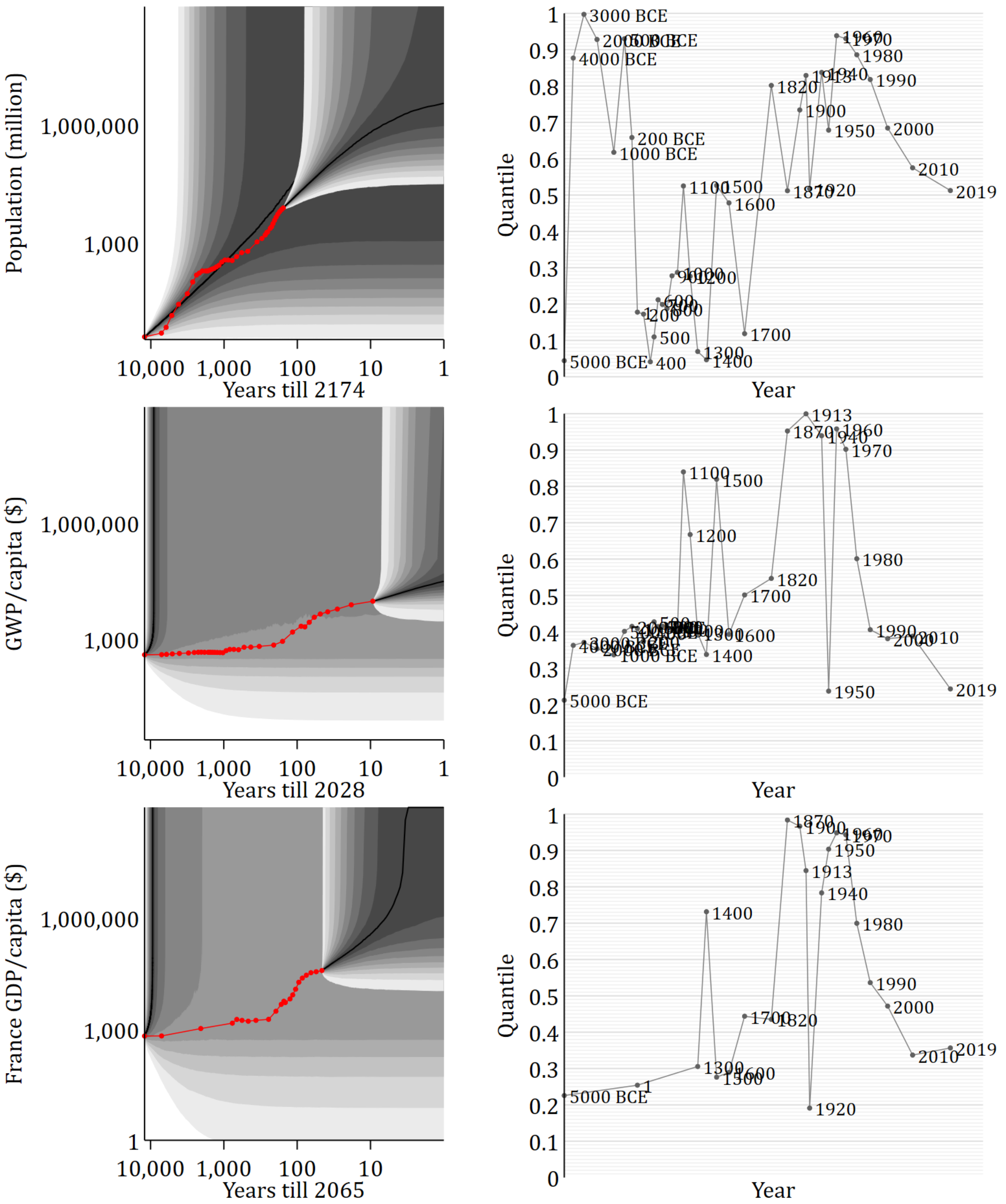

Note: Figures on the left and right constructed the same in the same ways as Figure 10 and Figure 11, respectively. For 5000 and 10,000 BCE, GDP/capita figures for “France” are the same as for the world.

**Figure 14. Factor stocks and output in simulated fully endogenous Cobb-Douglas economy with dynamics in resources**

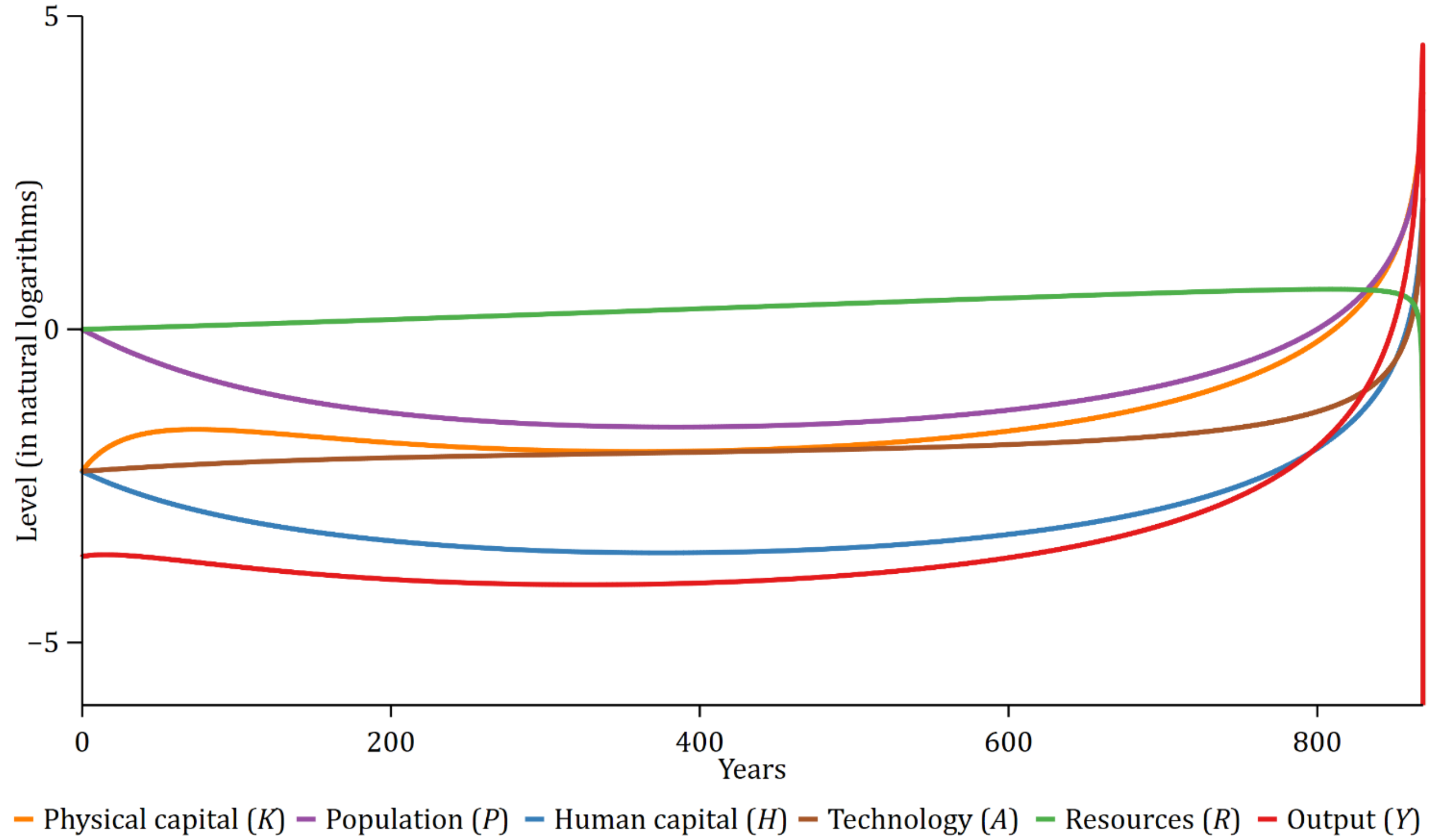

# A. Online Appendix: Derivation of formulas in sections 2.2 and 2.3

This appendix presents the algebra behind (16)–(18) and the computation of the equilibrium growth rate in the partially endogenous model and the conditions for its stability.

## A.1. Determinant of $-B$

We will need identities to compute the determinant and inverse of a matrix that is zero except in its diagonal and first column. Let $\mathbf{c}$ and $\mathbf{d}$ be column vectors and $[\![\mathbf{c}]\!]$ be defined as in (13). Since

$$\operatorname{diag}(\mathbf{d}) - [\![\mathbf{c}]\!] = \begin{bmatrix} d_0 - c_0 & 0 & \dots & 0 \\ -c_1 & d_1 & \dots & 0 \\ \vdots & \vdots & \ddots & \vdots \\ -c_k & 0 & \dots & d_k \end{bmatrix},$$

it is immediate that

$$|\operatorname{diag}(\mathbf{d}) - [\![\mathbf{c}]\!]| = (d_0 - c_0) \prod_{i>0} d_i. \tag{44}$$

And it can be checked by matrix multiplication that

$$(\operatorname{diag}(\mathbf{d}) - [\![\mathbf{c}]\!])^{-1} = \begin{bmatrix} \frac{1}{d_0 - c_0} & 0 & \dots & 0 \\ \frac{1}{d_0 - c_0}\frac{c_1}{d_1} & \frac{1}{d_1} & \dots & 0 \\ \vdots & \vdots & \ddots & \vdots \\ \frac{1}{d_0 - c_0}\frac{c_k}{d_k} & 0 & \dots & \frac{1}{d_k} \end{bmatrix} = \frac{1}{d_0 - c_0}\left[\!\!\left[\frac{\mathbf{c}}{\mathbf{d}}\right]\!\!\right] + \operatorname{diag}\left(\frac{\boldsymbol{\iota}}{\mathbf{d}}\right) \tag{45}$$

Define $\mathbf{A} \coloneqq \mathbf{I} - \boldsymbol{\Phi} = \mathbf{I} - [\![\boldsymbol{\phi}]\!]$, where, recall, $\boldsymbol{\phi}$ holds the elasticities of reinvestment with respect to the technology level. Then by (13), $\mathbf{B} = \boldsymbol{\iota}\boldsymbol{\alpha}' - \mathbf{A}$. Applying (44) and (45),

$$|\mathbf{A}| = 1 - \phi_0$$

$$\mathbf{A}^{-1} = \mathbf{I} + \frac{1}{1 - \phi_0}\boldsymbol{\Phi} \tag{46}$$

$$\operatorname{adj}(\mathbf{A}) = |\mathbf{A}|\mathbf{A}^{-1} = (1 - \phi_0)\mathbf{I} + \boldsymbol{\Phi}$$

Note that if technology has no role in the system, so that $\mathbf{A} = \mathbf{I}$, then the above formulas will be correct if we take $\boldsymbol{\phi} = \mathbf{0}$.

By the matrix determinant lemma,

$$
\begin{aligned}
|-\mathbf{B}| &= |\mathbf{I} - \mathbf{\Phi} - \boldsymbol{\iota}\boldsymbol{\alpha}'| \\
&= |\mathbf{A} - \boldsymbol{\iota}\boldsymbol{\alpha}'| \\
&= |\mathbf{A}| - \boldsymbol{\alpha}' \operatorname{adj}(\mathbf{A})\, \boldsymbol{\iota} \\
&= 1 - \phi_0 - \boldsymbol{\alpha}'[(1-\phi_0)\mathbf{I} + \mathbf{\Phi}]\boldsymbol{\iota} \\
&= 1 - \phi_0 - \boldsymbol{\alpha}'\mathbf{\Phi}\boldsymbol{\iota} - (1-\phi_0)\boldsymbol{\alpha}'\boldsymbol{\iota} \\
&= (1-\phi_0)(1-\boldsymbol{\alpha}'\boldsymbol{\iota}) - \boldsymbol{\alpha}'\boldsymbol{\phi}
\end{aligned}
\tag{47}
$$

Since $\phi_0$ means $\phi_A$, this confirms the characterization of $-|-\mathbf{B}|$ in (16).

## A.2. **Eigenvalues and eigenvectors of *B* in the single-output model**

We develop the characteristic equation for $\mathbf{B}$:

$$
0 = |\lambda\mathbf{I} - \mathbf{B}| = |(1+\lambda)\mathbf{I} - \boldsymbol{\iota}\boldsymbol{\alpha}' - \mathbf{\Phi}| = (1+\lambda)^{k+1}\left|\mathbf{I} - \frac{\mathbf{\Phi}}{1+\lambda} - \boldsymbol{\iota}\frac{\boldsymbol{\alpha}'}{1+\lambda}\right|.
$$

The determinant in the last expression is the same as that in (47), but with $\mathbf{\Phi}$ and $\boldsymbol{\alpha}$ pre-divided by $1+\lambda$. So this determinant equals the last line of (47) but with the same divisions applied. The characteristic equation becomes

$$
\begin{aligned}
0 &= (1+\lambda)^{k+1}\left[\left(1 - \frac{\phi_0}{1+\lambda}\right)\left(1 - \frac{\boldsymbol{\alpha}'\boldsymbol{\iota}}{1+\lambda}\right) - \frac{\boldsymbol{\alpha}'\boldsymbol{\phi}}{(1+\lambda)^2}\right] \\
&= (1+\lambda)^{k-1}[(1+\lambda-\phi_0)(1+\lambda-\boldsymbol{\alpha}'\boldsymbol{\iota}) - \boldsymbol{\alpha}'\boldsymbol{\phi}] \\
&= (1+\lambda)^{k-1}[(1+\lambda)^2 - (\boldsymbol{\alpha}'\boldsymbol{\iota} + \phi_0)(1+\lambda) + \phi_0\boldsymbol{\alpha}'\boldsymbol{\iota} - \boldsymbol{\alpha}'\boldsymbol{\phi}].
\end{aligned}
\tag{48}
$$

The roots of the quadratic subexpression are

$$
\lambda_\pm = \frac{\boldsymbol{\alpha}'\boldsymbol{\iota} + \phi_0}{2} - 1 \pm \sqrt{\left(\frac{\boldsymbol{\alpha}'\boldsymbol{\iota} + \phi_0}{2}\right)^2 + \boldsymbol{\alpha}'(\boldsymbol{\phi} - \phi_0\boldsymbol{\iota})}.
$$

Exactly one of these roots is positive when

$$
\left|\frac{\boldsymbol{\alpha}'\boldsymbol{\iota} + \phi_0}{2} - 1\right| < \sqrt{\left(\frac{\boldsymbol{\alpha}'\boldsymbol{\iota} + \phi_0}{2}\right)^2 + \boldsymbol{\alpha}'(\boldsymbol{\phi} - \phi_0\boldsymbol{\iota})}
$$

Successively restate that:

$$
\begin{gathered}
\left(\frac{\boldsymbol{\alpha}'\boldsymbol{\iota} + \phi_0}{2} - 1\right)^{\mathbf{2}} < \left(\frac{\boldsymbol{\alpha}'\boldsymbol{\iota} + \phi_0}{2}\right)^2 + \boldsymbol{\alpha}'(\boldsymbol{\phi} - \phi_0\boldsymbol{\iota}) \\
1 - (\boldsymbol{\alpha}'\boldsymbol{\iota} + \phi_0) + \left(\frac{\boldsymbol{\alpha}'\boldsymbol{\iota} + \phi_0}{2}\right)^{\mathbf{2}} < \left(\frac{\boldsymbol{\alpha}'\boldsymbol{\iota} + \phi_0}{2}\right)^2 + \boldsymbol{\alpha}'(\boldsymbol{\phi} - \phi_0\boldsymbol{\iota}) \\
1 - \boldsymbol{\alpha}'\boldsymbol{\iota} - \phi_0 < \boldsymbol{\alpha}'(\boldsymbol{\phi} - \phi_0\boldsymbol{\iota}) \\
(\phi_0 - 1)(\boldsymbol{\alpha}'\boldsymbol{\iota} - 1) < \boldsymbol{\alpha}'\boldsymbol{\phi} \\
-|-\mathbf{B}| > 0
\end{gathered}
$$

The last line uses (47). This confirms the characterization of the eigenvalues in and near (23).

Next we confirm that eigenvectors corresponding to $\lambda_\pm$ are $\mathbf{v}_\pm \coloneqq \boldsymbol{\phi} + (\lambda_\pm + 1 - \phi_0)\boldsymbol{\iota}$. For clarity, pick one of $\lambda_\pm$ and define $\kappa \coloneqq 1 + \lambda_\pm$. By (48), $\kappa$ satisfies

$$\kappa^2 - \kappa(\boldsymbol{\alpha}'\boldsymbol{\iota} + \phi_0) + \phi_0\boldsymbol{\alpha}'\boldsymbol{\iota} - \boldsymbol{\alpha}'\boldsymbol{\phi} = 0.$$

It follows that

$$\begin{aligned}(\mathbf{I}+\mathbf{B})\mathbf{v}_\pm - \kappa\mathbf{v}_\pm &= (\boldsymbol{\iota}\boldsymbol{\alpha}' + [\![\boldsymbol{\phi}]\!])(\boldsymbol{\phi} + (\kappa-\phi_0)\boldsymbol{\iota}) - \kappa(\boldsymbol{\phi} + (\kappa-\phi_0)\boldsymbol{\iota}) \\ &= \boldsymbol{\iota}\boldsymbol{\alpha}'\boldsymbol{\phi} + (\kappa-\phi_0)\boldsymbol{\iota}\boldsymbol{\alpha}'\boldsymbol{\iota} + [\![\boldsymbol{\phi}]\!]\boldsymbol{\phi} + [\![\boldsymbol{\phi}]\!](\kappa-\phi_0)\boldsymbol{\iota} - \kappa\boldsymbol{\phi} - \kappa^2\boldsymbol{\iota} + \kappa\phi_0\boldsymbol{\iota} \\ &= \boldsymbol{\iota}\boldsymbol{\alpha}'\boldsymbol{\phi} + (\kappa-\phi_0)\boldsymbol{\iota}\boldsymbol{\alpha}'\boldsymbol{\iota} + \phi_0\boldsymbol{\phi} + (\kappa-\phi_0)\boldsymbol{\phi} - \kappa\boldsymbol{\phi} - \kappa^2\boldsymbol{\iota} + \kappa\phi_0\boldsymbol{\iota} \\ &= \boldsymbol{\iota}\boldsymbol{\alpha}'\boldsymbol{\phi} + (\kappa-\phi_0)\boldsymbol{\iota}\boldsymbol{\alpha}'\boldsymbol{\iota} - \kappa^2\boldsymbol{\iota} + \kappa\phi_0\boldsymbol{\iota} \\ &= [\boldsymbol{\alpha}'\boldsymbol{\phi} + (\kappa-\phi_0)\boldsymbol{\alpha}'\boldsymbol{\iota} - \kappa^2 + \kappa\phi_0]\boldsymbol{\iota} \\ &= 0\boldsymbol{\iota}\end{aligned}$$

Thus $\mathbf{v}_\pm$is an eigenvector of $\mathbf{I}+\mathbf{B}$ corresponding to eigenvalue $\kappa$, and an eigenvector of $\mathbf{B}$ corresponding to $\kappa - 1 = \lambda_\pm$.

## A.3. **The characteristic equation of $-\boldsymbol{\delta} \circ \boldsymbol{B}$**

By (13),

$$\begin{aligned}-\boldsymbol{\delta}\circ\mathbf{B} &= \boldsymbol{\delta}\circ\mathbf{I} - \boldsymbol{\delta}\circ\boldsymbol{\Phi} - \boldsymbol{\delta}\circ\boldsymbol{\iota}\boldsymbol{\alpha}' \\ &= \operatorname{diag}(\boldsymbol{\delta}) - [\![\boldsymbol{\delta}\circ\boldsymbol{\phi}]\!] - \boldsymbol{\delta}\boldsymbol{\alpha}'.\end{aligned}$$

The characteristic equation of this matrix can be written

$$0 = |\operatorname{diag}(\boldsymbol{\delta} - \lambda\boldsymbol{\iota}) - [\![\boldsymbol{\delta}\circ\boldsymbol{\phi}]\!] - \boldsymbol{\delta}\boldsymbol{\alpha}'|.$$

Applying the matrix determinant lemma and the determinant and adjoint identities (44) and (45),

$$\begin{aligned}0 &= |\operatorname{diag}(\boldsymbol{\delta} - \lambda\boldsymbol{\iota}) - [\![\boldsymbol{\delta}\circ\boldsymbol{\phi}]\!] - \boldsymbol{\delta}\boldsymbol{\alpha}'| \\ &= (1 - \boldsymbol{\alpha}'(\operatorname{diag}(\boldsymbol{\delta} - \lambda\boldsymbol{\iota}) - [\![\boldsymbol{\delta}\circ\boldsymbol{\phi}]\!])^{-1}\boldsymbol{\delta}) \cdot |\operatorname{diag}(\boldsymbol{\delta} - \lambda\boldsymbol{\iota}) - [\![\boldsymbol{\delta}\circ\boldsymbol{\phi}]\!]| \\ &= \left(1 - \boldsymbol{\alpha}'\left(\frac{1}{\delta_0 - \lambda - \delta_0\phi_0}\left[\!\!\left[\frac{\boldsymbol{\delta}\circ\boldsymbol{\phi}}{\boldsymbol{\delta} - \lambda\boldsymbol{\iota}}\right]\!\!\right] + \operatorname{diag}\left(\frac{\boldsymbol{\iota}}{\boldsymbol{\delta} - \lambda\boldsymbol{\iota}}\right)\right)\boldsymbol{\delta}\right) \cdot (\delta_0 - \lambda - \delta_0\phi_0) \cdot \prod_{i>0}(\delta_i - \lambda) \\ &= \left(1 - \boldsymbol{\alpha}'\left(\frac{\boldsymbol{\delta}}{\boldsymbol{\delta} - \lambda\boldsymbol{\iota}} \circ \left(\frac{\delta_0}{\delta_0(1-\phi_0) - \lambda}\boldsymbol{\phi} + \boldsymbol{\iota}\right)\right)\right) \cdot (\delta_0(1-\phi_0) - \lambda) \cdot \prod_{i>0}(\delta_i - \lambda)\end{aligned}$$

Because the second and third factors appear in the denominators of the first factors, the right side can only be 0 by virtue of the second and third factors being 0 if $\lambda$ is at least a *double* root of $\prod_{i>0}(\delta_i - \lambda)$. For example, if $\delta_1 = \delta_2$, then $\lambda = \delta_1 = \delta_2$ would be a root. However, an infinitesimal change in the elements of $\boldsymbol{\delta}$ could break such solutions.

For non-degenerate solution, we drop the second and third terms. The characteristic equation distills to

$$1 = \boldsymbol{\alpha}'\left(\frac{\boldsymbol{\delta}}{\boldsymbol{\delta} - \lambda\boldsymbol{\iota}} \circ \left(\boldsymbol{\iota} + \frac{\delta_0}{\delta_0(1-\phi_0) - \lambda}\boldsymbol{\phi}\right)\right)$$

If $\boldsymbol{\delta}$ is entirely non-zero, this can be written somewhat more intuitively as

$$1 = \boldsymbol{\alpha}'\left(\frac{\boldsymbol{\iota}}{\boldsymbol{\iota} + \frac{\lambda}{-\boldsymbol{\delta}}} \circ \left(\boldsymbol{\iota} + \frac{\boldsymbol{\phi}}{1 + \frac{\lambda}{-\delta_0} - \phi_0}\right)\right)$$

$$= \left(\frac{\boldsymbol{\alpha}}{\boldsymbol{\iota} + \frac{\lambda}{-\boldsymbol{\delta}}}\right)' \left(\boldsymbol{\iota} + \frac{\boldsymbol{\phi}}{1 + \frac{\lambda}{-\delta_A} - \phi_A}\right).$$

This is (18).

## A.4. **Steady-state growth in the partially endogenous model**

The equations for the equilibrium growth rates take somewhat different form when technology is an exogenous factor than when it is endogenous.

If technology is exogenous, $\boldsymbol{\Phi}_{en,en} = \mathbf{0}$, $\mathbf{B}_{en,en} = \boldsymbol{\iota}_{en}\boldsymbol{\alpha}'_{en} - \mathbf{I}_{en,en}$, and $\mathbf{B}_{en,ex} = \boldsymbol{\iota}_{en}\boldsymbol{\alpha}'_{ex} + \boldsymbol{\Phi}_{en,ex}$. By the Sherman-Morrison formula,

$$-\mathbf{B}_{en,en}^{-1} = \mathbf{I}_{en,en} + \frac{\boldsymbol{\iota}_{en}\boldsymbol{\alpha}'_{en}}{1 - \boldsymbol{\alpha}'_{en}\boldsymbol{\iota}_{en}}.$$

So the growth rates of endogenous factors are

$$\begin{aligned}
\mathbf{z}_{en}^* &= -\mathbf{B}_{en,en}^{-1}\mathbf{B}_{en,ex}\boldsymbol{\delta}_{ex} \\
&= \left(\mathbf{I}_{en,en} + \frac{\boldsymbol{\iota}_{en}\boldsymbol{\alpha}'_{en}}{1 - \boldsymbol{\alpha}'_{en}\boldsymbol{\iota}_{en}}\right)(\boldsymbol{\iota}_{en}\boldsymbol{\alpha}'_{ex} + \boldsymbol{\Phi}_{en,ex})\boldsymbol{\delta}_{ex} \\
&= \left(\mathbf{I}_{en,en} + \frac{\boldsymbol{\iota}_{en}\boldsymbol{\alpha}'_{en}}{1 - \boldsymbol{\alpha}'_{en}\boldsymbol{\iota}_{en}}\right)(\boldsymbol{\iota}_{en}\boldsymbol{\alpha}'_{ex}\boldsymbol{\delta}_{ex} + \delta_0\boldsymbol{\phi}_{en}) \\
&= \boldsymbol{\iota}_{en}\boldsymbol{\alpha}'_{ex}\boldsymbol{\delta}_{ex} + \delta_0\boldsymbol{\phi}_{en} + \frac{\boldsymbol{\alpha}'_{en}\boldsymbol{\iota}_{en}\boldsymbol{\alpha}'_{ex}\boldsymbol{\delta}_{ex} + \delta_0\boldsymbol{\alpha}'_{en}\boldsymbol{\phi}_{en}}{1 - \boldsymbol{\alpha}'_{en}\boldsymbol{\iota}_{en}}\boldsymbol{\iota}_{en} \\
&= \delta_0\boldsymbol{\phi}_{en} + \frac{\boldsymbol{\alpha}'_{ex}\boldsymbol{\delta}_{ex}(1 - \boldsymbol{\alpha}'_{en}\boldsymbol{\iota}_{en}) + \boldsymbol{\alpha}'_{en}\boldsymbol{\iota}_{en}\boldsymbol{\alpha}'_{ex}\boldsymbol{\delta}_{ex} + \delta_0\boldsymbol{\alpha}'_{en}\boldsymbol{\phi}_{en}}{1 - \boldsymbol{\alpha}'_{en}\boldsymbol{\iota}_{en}}\boldsymbol{\iota}_{en} \\
&= \delta_0\boldsymbol{\phi}_{en} + \frac{\delta_0\boldsymbol{\alpha}'_{en}\boldsymbol{\phi}_{en} + \boldsymbol{\alpha}'_{ex}\boldsymbol{\delta}_{ex}}{1 - \boldsymbol{\alpha}'_{en}\boldsymbol{\iota}_{en}}\boldsymbol{\iota}_{en}.
\end{aligned}$$

Output growth is

$$\begin{aligned}
Z^* &= \boldsymbol{\alpha}'_{en}\mathbf{z}_{en}^* + \boldsymbol{\alpha}'_{ex}\mathbf{z}_{ex}^* \\
&= \boldsymbol{\alpha}'_{en}\left(\delta_0\boldsymbol{\phi}_{en} + \frac{\delta_0\boldsymbol{\alpha}'_{en}\boldsymbol{\phi}_{en} + \boldsymbol{\alpha}'_{ex}\boldsymbol{\delta}_{ex}}{1 - \boldsymbol{\alpha}'_{en}\boldsymbol{\iota}_{en}}\boldsymbol{\iota}_{en}\right) + \boldsymbol{\alpha}'_{ex}\boldsymbol{\delta}_{ex} \\
&= \delta_0\boldsymbol{\alpha}'_{en}\boldsymbol{\phi}_{en} + \frac{\boldsymbol{\alpha}'_{en}\boldsymbol{\iota}_{en}(\delta_0\boldsymbol{\alpha}'_{en}\boldsymbol{\phi}_{en} + \boldsymbol{\alpha}'_{ex}\boldsymbol{\delta}_{ex})}{1 - \boldsymbol{\alpha}'_{en}\boldsymbol{\iota}_{en}} + \boldsymbol{\alpha}'_{ex}\boldsymbol{\delta}_{ex} \\
&= \frac{\delta_0\boldsymbol{\alpha}'_{en}\boldsymbol{\phi}_{en} + \boldsymbol{\alpha}'_{ex}\boldsymbol{\delta}_{ex}}{1 - \boldsymbol{\alpha}'_{en}\boldsymbol{\iota}_{en}}.
\end{aligned}$$

On the other hand, if technology is endogenous,

$$\mathbf{B}_{en,en} = \mathbf{A}_{en,en} = \mathbf{I}_{en,en} - \boldsymbol{\Phi}_{en,en}$$

The inverse formula (46) for $\mathbf{A}$ carries over *mutatis mutandis*:

$$\mathbf{A}_{en,en}^{-1} = \mathbf{I}_{en,en} + \frac{1}{1 - \phi_0}\boldsymbol{\Phi}_{en,en}$$

Again using the Sherman-Morrison formula,

$$-\mathbf{B}_{en,en}^{-1} = \left(\mathbf{A}_{en,en} - \boldsymbol{\iota}_{en}\boldsymbol{\alpha}'_{en}\right)^{-1}$$

$$= \mathbf{A}_{en,en}^{-1} + \frac{\mathbf{A}_{en,en}^{-1}\boldsymbol{\iota}_{en}\boldsymbol{\alpha}'_{en}\mathbf{A}_{en,en}^{-1}}{1 - \boldsymbol{\alpha}'_{en}\mathbf{A}_{en,en}^{-1}\boldsymbol{\iota}_{en}}$$

$$= \mathbf{I}_{en,en} + \frac{\boldsymbol{\Phi}_{en,en}}{1-\phi_0} + \frac{\left(\mathbf{I}_{en,en} + \frac{\boldsymbol{\Phi}_{en,en}}{1-\phi_0}\right)\boldsymbol{\iota}_{en}\boldsymbol{\alpha}'_{en}\left(\mathbf{I}_{en,en} + \frac{\boldsymbol{\Phi}_{en,en}}{1-\phi_0}\right)}{1 - \boldsymbol{\alpha}'_{en}\left(\mathbf{I}_{en,en} + \frac{\boldsymbol{\Phi}_{en,en}}{1-\phi_0}\right)\boldsymbol{\iota}_{en}}$$

$$= \mathbf{I}_{en,en} + \frac{\boldsymbol{\Phi}_{en,en}}{1-\phi_0} + \frac{\left(\boldsymbol{\iota}_{en} + \frac{\boldsymbol{\phi}_{en}}{1-\phi_0}\right)\left(\boldsymbol{\alpha}'_{en} + \frac{\boldsymbol{\alpha}'_{en}\boldsymbol{\Phi}_{en,en}}{1-\phi_0}\right)}{1 - \boldsymbol{\alpha}'_{en}\boldsymbol{\iota}_{en} - \frac{\boldsymbol{\alpha}'_{en}\boldsymbol{\phi}_{en}}{1-\phi_0}}.$$

Meanwhile, since now $\boldsymbol{\Phi}_{en,ex} = \mathbf{0}$,

$$\mathbf{B}_{en,ex} = \boldsymbol{\iota}_{en}\boldsymbol{\alpha}'_{ex}.$$

Combining,

$$\mathbf{z}_{en}^{*} = -\mathbf{B}_{en,en}^{-1}\mathbf{B}_{en,ex}\boldsymbol{\delta}_{ex}$$

$$= \left(\mathbf{I}_{en,en} + \frac{\boldsymbol{\Phi}_{en,en}}{1-\phi_0} + \frac{\left(\boldsymbol{\iota}_{en} + \frac{\boldsymbol{\phi}_{en}}{1-\phi_0}\right)\left(\boldsymbol{\alpha}'_{en} + \frac{\boldsymbol{\alpha}'_{en}\boldsymbol{\Phi}_{en,en}}{1-\phi_0}\right)}{1 - \boldsymbol{\alpha}'_{en}\boldsymbol{\iota}_{en} - \frac{\boldsymbol{\alpha}'_{en}\boldsymbol{\phi}_{en}}{1-\phi_0}}\right)\boldsymbol{\iota}_{en}\boldsymbol{\alpha}'_{ex}\boldsymbol{\delta}_{ex}$$

$$= \boldsymbol{\alpha}'_{ex}\boldsymbol{\delta}_{ex}\boldsymbol{\iota}_{en} + \frac{\boldsymbol{\alpha}'_{ex}\boldsymbol{\delta}_{ex}\boldsymbol{\phi}_{en}}{1-\phi_0} + \frac{\left(\boldsymbol{\iota}_{en} + \frac{\boldsymbol{\phi}_{en}}{1-\phi_0}\right)\left(\boldsymbol{\alpha}'_{en}\boldsymbol{\iota}_{en} + \frac{\boldsymbol{\alpha}'_{en}\boldsymbol{\phi}_{en}}{1-\phi_0}\right)\boldsymbol{\alpha}'_{ex}\boldsymbol{\delta}_{ex}}{1 - \boldsymbol{\alpha}'_{en}\boldsymbol{\iota}_{en} - \frac{\boldsymbol{\alpha}'_{en}\boldsymbol{\phi}_{en}}{1-\phi_0}}$$

$$= \frac{\left(\boldsymbol{\iota}_{en} + \frac{\boldsymbol{\phi}_{en}}{1-\phi_0}\right)\left(1 - \boldsymbol{\alpha}'_{en}\boldsymbol{\iota}_{en} - \frac{\boldsymbol{\alpha}'_{en}\boldsymbol{\phi}_{en}}{1-\phi_0}\right)\boldsymbol{\alpha}'_{ex}\boldsymbol{\delta}_{ex} + \left(\boldsymbol{\iota}_{en} + \frac{\boldsymbol{\phi}_{en}}{1-\phi_0}\right)\left(\boldsymbol{\alpha}'_{en}\boldsymbol{\iota}_{en} + \frac{\boldsymbol{\alpha}'_{en}\boldsymbol{\phi}_{en}}{1-\phi_0}\right)\boldsymbol{\alpha}'_{ex}\boldsymbol{\delta}_{ex}}{1 - \boldsymbol{\alpha}'_{en}\boldsymbol{\iota}_{en} - \frac{\boldsymbol{\alpha}'_{en}\boldsymbol{\phi}_{en}}{1-\phi_0}}$$

$$= \frac{\boldsymbol{\iota}_{en} + \frac{\boldsymbol{\phi}_{en}}{1-\phi_0}}{1 - \boldsymbol{\alpha}'_{en}\boldsymbol{\iota}_{en} - \frac{\boldsymbol{\alpha}'_{en}\boldsymbol{\phi}_{en}}{1-\phi_0}}\boldsymbol{\alpha}'_{ex}\boldsymbol{\delta}_{ex}.$$

$$Z^{*} = \boldsymbol{\alpha}'_{en}\mathbf{z}_{en}^{*} + \boldsymbol{\alpha}'_{ex}\mathbf{z}_{ex}^{*}$$

$$= \boldsymbol{\alpha}'_{en}\left(\frac{\boldsymbol{\iota}_{en} + \frac{\boldsymbol{\phi}_{en}}{1-\phi_0}}{1 - \boldsymbol{\alpha}'_{en}\boldsymbol{\iota}_{en} - \frac{\boldsymbol{\alpha}'_{en}\boldsymbol{\phi}_{en}}{1-\phi_0}}\boldsymbol{\alpha}'_{ex}\boldsymbol{\delta}_{ex}\right) + \boldsymbol{\alpha}'_{ex}\boldsymbol{\delta}_{ex}$$

$$= \frac{\boldsymbol{\alpha}'_{en}\boldsymbol{\iota}_{en} + \frac{\boldsymbol{\alpha}'_{en}\boldsymbol{\phi}_{en}}{1-\phi_0}}{1 - \boldsymbol{\alpha}'_{en}\boldsymbol{\iota}_{en} - \frac{\boldsymbol{\alpha}'_{en}\boldsymbol{\phi}_{en}}{1-\phi_0}}\boldsymbol{\alpha}'_{ex}\boldsymbol{\delta}_{ex} + \boldsymbol{\alpha}'_{ex}\boldsymbol{\delta}_{ex}$$

$$= \frac{\boldsymbol{\alpha}'_{ex}\boldsymbol{\delta}_{ex}}{1 - \boldsymbol{\alpha}'_{en}\boldsymbol{\iota}_{en} - \frac{\boldsymbol{\alpha}'_{en}\boldsymbol{\phi}_{en}}{1-\phi_0}}$$

$$= \frac{\boldsymbol{\alpha}'_{ex}\boldsymbol{\delta}_{ex}(1-\phi_0)}{-(\boldsymbol{\alpha}'_{en}\boldsymbol{\phi}_{en} + (1-\phi_0)\boldsymbol{\alpha}'_{en}\boldsymbol{\iota}_{en})}.$$

Using (21), the results for output growth can be consolidated as

$$Z^* = \frac{1}{|-\mathbf{B}_{en,en}|} \cdot \begin{cases} \boldsymbol{\alpha}'_{ex}\boldsymbol{\delta}_{ex} + \delta_A \boldsymbol{\alpha}'_{en}\boldsymbol{\phi}_{en} & \text{if technology is exogenous} \\ \boldsymbol{\alpha}'_{ex}\boldsymbol{\delta}_{ex}(1-\phi_A) & \text{if technology is endogenous} \end{cases} \tag{49}$$

If the sufficient condition for instability (21) holds, then the denominator in (49) is negative. For positive growth, the numerators must be negative too, which would require, in the exogenous case, that $\boldsymbol{\alpha}'_{en}\boldsymbol{\phi}_{en}$ be adequately large and negative (technological growth substantially damps investment in endogenous factors); in the endogenous case, $\phi_A > 1$ (returns to R&D are strongly increasing). Conceding those restrictions results in a picture of a system that operates in one of two main modes, depending on the prominence of exogenous factors in production. With enough exogeneity, as in the Solow-Swan model, the equilibrium growth rate is positive and stable. Otherwise it is negative and unstable.

# B. Online Appendix: Confirming and characterizing the solutions of the Feller/CIR diffusion

The reflecting- and absorbing-barrier solutions for the Feller/CIR diffusion mentioned in section 3.1 are long-established. However, available derivations and confirmations of the solutions are complex and do not treat the two and a unified way. This appendix aims to confirm that the asserted solutions indeed satisfy the corresponding Kolmogorov forward equation, and derive some of their properties, using little more than ordinary calculus.

## B.1. Problem

We aim to inventory the solutions of the stochastic differential equation,

$$dX_t = (bX_t + c)dt + \sqrt{2aX_t}dW_t \tag{50}$$

where $a > 0$ and $W_t$ is standard Brownian (Wiener) motion. The process is singular at $X_t = 0$, in that the second (diffusion) term in vanishes there. By "inventory solutions," I mean to state the transition densities compatible with this stochastic equation of motion as well as with associated Markovian boundary conditions.

The corresponding Kolmogorov forward equation, which Feller (1951b) solves, is

$$\frac{\partial u}{\partial t} = -\frac{\partial}{\partial X}\big((bX + c)u\big) + \frac{\partial^2}{\partial X^2}(aXu),$$

in which $u(t, X)$ is the transition probability density at time $t$ conditional on a starting value $X_0$. As a general matter, the first-order term captures the effect on the time derivative of the density of the deterministic component of the SDE and the second-order term the effect of random diffusion. Put another way,

$$\frac{\partial u}{\partial t} = -\frac{\partial J}{\partial X} \tag{51}$$

where

$$\begin{aligned} J &:= (bX + c)u - \frac{\partial}{\partial X}(aXu) \\ &= (a\nu + bX)u - aX\frac{\partial u}{\partial X} \end{aligned} \tag{52}$$

with $\nu := c/a - 1$. $J(t, X)$ is the *flux* of the diffusion.

## B.2. Reducing the problem

Just as the transformation $Y_t = X_t^{-1/B}$ reduces the four-parameter diffusion (33) to the three-parameter Feller/CIR (50), further transformations can reduce the Feller/CIR to a one-parameter model, the half squared Bessel process (Göing-Jaeschke and Yor 2003).[39] First, we normalize scale in (50) by dividing $X_t$ by $a$. Second, we observe this SDE's tendency to exponential growth or decay—for the sake of exposition, assume growth—which is governed by the drift term $bX_t dt$. A second rescaling, by $e^{-bt}$, might remove that. To combine these steps, we define a new stochastic variable:

$$Z_t := \frac{e^{-bt}}{a} X_t \tag{53}$$

Differentiating both sides of that with ordinary calculus gives

$$dZ_t = \frac{e^{-bt}}{a}(dX_t - bX_t dt).$$

Substituting those two equations into (50) and rearranging,

$$dZ_t = \tilde{c}e^{-bt}dt + \sqrt{2Z_t e^{-bt}}dW_t, \text{ with } \tilde{c} := c/a. \tag{54}$$

The rescaling indeed simplifies by removing the variance coefficient $a$ and the exponential growth term $bX_t dt$. But it complicates by introducing exponential decay into both the drift coefficient ($\tilde{c}e^{-bt}$) and the variance of the diffusion term ($2Z_t e^{-bt}$). Yet since the cumulative variance of a Wiener process equals elapsed time, these two appearances of $e^{-bt}$ are jointly equivalent to a time change (Oksendal 2013, Theorem 8.5.7).[40] That is, to observe the evolution of sample paths according to (54) is equivalent to observing paths according to (54) with the $e^{-bt}$ factors deleted, but subject to a slow-motion effect that causes playback speed to decay exponentially. To express that notion, we define a new, slowing clock $\tilde{t}$ subject to $\tilde{t}(0) = 0$ and $\partial\tilde{t}/\partial t = e^{-bt}$. That is,

$$\tilde{t} := \int_0^t e^{-bs} ds = \begin{cases} (1 - e^{-bt})/b & \text{if } b \neq 0 \\ t & \text{if } b = 0 \end{cases}. \tag{55}$$

The time-transformed variable is

$$\tilde{Z}_{\tilde{t}} := Z_t.$$

Then

$$d\tilde{Z}_{\tilde{t}} = \tilde{c}d\tilde{t} + \sqrt{2\tilde{Z}_{\tilde{t}}}d\widetilde{W}_{\tilde{t}}, \tag{56}$$

where $d\widetilde{W}_{\tilde{t}}$ is another standard Wiener process. This SDE is a *half* squared Bessel process. $2\tilde{Z}_t$ can be realized as the sum of $2\tilde{c}$ squared, standard independent Brownian motions, at least when $2\tilde{c}$ is a positive integer. But we allow $\tilde{c}$ to be any real.

[39] I thank Goran Peskir for pointing out the connection to the squared Bessel. I halve the squared Bessel for beauty.
[40] This would not be the case if $X_t$ entered the diffusion coefficient of (50) with a power other than 1/2.

The Kolmogorov forward equation corresponding to the simpler SDE (56) is:

$$\frac{\partial u}{\partial \tilde{t}} = -\frac{\partial}{\partial \tilde{Z}_{\tilde{t}}}(\tilde{c}u) + \frac{\partial^2}{\partial \tilde{Z}_{\tilde{t}}^2}(\tilde{Z}_{\tilde{t}}u). \tag{57}$$

Solutions for this diffusion equation can be transformed into ones for $X_t$ in the Feller/CIR diffusion by inverting (53), and then on to the "superexponential" diffusion via $Y_t = X_t^{-1/B}$.

## B.3. Constructing solutions for the half squared Bessel

The solutions for the half squared Bessel are constructed as follows. First define the standard gamma density function,

$$f_\Gamma(x;\alpha) := \frac{e^{-x}x^{\alpha-1}}{\Gamma(\alpha)}.$$

When $\alpha > 0$, this gives rise to a proper probability distribution over $(0,\infty)$. In using it as the basis for solutions to Feller's diffusion equation, we will extend the function beyond those domains—to $\alpha \leq 0$ and to $x = 0$. A remarkable feature of $f_\Gamma$ is that derivatives and integrals with respect to the argument are differences and countable sums in the parameter. As for derivatives,

$$\frac{\partial}{\partial x}f_\Gamma(x;\alpha) = \frac{(\alpha-1)e^{-x}x^{\alpha-2}}{\Gamma(\alpha)} - \frac{e^{-x}x^{\alpha-1}}{\Gamma(\alpha)} = f_\Gamma(x;\alpha-1) - f_\Gamma(x;\alpha) = -\nabla f_\Gamma(x;\alpha), \tag{58}$$

where $\nabla$ is the unit-interval backward difference operator with respect to the parameter. Meanwhile, repeated application of integration by parts to the integral definition of the cumulative gamma distribution function, $F_\Gamma$, gives the identity

$$F_\Gamma(x;\alpha) = \sum_{m=0}^{\infty} f_\Gamma(x;\alpha+m+1) \tag{59}$$

when $\alpha > 0$. In fact, to extend to the domain of $F_\Gamma$, we will take (59) to define the function for negative, non-integer $\alpha$. (For $\alpha = 0$, we take the limiting value, which is 1. For negative integer $\alpha$, the formula is indeterminate.)

The next stepping-stone is an expression for the noncentral $\chi^2$ density:

$$f_{\chi^2}(x;\lambda,\nu) := \sum_{m=0}^{\infty} f_\Gamma(\lambda;m+1)f_\Gamma(x;m+\nu+1). \tag{60}$$

We restrict to $\lambda, x > 0$. If $\nu \geq 0$, all the terms in the sum are positive, and we have

$$f_{\chi^2}(x;\lambda,\nu) < \left(\sum_{m=0}^{\infty} f_\Gamma(\lambda;m+1)\right)\left(\sum_{m=0}^{\infty} f_\Gamma(x;m+\nu+1)\right) = F_\Gamma(\lambda;0)F_\Gamma(x;\nu) = F_\Gamma(x;\nu) < 1.$$

Thus the sum in (60) is bounded above. If $\nu < 0$, a finite number of early terms may be negative. But all terms for $m \geq |\nu|$ will be positive, so the series converges.

This formulation of the noncentral $\chi^2$ function is nonstandard. Let $k$ be the familiar degrees-of-freedom index associated with the distribution; then the parameter $\nu$ used here equals $k/2 - 1$. If $k$ is a positive integer, and if $x_i$ are $k$ normal variates with variance 1 and means $\mu_i$ satisfying $\lambda = \frac{1}{2}\sum_i \mu_i^2$, then $f_{\chi^2}(x;\lambda,\nu)$ is the density of $\frac{1}{2}\sum_i x_i^2$. The usual definition drops the factors of 1/2.

The formulation (60) maps directly to a more familiar presentation of the noncentral $\chi^2$ distribution, as a Poisson mixture of central $\chi^2$ distributions. Within (60), $f_\Gamma(\lambda; m+1)$ is the Poisson probability $f_P(m;\lambda)$ while $f_\Gamma(x; m+\nu+1)$ is the density of $x$ when $2x$ has the conventional central $\chi^2$ distribution with $k+2m$ degrees of freedom, which we could call the central $\chi^2/2$ distribution.

Alongside the noncentral $\chi^2$, I define what I call the Feller density function:

$$f_{-\chi^2}(x;\lambda,\nu) \coloneqq \sum_{m=0}^{\infty} f_\Gamma(\lambda; m-\nu+1) f_\Gamma(x; m+1), \tag{61}$$

again restricting to $\lambda, x > 0$. With this notation, it is natural to write $f_{\pm\chi^2}$ to represent the noncentral $\chi^2$ and Feller density functions as a pair. The two are connected by

$$f_{\chi^2}(x;\lambda,\nu) = f_{-\chi^2}(\lambda; x, -\nu). \tag{62}$$

In fact, the two functions coincide when $\nu$ is an integer. The match is immediate when $\nu = 0$. When $\nu$ is a negative integer, the first $|\nu|$ terms of (60) are 0. So then

$$\begin{aligned} f_{\chi^2}(x;\lambda,\nu) &= \sum_{m=-\nu}^{\infty} f_\Gamma(\lambda; m+1) f_\Gamma(x; m+\nu+1) \\ &= \sum_{m=0}^{\infty} f_\Gamma(\lambda; m-\nu+1) f_\Gamma(x; m+1) \\ &= f_{-\chi^2}(x;\lambda,\nu). \end{aligned}$$

And when $\nu$ is a positive integer, the first $\nu$ terms of (61) are 0, leading again to equality.

A more common way of writing these density functions is

$$f_{\pm\chi^2}(x;\lambda,\nu) = e^{-\lambda-x}\left(\frac{x}{\lambda}\right)^{\nu/2} I_{\pm\nu}\left(2\sqrt{\lambda x}\right)$$

where $I_{\pm\nu}$ is the modified Bessel function of the first kind:

$$I_{\pm\nu}(z) \coloneqq \sum_{m=0}^{\infty} \frac{(z/2)^{2m\pm\nu}}{\Gamma(m+1)\Gamma(m\pm\nu+1)}.$$

The modified Bessel function of the second kind is

$$K_{\pm\nu}(z) \coloneqq \frac{\pi}{2\sin\nu\pi}\left(I_{-\nu}(z) - I_\nu(z)\right),$$

which is known to go to 0 as $z \to \infty$. It follows that $f_{\chi^2} - f_{-\chi^2}$ converges to 0 as $x \to \infty$. The two densities differ, rather, in their behavior as $x \downarrow 0$.

Figure 4 in the main text plots $f_{\chi^2}$ and $f_{-\chi^2}$ for $\lambda = 1$ and $\nu = -3.0, -2.5, \dots, +3.0$. As noted, when $\nu$ is an integer, the two coincide. Otherwise, the two fork toward the left; in some of these cases $f_{\chi^2}$ diverges to infinity. Notably, some of the plots exhibit features inadmissible for a probability distribution: they take negative values, or perhaps diverge rapidly enough to have infinite total integral.

We use the densities $f_{\pm\chi^2}$ to define diffusions that we will show satisfy the half squared Bessel process (56) and, after transformation, the Feller process (50). We construct the diffusions by expressing the inputs to $f_{\pm\chi^2}$ as functions of time and the primary parameters in (50):

$$\begin{aligned} x &:= \tilde{Z}_{\tilde{t}}/\tilde{t} \\ \lambda &:= \tilde{Z}_0/\tilde{t} \\ \nu &:= \tilde{c} - 1 \end{aligned} \tag{63}$$

Adjusting for the change in variables from $\tilde{Z}_t$ to $x$, whose Jacobian is $1/\tilde{t}$, our two asserted fundamental solutions to the squared Bessel diffusion equation are:

$$f^*_{\pm\chi^2}\left(\tilde{Z}_{\tilde{t}}; \tilde{Z}_0, \tilde{t}, \tilde{c}\right) := \frac{1}{\tilde{t}} f_{\pm\chi^2}\left(\frac{\tilde{Z}_{\tilde{t}}}{\tilde{t}}; \frac{\tilde{Z}_0}{\tilde{t}}, \nu\right) \tag{64}$$

Incorporating the scale and time transformations (53) and (55), and the Jacobian thereof, namely $e^{-bt}/a$, the fundamental solutions for the Feller diffusion (50) are in (42). Algebraic manipulations confirm that $f^*_{-\chi^2}$ coincides with Feller's (1951b, eq. 6.2) explicit solution for $\nu \leq 0$, except that Feller erroneously includes a factor of $(2b)^\nu$.

The transition densities for $Y_t$ are in (43). Figure 15 shows how the Feller/CIR is connected to other common stochastic models through parameter restrictions and the power transform. (The figure uses the definition $\gamma := -1/B$.) The figure shows how any solution for the transition density under the Feller/CIR diffusion (50) is bequeathed to all the other diffusions via these links.

**Figure 15. Relationships among univariate stochastic models**

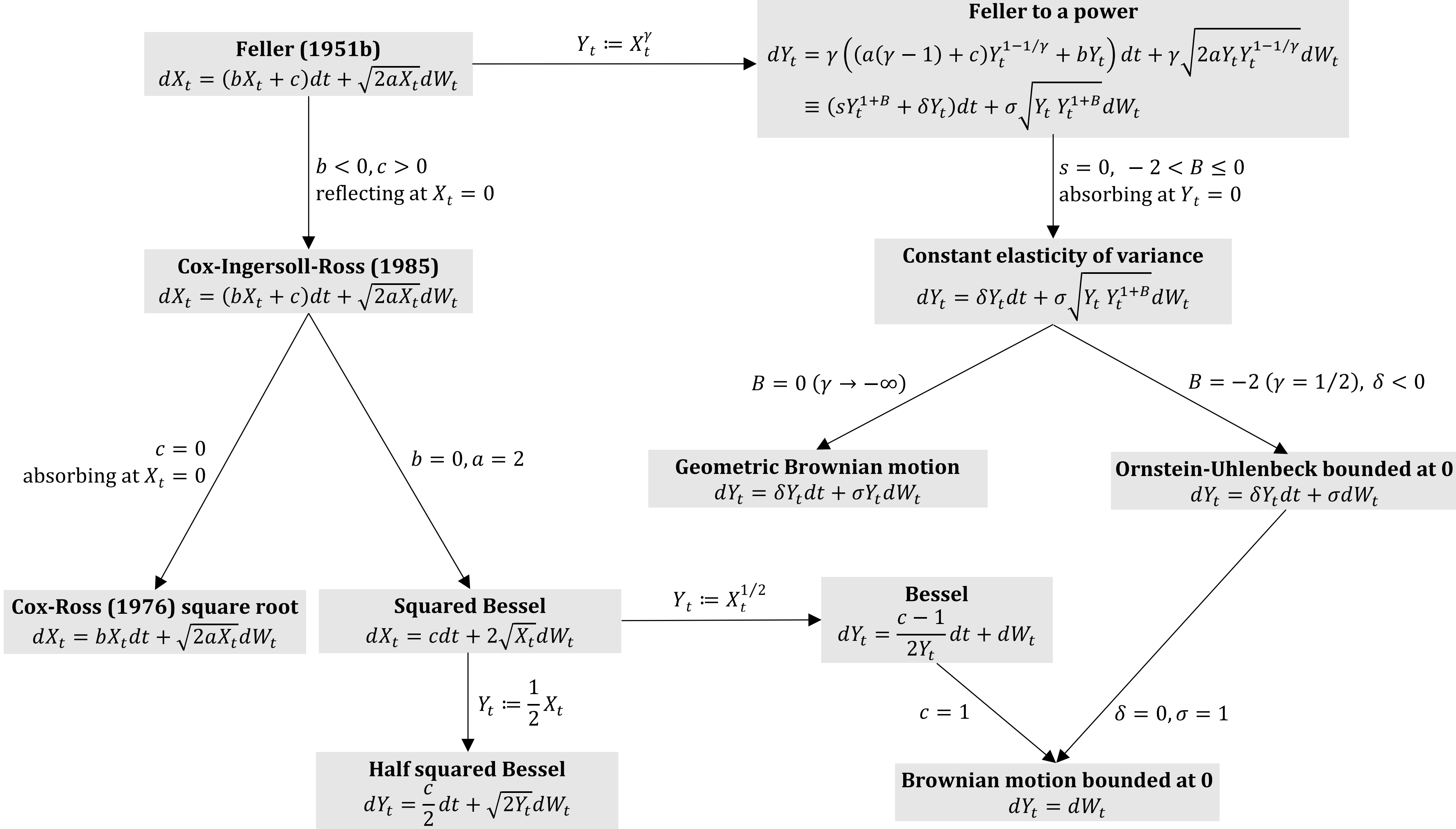

## B.4. Confirming the solutions

To confirm the asserted solutions, we need only consider the half squared Bessel subcase. Our task is to show the asserted solutions (64) indeed solve the forward diffusion equation (57).

## B.5. Recurrence relation

Let $L$ be the lag operator with respect to the last parameter of any of the above densities and diffusions, with a step interval of 1. For example, $\left(Lf_{\chi^2}\right)(x;\lambda,\nu) = f_{\chi^2}(x;\lambda,\nu-1)$. Then the gamma density function obeys

$$(aL^{-1}f_\Gamma)(x;\alpha) = xf_\Gamma(x;\alpha).$$

$f_{\pm\chi^2}$ inherit a form of this relation. For $f_{\chi^2}$,

$$\begin{aligned}
xf_{\chi^2}(x;\lambda,\nu-1) &= \sum_{m=0}^{\infty} f_\Gamma(\lambda;m+1)xf_\Gamma(x;m+\nu) \\
&= \sum_{m=0}^{\infty} f_\Gamma(\lambda;m+1)(m+\nu)f_\Gamma(x;m+\nu+1) \\
&= \sum_{m=0}^{\infty} mf_\Gamma(\lambda;m+1)f_\Gamma(x;m+\nu+1) + \nu\sum_{m=0}^{\infty} f_\Gamma(\lambda;m+1)f_\Gamma(x;m+\nu+1) \\
&= \sum_{m=1}^{\infty} mf_\Gamma(\lambda;m+1)f_\Gamma(x;m+\nu+1) + \nu f_{\chi^2}(x;\lambda,\nu) \\
&= \sum_{m=0}^{\infty} (m+1)f_\Gamma(\lambda;m+2)f_\Gamma(x;m+\nu+2) + \nu f_{\chi^2}(x;\lambda,\nu) \\
&= \sum_{m=0}^{\infty} \lambda f_\Gamma(\lambda;m+1)f_\Gamma(x;m+\nu+2) + \nu f_{\chi^2}(x;\lambda,\nu) \\
&= \lambda f_{\chi^2}(x;\lambda,\nu+1) + \nu f_{\chi^2}(x;\lambda,\nu)
\end{aligned}$$

Swapping the symbols $x$ and $\lambda$ in the above, replacing $\nu$ with $-\nu$, and applying (62) produces the same relation for $f_{-\chi^2}$. To encapsulate, we write,

$$\nu f_\pm = [xL - \lambda L^{-1}]f_\pm.$$

By way of (42), the recurrence applies to the asserted Feller diffusion solutions as well:

$$\nu f_\pm^* = \nu\frac{\partial x}{\partial X_t}f_\pm = \frac{\partial x}{\partial X_t}[xL - \lambda L^{-1}]f_\pm^*. \tag{65}$$

### B.5.1 Cross-sectional derivative relation

The derivative identity (58) for the gamma function transfers directly to $f_{\chi^2}$ via linearity in (60):

$$\frac{\partial}{\partial x}f_{\chi^2}(x;\lambda,\nu) = -\nabla f_{\chi^2}(x;\lambda,\nu).$$

With more work, we get something similar with respect to $\lambda$:

$$\begin{aligned}\frac{\partial}{\partial\lambda}f_{\chi^2}(x;\lambda,\nu) &= \sum_{m=0}^{\infty}(-\nabla)f_\Gamma(\lambda;m+1)f_\Gamma(x;m+\nu+1)\\ &= \sum_{m=0}^{\infty}\big(f_\Gamma(\lambda;m)-f_\Gamma(\lambda;m+1)\big)f_\Gamma(x;m+\nu+1)\\ &= \sum_{m=0}^{\infty}f_\Gamma(\lambda;m)f_\Gamma(x;m+\nu+1)-\sum_{m=0}^{\infty}f_\Gamma(\lambda;m+1)f_\Gamma(x;m+\nu+1)\\ &= \sum_{m=1}^{\infty}f_\Gamma(\lambda;m)f_\Gamma(x;m+\nu+1)-f_{\chi^2}(x;\lambda,\nu)\\ &= \sum_{m=0}^{\infty}f_\Gamma(\lambda;m+1)f_\Gamma(x;m+\nu+2)-f_{\chi^2}(x;\lambda,\nu)\\ &= \Delta f_{\chi^2}(x;\lambda,\nu),\end{aligned}$$

in which $\Delta$ is the one-unit forward difference operator with respect to $\nu$. Once more, applying (62) and replacing $-\nu$ with $\nu$ throughout produces the same identities for $f_{-\chi^2}$. Thus we will write more compactly,

$$\begin{aligned}\frac{\partial}{\partial x}f_\pm &= -\nabla f_\pm\\ \frac{\partial}{\partial\lambda}f_\pm &= \Delta f_\pm.\end{aligned}$$

These relations too bootstrap to the asserted Feller diffusion solutions, including the half squared Bessel; the first bootstrapped relation we will use. Starting from (42),

$$\begin{aligned}\frac{\partial}{\partial X_t}f_\pm^* &= \frac{\partial}{\partial X_t}\left(\frac{\partial x}{\partial X_t}f_\pm\right)\\ &= \frac{\partial x}{\partial X_t}\frac{\partial x}{\partial X_t}\frac{\partial}{\partial x}f_\pm\\ &= -\frac{\partial x}{\partial X_t}\frac{\partial x}{\partial X_t}\nabla f_\pm\\ &= -\frac{\partial x}{\partial X_t}\nabla f_\pm^*.\end{aligned} \tag{66}$$

### B.5.2 Flux

Note that

$$\begin{aligned}x &= \frac{\partial x}{\partial X_t}X_t = \frac{e^{-bt}}{a\tilde{t}}X_t = \frac{be^{-bt}}{a(1-e^{-bt})}X_t\\ \lambda &= \frac{\partial\lambda}{\partial X_0}X_0 = \frac{1}{a\tilde{t}}X_0 = \frac{b}{a(1-e^{-bt})}X_t\\ \frac{\partial\lambda}{\partial X_0} &= \frac{\partial x}{\partial X_t}e^{bt} = \frac{\partial x}{\partial X_t}+\frac{b}{a}\end{aligned}$$

Starting with the statement of the flux in (52), applying the recurrence relation (65), derivative rule (66), and then identities just above, the flux of the asserted solutions is

$$
\begin{aligned}
J &= (av + bX_t)f_\pm^* - aX_t\frac{\partial}{\partial X_t}f_\pm^* \\
&= a\left[vf_\pm^* + \frac{b}{a}X_tf_\pm^* - X_t\left(-\frac{\partial x}{\partial X_t}\nabla f_\pm^*\right)\right] \\
&= a\left[xL - \lambda L^{-1} + \frac{b}{a}X_t + \frac{\partial x}{\partial X_t}X_t\nabla\right]f_\pm^* \\
&= a\left[\frac{\partial x}{\partial X_t}X_tL - \frac{\partial x}{\partial X_t}e^{bt}X_0L^{-1} + \frac{b}{a}X_t + \frac{\partial x}{\partial X_t}X_t - \frac{\partial x}{\partial X_t}X_tL\right]f_\pm^* \\
&= a\left[-\frac{\partial x}{\partial X_t}e^{bt}X_0L^{-1} + \frac{\partial x}{\partial X_t}X_te^{bt}\right]f_\pm^* \\
&= \frac{1}{\tilde{t}}[X_t - X_0L^{-1}]f_\pm^*
\end{aligned}
\tag{67}
$$

In the special case of the half squared Bessel, $a = 1$ and $b = 0$, so this simplifies to

$$
J = \frac{1}{t}[\tilde{Z}_t - \tilde{Z}_0L^{-1}]f_\pm^*. \tag{68}
$$

### B.5.3 Plugging into Feller's diffusion equation

At last, to confirm the asserted solutions, we check that for the half squared Bessel, they satisfy the requirement (51) that the spatial derivative of the flux is the negative of the time derivative of the density. The spatial derivative of the flux (68) is

$$
\begin{aligned}
\frac{\partial J}{\partial \tilde{Z}_t} &= \frac{1}{t}\frac{\partial}{\partial \tilde{Z}_t}[\tilde{Z}_t - \tilde{Z}_0L^{-1}]f_\pm^* \\
&= \frac{1}{t}\left[1 + \tilde{Z}_t\frac{\partial}{\partial \tilde{Z}_t} - \tilde{Z}_0L^{-1}\frac{\partial}{\partial \tilde{Z}_t}\right]\frac{1}{t}f_\pm \\
&= \frac{1}{t}\left[1 + \tilde{Z}_t\left(-\frac{1}{t}\nabla\right) - \tilde{Z}_0L^{-1}\left(-\frac{1}{t}\nabla\right)\right]\frac{1}{t}f_\pm \\
&= \frac{1}{t^2}[t - \tilde{Z}_t\nabla + \tilde{Z}_0L^{-1}\nabla]f_\pm^* \\
&= \frac{1}{t^2}[t - (\tilde{Z}_t\nabla - \tilde{Z}_0\Delta)]f_\pm^*.
\end{aligned}
$$

The time derivate of the density is

$$
\begin{aligned}
\frac{\partial f_\pm^*}{\partial t} &= \frac{\partial}{\partial t}\left(\frac{1}{t}f_\pm(x;\lambda,v)\right) \\
&= \left[-\frac{1}{t^2} + \frac{1}{t}\left(\frac{\partial x}{\partial t}\frac{\partial}{\partial x} + \frac{\partial \lambda}{\partial t}\frac{\partial}{\partial \lambda}\right)\right]f_\pm \\
&= -\frac{1}{t^2}\left[1 - t\left(-\frac{\tilde{Z}_t}{t^2}(-\nabla) - \frac{\tilde{Z}_0}{t^2}\Delta\right)\right]f_\pm \\
&= -\frac{1}{t^2}[t - (\tilde{Z}_t\nabla - \tilde{Z}_0\Delta)]f_\pm^*,
\end{aligned}
$$

as desired.

## B.6. Consistency with the initial condition

Confirming that these diffusions satisfy the initial condition $Y_t|_{t=0} = Y_0$ is complicated by the change in the mathematical character of the diffusion in the instant after $t = 0$. At $t = 0$, the distribution is by assumption concentrated with infinite density at a single point; after, it is spread across all positive reals. This change in character manifests in the fact that the mapping (63) of $\tilde{Z}_0, \tilde{Z}_t \mapsto \lambda, x$ is not defined at $t = 0$. So we must instead investigate the behavior of the diffusions in the limit as $t \downarrow 0$.

Starting with the noncentral $\chi^2$ diffusion,

$$\lim_{t\downarrow 0} f^*_{\chi^2}(\tilde{Z}_t; \tilde{Z}_0, t, \tilde{c}) = \lim_{t\downarrow 0} \frac{1}{t} \sum_{m=0}^{\infty} f_\Gamma\left(\frac{\tilde{Z}_0}{t}; m+1\right) f_\Gamma\left(\frac{\tilde{Z}_t}{t}; m+\nu+1\right)$$
$$= \lim_{t\downarrow 0} \sum_{m=0}^{\infty} \frac{1}{\sqrt{t}} f_P\left(m; \frac{\tilde{Z}_0}{t}\right) \frac{1}{\sqrt{t}} f_P\left(m+\nu; \frac{\tilde{Z}_t}{t}\right),$$

where $f_P(z; \beta)$ is again the Poisson density function.

In general, if $z \sim f_P(\cdot; \beta)$, then as $\beta \to \infty$, $f_P(z; \beta)$ becomes well approximated by the normal density with the same mean and variance as $f_P$, namely $\beta$ and $\beta$. (More precisely, $z/\sqrt{\beta}$, which has density $\sqrt{\beta} f_P(z; \beta)$, converges in distribution to $\mathcal{N}(\sqrt{\beta}, 1)$.) So, using $f_\mathcal{N}$ to represent the normal density parameterized by mean and variance, we develop the above as

$$= \lim_{t\downarrow 0} \sum_{m=0}^{\infty} \frac{1}{\sqrt{t}} f_\mathcal{N}\left(m; \frac{\tilde{Z}_0}{t}, \frac{\tilde{Z}_0}{t}\right) \frac{1}{\sqrt{t}} f_\mathcal{N}\left(m+\nu; \frac{\tilde{Z}_t}{t}, \frac{\tilde{Z}_t}{t}\right)$$
$$= \lim_{t\downarrow 0} \sum_{m=0}^{\infty} f_\mathcal{N}\left(m\sqrt{t}; \frac{\tilde{Z}_0}{\sqrt{t}}, \tilde{Z}_0\right) f_\mathcal{N}\left(m\sqrt{t}; \frac{\tilde{Z}_t}{\sqrt{t}} - \nu\sqrt{t}, \tilde{Z}_t\right)$$
$$= \lim_{t\downarrow 0} \frac{1}{\sqrt{t}} \sum_{m=0}^{\infty} \sqrt{t} f_\mathcal{N}\left(m\sqrt{t}; \frac{\tilde{Z}_0}{\sqrt{t}}, \tilde{Z}_0\right) f_\mathcal{N}\left(m\sqrt{t}; \frac{\tilde{Z}_t}{\sqrt{t}} - \nu\sqrt{t}, \tilde{Z}_t\right).$$

The limit of the sum is a Riemann integral of the pointwise product of two normal curves. So we have

$$= \lim_{t\downarrow 0} \frac{1}{\sqrt{t}} \int_0^\infty f_\mathcal{N}\left(m\sqrt{t}; \frac{\tilde{Z}_0}{\sqrt{t}}, \tilde{Z}_0\right) f_\mathcal{N}\left(m\sqrt{t}; \frac{\tilde{Z}_t}{\sqrt{t}} - \nu\sqrt{t}, \tilde{Z}_t\right) dm$$

As $t \downarrow 0$ the centers of the normal distributions go to $+\infty$ even as their variances hold constant. So we may change the lower bound of the integral to $-\infty$. Moreover, if $\tilde{Z}_0 \neq \tilde{Z}_t$, then the means of the two normal curves in the last version become more distant from each other even as the variances hold constant, driving the integral of their pointwise product to zero. Thus the limit works out to the Dirac delta function $\delta_{\tilde{Z}_0}(\tilde{Z}_t)$, which is the initial condition.

The same holds for $f^*_{-\chi^2}(X_t; X_0, t, a, b, \nu)$.

## B.7. Characteristics of the solutions

Although the functions $f^*_{\pm\chi^2}$ indeed solve the Kolmogorov forward equation for the half squared Bessel process, they do not behave like proper diffusions for all parameter values. Sometimes, for example, they take negative values. Here we review such properties and their dependence on $\nu$.

### B.7.1 Positivity

When $\alpha \geq 0$, $f_\Gamma(x;\alpha)$ never takes negative values over $[0,\infty)$. As a result, $f^*_{\pm\chi^2}$ are everywhere non-negative, i.e., positivity preserving, if their evaluation requires passing only non-negative parameter values to $f_\Gamma$. In particular, examining the definitions (60), (61), and (42), we see that $f^*_{\chi^2}$ is positive preserving when $\nu \geq -1$, as is $f^*_{-\chi^2}$ when $\nu \leq 1$.

Integer values for $\nu$ generate one class of exceptions to those generalizations—but not a very interesting one. Since $f^*_{\chi^2} = f^*_{-\chi^2}$ for integer $\nu$, if one diffusion is positive—and one always is—then both are.

Summary:

$f^*_{\chi^2}$ preserves positivity if $\nu \geq -1$ or $\nu$ is an integer

$f^*_{-\chi^2}$ preserves positivity if $\nu \leq 1$ or $\nu$ is an integer

Note that we have not asserted that these diffusions fail to preserve positivity for all values of $\nu$ outside the indicated sets. But that is largely moot because outside these sets, the diffusions are poorly behaved in other respects, as we will see.

### B.7.2 Density near zero

The density of $f_{\chi^2}$ in the $x \downarrow 0$ limit is

$$\lim_{x\downarrow 0}\sum_{m=0}^{\infty} f_\Gamma(\lambda;m+1)f_\Gamma(x;m+\nu+1) = \sum_{m=0}^{\infty} f_\Gamma(\lambda;m+1)\lim_{x\downarrow 0} f_\Gamma(x;m+\nu+1)$$
$$= \sum_{m=0}^{\infty} f_\Gamma(\lambda;m+1)\frac{1}{\Gamma(m+\nu+1)}\lim_{x\downarrow 0} x^{m+\nu}.$$

If $\nu \in \mathbb{Z}$, $f_{\chi^2}$ equals $f_{-\chi^2}$, which we analyze just below. Otherwise, the $\Gamma(\cdot)$ denominator above is always bounded. As a result, $f_{\chi^2}$ has a zero or pole at 0 of order $|\nu|$: if $\nu > 0$, the $m = 0$ term converges to 0 as $x^\nu$, and later terms do so more rapidly; if $\nu$ is a negative non-integer, the same term diverges most rapidly, as $x^\nu$. In the latter case, the sign of that dominating divergent term is that of $\Gamma(\nu+1)$, namely $(-1)^{\lceil\nu\rceil}$, as seen in Figure 4.

Under $f_{-\chi^2}$, the density near 0 is finite for all $\nu$. For we have

$$\lim_{x\downarrow 0} f_{-\chi^2} = \sum_{m=0}^{\infty} f_\Gamma(\lambda;m-\nu+1)\lim_{x\downarrow 0} f_\Gamma(x;m+1);$$

and observing that the limit in the right expression is 1 for $m = 0$, and zero otherwise, leads to

$$= f_\Gamma(\lambda;1-\nu).$$

This is non-negative for $\nu \leq 1$ and equals zero only when $\nu$ is a positive integer—features also visible in Figure 4.

Shifting from distribution to the full Feller diffusion, under $f^*_{-\chi^2}$, the density near the zero boundary at a given time is

$$\lim_{X_t\downarrow 0} f^*_{-\chi^2} = \frac{e^{-bt}}{a\tilde{t}}\lim_{x\downarrow 0} f_{-\chi^2} = \frac{e^{-bt}}{a\tilde{t}} f_\Gamma(\lambda;1-\nu) = \frac{e^{-bt}}{a\tilde{t}} f_\Gamma\left(\frac{\tilde{Z}_0}{\tilde{t}};1-\nu\right) = \frac{e^{-bt}}{a\tilde{t}} f_\Gamma\left(\frac{X_0}{a\tilde{t}};1-\nu\right) = \frac{1-\nu}{X_0\tilde{t}} f_\Gamma(\lambda;2-\nu),$$

which is finite and non-zero except when $\nu$ is a positive integer, when the limit is 0.

Summary:

$$\lim_{x\downarrow 0} f_{\chi^2} = \begin{cases} 0 & \text{if } \nu > 0 \\ f_\Gamma(\lambda; 1-\nu) & \text{if } \nu \in \{0,-1,\dots\} \\ \pm\infty & \text{otherwise} \end{cases}$$
$$\lim_{x\downarrow 0} f_{-\chi^2} = f_\Gamma(\lambda; 1-\nu) \tag{69}$$

(The latter is zero if $\nu \in \{0,-1,\dots\}$.)

$$\lim_{X_t\downarrow 0} f^*_{\chi^2} = \frac{e^{-bt}}{a\tilde{t}} \cdot \begin{cases} 0 & \text{if } \nu > 0 \\ f_\Gamma(\lambda; 1-\nu) & \text{if } \nu \in \{0,-1,\dots\} \\ \pm\infty & \text{otherwise} \end{cases}$$
$$\lim_{X_t\downarrow 0} f^*_{-\chi^2} = \frac{e^{-bt}}{a\tilde{t}} \cdot f_\Gamma(\lambda; 1-\nu) \tag{70}$$

### B.7.3 Flux near 0

By (67), the flux under the Feller diffusion solutions is

$$J = \frac{1}{\tilde{t}}[X_t - X_0 L^{-1}] f^*_\pm = \frac{1}{\tilde{t}}[x - e^{-bt}\lambda L^{-1}] f_\pm. \tag{71}$$

We are interested in the $X_t \downarrow 0$ ($x \downarrow 0$) limit.

As noted in the previous subsection, under $f^*_+$, if $\nu \notin \mathbb{Z}$, the density behaves as $x \to 0$ as $(-1)^{\lceil \nu \rceil} x^\nu$. Thus, in both $x f_+$ and $-L^{-1} f_+$ behave there as $(-1)^{\lceil \nu \rceil} x^{\nu+1}$; and $J$ does as well. That in this limit the density and flux have the same sign implies that when mass is accumulating near $X_t = 0$ with unbounded, positive density—when $\lceil \nu \rceil$ is even—it is arriving there at least in part from below. For a positive flux indicates movement from smaller to larger $X_t$ coordinates. In other words, the singularity at $X_t = 0$ is a source. The odd-$\lceil \nu \rceil$ behavior is perhaps best interpreted in the same way: when negative mass is accumulating near $X_t = 0$, the associated negative values for $J$ indicate movement of such negative mass up and out of the singularity.

A singularity emitting mass, positive or negative, may be unrealistic in many modeling contexts. So we typically consider $\nu < -1$ inadmissible for the $f^*_{\chi^2}$ diffusion; then $x = 0$ is a zero rather than pole of the density.

Turning to the $f^*_{-\chi^2}$ diffusion, now $f_-$ and $L^{-1} f_-$ in (71) are, by (69), bounded. As a result, we can compute the $X_t \downarrow 0$ limit by plugging $X_t = 0$ into (71)—or into another expression for the flux. Doing so in (52) yields

$$\lim_{X_t\downarrow 0} J = a\nu f^*_{-\chi^2}(0; Y_0, t, a, b, \nu) = a\nu \frac{e^{-bt}}{a\tilde{t}} f_\Gamma(\lambda; 1-\nu) = -\frac{e^{-bt}}{\tilde{t}} \lambda f_\Gamma(\lambda; -\nu),$$

the second equality using (70). The implications are akin to those for $f^*_{\chi^2}$ when $\nu < 0$. Under $f^*_{-\chi^2}$, if $\nu > 0$, then the positive density near $X_t = 0$ is arriving there, on net, from below: the singularity is again a source. If $\nu < 0$ then the flux of $f^*_{-\chi^2}$ near $X_t = 0$ is negative, making the singularity a sink: mass reaching the boundary exits the system, if by system we mean the mass diffusing in the range $(0, \infty)$.

Summary:

$$\lim_{X_t \downarrow 0} J = \begin{cases} \pm\infty & \text{under } f^*_{\chi^2} \text{ if } \nu < -1, \nu \notin \mathbb{Z} \\ 0 & \text{under } f^*_{\chi^2} \text{ if } \nu > -1; \text{ or } f^*_{-\chi^2} \text{ with } \nu \in \{0,1,\dots\} \\ -\frac{e^{-bt}}{\tilde{t}} \lambda f_\Gamma(\lambda; -\nu) & \text{under } f^*_{-\chi^2}; \text{ or } f^*_{\chi^2} \text{ with } \nu \in \{-1,-2,\dots\} \end{cases}$$

### B.7.4 Norm

The discovery that some diffusions satisfying Feller's equation gain or lose mass at the boundary motivates interest in the total mass away from the boundary. More precisely, we investigate

$$\|f_\pm\| \coloneqq \lim_{\underline{x} \downarrow 0} \int_{\underline{x}}^{\infty} f_\pm dx, \tag{72}$$

which we will call the norm of $f_\pm$. The definition extends from distributions to diffusions in the obvious way, becoming time-dependent. When (positive) mass exits the system, the norm should decrease.

Again focusing first on $f_{\chi^2}$, we have, by substituting (60) into (72),

$$\|f_{\chi^2}\| = \lim_{\underline{x} \downarrow 0} \int_{\underline{x}}^{\infty} f_{\chi^2}(x; \lambda, \nu) dx = \sum_{m=0}^{\infty} f_\Gamma(\lambda; m+1) \lim_{\underline{x} \downarrow 0} \int_{\underline{x}}^{\infty} f_\Gamma(x; m+\nu+1) dx. \tag{73}$$

If $\nu > -1$, then for all $m$, $m+\nu+1 > 0$ and $f_\Gamma(x; m+\nu+1)$ is a proper gamma distribution, with total integral 1. Then the norm of $f_{\chi^2}$ is an exhaustive sum of Poisson probabilities:

$$= \sum_{m=0}^{\infty} f_\Gamma(\lambda; m+1) 1 = \sum_{m=0}^{\infty} f_P(m; \lambda) = 1.$$

Thus for $\nu > -1$, $f_{\chi^2}$ as defined in (60) is a proper distribution over $(0, \infty)$.

But if $\nu = -1$, the norm of $f_{\chi^2}$ falls below 1. In particular, segregating the $m = 0$ term in (55), we now get

$$\|f_{\chi^2}\| = f_\Gamma(\lambda; 1) \lim_{\underline{x} \downarrow 0} \int_{\underline{x}}^{\infty} f_\Gamma(x; 0) dx + \sum_{m=1}^{\infty} f_\Gamma(\lambda; m+1).$$

Since the integrand is identically 0, the whole first term is too. The second term is a non-exhaustive sum of Poisson probabilities, $\sum_{m=1}^{\infty} f_P(m; \lambda) = 1 - e^{-\lambda}$. Thus now $\|f_{\chi^2}\| = 1 - e^{-\lambda} = F_\Gamma(\lambda; 1)$.

More generally, if $\nu$ is a negative integer,

$$\|f_{\chi^2}\| = \sum_{m=0}^{-\nu-1} f_\Gamma(\lambda; m+1) \lim_{\underline{x} \downarrow 0} \int_{\underline{x}}^{\infty} f_\Gamma(x; m+\nu+1) dx + \sum_{m=-\nu}^{\infty} f_\Gamma(\lambda; m+1)$$

The first sum is zero and the second is

$$= \sum_{m=0}^{\infty} f_\Gamma(\lambda; m-\nu+1) = F_\Gamma(\lambda; -\nu)$$

using (59).

If $\nu$ is not an integer yet less than $-1$, then the integrals in the early terms of (73) are unbounded, so that in general the norm is too. In particular, the $m = 0$ term dominates, and diverges with $\int x^\nu dx = x^{\nu+1}$. This behavior is consistent with the earlier finding that, under the same conditions, the singularity is an unbounded source of positive or negative mass.

$f_{-\chi^2}$ behaves differently in this regard. From the definition (61), we calculate

$$\|f_{-\chi^2}\| = \sum_{m=0}^{\infty} f_\Gamma(\lambda; m-\nu+1) \lim_{\underline{x}\downarrow 0} \int_{\underline{x}}^{\infty} f_\Gamma(x; m+1)dx = \sum_{m=0}^{\infty} f_\Gamma(\lambda; m-\nu+1). \tag{74}$$

When $\nu < 0$, (59) equates this to $F_\Gamma(\lambda; -\nu)$. When $\nu \geq 0$ is an integer, $f_{-\chi^2} = f_{\chi^2}$, which we have already reviewed. Finally, when $\nu > 0$ is not an integer, (74) decomposes as

$$= \sum_{m=0}^{\lfloor \nu \rfloor} f_\Gamma(\lambda; m-\nu+1) + \sum_{m=\lceil \nu \rceil}^{\infty} f_\Gamma(\lambda; m-\nu+1) = \sum_{m=0}^{\lfloor \nu \rfloor} f_\Gamma(\lambda; m-\nu+1) + F_\Gamma(\lambda; \lceil \nu \rceil - \nu)$$

This is finite given $\lambda$ and $\nu$. But by choosing $X_0$, thus $\lambda$, close to zero, the dominant $m = 0$ term can be made arbitrarily large, growing with $\lambda^{-\nu}$. In general, the norm is unbounded.

Summary:

$$\left\|f_{\chi^2}^*\right\| = \begin{cases} 1 & \text{if } \nu > -1 \\ F_\Gamma(\lambda; -\nu) & \text{if } \nu \in \{-1, -2, \ldots\} \\ \pm\infty & \text{otherwise} \end{cases}$$

$$\left\|f_{-\chi^2}^*\right\| = \begin{cases} F_\Gamma(\lambda; -\nu) & \text{if } \nu < 0 \\ 1 & \text{if } \nu \in \{0, 1, \ldots\} \\ F_\Gamma(\lambda; \lceil \nu \rceil - \nu) + \sum_{m=0}^{\lfloor \nu \rfloor} f_\Gamma(\lambda; m-\nu+1) & \text{otherwise} \end{cases} \tag{75}$$

According to (63), $\lambda$ decreases monotonically with $t$; it converges to $X_0/a$ when $b > 0$ and to 0 otherwise. As result, in those cases where the norm is finite but less than 1, equaling $F_\Gamma(\lambda; -\nu)$, the diffusion is also norm decreasing: $\left\|f_{\pm\chi^2}^*\right\|$ declines toward $F_\Gamma(X_0/a\,; -\nu)$ if $b < 0$ and toward 0 otherwise. On the other hand, the final case of (75) passes negative arguments to $f_\Gamma(\cdot)$. This can produce large values of sign $(-1)^{\lceil \nu \rceil}$; and the magnitudes rise—unboundedly so when $b \leq 0$ and thus $\lambda \to 0$.

## B.8. Completing the statement of the solutions

This review of the characteristics of $f_{\chi^2}$ and $f_{-\chi^2}$, $f_{\chi^2}^*$ and $f_{-\chi^2}^*$, summarized in the table below, equips us to address some unfinished business. We have found that $f_{\chi^2}^*$ possesses traits of a plausible physical or economic model—positivity, a cross-sectional norm of 1—when $\nu > -1$ or $\nu$ is an integer. Notably, the density defined for $f_{\chi^2}$ does integrate to 1 over $(0, \infty)$. In partial contrast, for $\nu \leq 0$, $f_{-\chi^2}^*$ is positive and possesses a cross-sectional norm in the range $[0,1]$—but not necessarily equal to 1. Definition (61) therefore does not constitute a proper distribution.

How to fix this deficiency is now clear. Under the most meaningful parameter range, $\nu \leq 0$, the cross-sectional norm of $f_{-\chi^2}^*$ starts at 1 and then declines as $F_\Gamma(\lambda; -\nu)$. Meanwhile, near $X_t = 0$, the flux is negative: the singularity is a sink. So we know where the mass goes, and how much has gone at any given time.

We therefore define $f_{-\chi^2}$ more rigorously with a cumulative distribution function over $[0, \infty)$. It accumulates the mass at $X_t = 0$, where sample paths enter a cemetery or coffin state. The full Feller distribution thus has a discrete and a diffuse component:

$$F_{-\chi^2}(x;\lambda,\nu) = \begin{cases} 1 - F_\Gamma(\lambda;-\nu) & \text{if } x = 0 \\ 1 - F_\Gamma(\lambda;-\nu) + \sum_{m=0}^{\infty} f_\Gamma(\lambda; m-\nu+1) \int_0^x f_\Gamma(x'; m+1)dx' & \text{if } x > 0 \end{cases} \tag{76}$$

When $\nu$ is an integer, we transfer this definition to the noncentral $\chi^2$ cdf, $F_{\chi^2}$. In general, the augmentations carry over to $f_{\chi^2}^*$ and $f_{-\chi^2}^*$.

A special case is $\nu = -1$, when $f_{\chi^2}$ and $f_{-\chi^2}$ both give the noncentral $\chi^2$ distribution with $k = 2(\nu + 1) = 0$ degrees of freedom. Under this distribution, the mass at the zero boundary is $1 - F_\Gamma(\lambda; 1) = e^{-\lambda}$ (Siegel 1979; Hjort 1988).

By the definition of $\lambda$ in (63), as $t \to \infty$, $\lambda \downarrow b$ when $b > 0$; otherwise, $\lambda \downarrow 0$. As a result, if $b \leq 0$, essentially all of the mass eventually enters the singularity. But if $b > 0$, the levitating component of exponential growth, deriving from the $bX_t dt$ term in (50), rescues $F_\Gamma(b; -\nu)$ of the mass from this fate.

**Characteristics of noncentral $\chi^2$ and Feller diffusions, $f_\pm^*$**

| | $\nu < -1$ | | $\nu = -1$ | $-1 < \nu < 0$ | | $\nu = 0$ | $0 < \nu < 1$ | | $1 \leq \nu$ | |
|---|---|---|---|---|---|---|---|---|---|---|
| | $f_{\chi^2}^*, \nu \notin \mathbb{Z}$ | $f_{-\chi^2}^*$ | $f_{\chi^2}^* = f_{-\chi^2}^*$ | $f_{\chi^2}^*$ | $f_{-\chi^2}^*$ | $f_{\chi^2}^* = f_{-\chi^2}^*$ | $f_{\chi^2}^*$ | $f_{-\chi^2}^*$ | $f_{\chi^2}^*$ | $f_{-\chi^2}^*, \nu \notin \mathbb{Z}$ |
| Positivity preserving? | N | Y | Y | Y | Y | Y | Y | Y | Y | N |
| Density near 0 | $\pm\infty$ | $(0,\infty)$ | $(0,\infty)$ | $\infty$ | $(0,\infty)$ | $(0,\infty)$ | 0 | $(0,\infty)$ | 0 | $(-\infty,\infty)$ |
| Flux near 0 | $\pm\infty$ | $(-\infty,0)$ | $(-\infty,0)$ | 0 | $(-\infty,0)$ | 0 | 0 | $(0,\infty)$ | 0 | $(-\infty,\infty)$ |
| Norm | $\pm\infty$ | $< 1$, decreasing | $< 1$, decreasing | 1 | $< 1$, decreasing | 1 | 1 | $> 1$, increasing, unbounded if $b \leq 0$ | 1 | $\pm$, increasing magnitude, unbounded if $b \leq 0$ |

## B.9. Further characteristics of the solutions

### B.9.1 Time to zero

By (76), under the Feller diffusion, the fraction of paths that go 0 by time $t$ is

$$q = 1 - F_\Gamma(\lambda; -\nu) = 1 - F_\Gamma\left(\frac{X_0}{a\tilde{t}}; 1 - \tilde{c}\right)$$

This is also the fraction of paths $X_t^\gamma$, with $\gamma < 0$, that have exploded. Solving for $\tilde{t}$,

$$\tilde{t} = \frac{X_0}{aF_\Gamma^{-1}(1 - q; 1 - \tilde{c})}$$

Inverting the definition of $\tilde{t}$ in (41),

$$t = \begin{cases} -\frac{1}{b}\ln(1 - b\tilde{t}) & \text{if } b \neq 0 \\ \tilde{t} & \text{if } b = 0 \end{cases}.$$

Combining the above gives the time when a fraction $q$ of the Feller-diffusion paths have gone to 0 or negative powers thereof have exploded:

$$t = \begin{cases} -\frac{1}{b}\ln\left(1 - \frac{bX_0}{aF_\Gamma^{-1}(1 - q; -\nu)}\right) & \text{if } b \neq 0 \\ \frac{X_0}{aF_\Gamma^{-1}(1 - q; -\nu)} & \text{if } b = 0 \end{cases}$$

For example, if $q = 0.5$, this is the median hitting time of $X_t$ at zero and the median explosion time of $X_t^\gamma$.

## B.9.2 Moments

### Moments of $x$

The $r^{th}$ raw moment of the standard gamma density is

$$\int_0^\infty x^r f_\Gamma(x;\alpha)dx = \int_0^\infty x^r \frac{x^{\alpha-1}}{\Gamma(\alpha)} e^{-x}dx = \frac{\Gamma(r+\alpha)}{\Gamma(\alpha)} \int_0^\infty \frac{x^{r+\alpha-1}}{\Gamma(r+\alpha)} e^{-x}dx$$

If $r + \alpha > 0$ the integral is 1, and the raw moment is

$$\frac{\Gamma(r+\alpha)}{\Gamma(\alpha)}.$$

Otherwise the integral and the moment are infinite.

It follows that the $r^{th}$ raw moment of the noncentral $\chi^2$ density is

$$\begin{aligned}\int_0^\infty x^r f_{\chi^2}(x;\lambda,\nu)dx &= \int_0^\infty \sum_{m=0}^\infty f_\Gamma(\lambda;m+1)x^r f_\Gamma(x;m+\nu+1)\,dx \\ &= \sum_{m=0}^\infty f_\Gamma(\lambda;m+1)\int_0^\infty x^r f_\Gamma(x;m+\nu+1)dx \\ &= \sum_{m=0}^\infty f_\Gamma(\lambda;m+1)\frac{\Gamma(r+m+\nu+1)}{\Gamma(m+\nu+1)}.\end{aligned}$$

This holds if $\nu + r + 1 > 0$. Otherwise the moment does not exist.

In particular, if $\nu > -2$, the mean of the density $f_{\chi^2}(x;\lambda,\nu)$ is

$$\begin{aligned}\mathrm{E}_x[f_{\chi^2}] = \sum_{m=0}^\infty f_\Gamma(\lambda;m+1)\frac{\Gamma(m+\nu+2)}{\Gamma(m+\nu+1)} &= \sum_{m=0}^\infty f_\Gamma(\lambda;m+1)(m+\nu+1) \\ &= \sum_{m=0}^\infty f_\Gamma(\lambda;m+1)m + (1+\nu)\sum_{m=0}^\infty f_\Gamma(\lambda;m+1) \\ &= \sum_{m=0}^\infty f_P(m;\lambda)m + (1+\nu)\sum_{m=0}^\infty f_\Gamma(m;\lambda)\end{aligned}$$

The first sum in the last expression is the 1st moment of the Poisson distribution, which is $\lambda$, while the second is the 0th moment, which is 1. Thus the mean of the noncentral $\chi^2$ density is $\lambda + \nu + 1$.

Similarly, if $\nu > -3$, the 2nd raw moment of $f_{\chi^2}$ is

$$\begin{aligned}\mathrm{E}_{x^2}[f_{\chi^2}] &= \sum_{m=0}^\infty f_\Gamma(\lambda;m+1)\frac{\Gamma(m+\nu+3)}{\Gamma(m+\nu+1)} \\ &= \sum_{m=0}^\infty f_P(m;\lambda)(m+\nu+1)(m+\kappa+2) \\ &= \sum_{m=0}^\infty f_P(m;\lambda)m^2 + (\nu+2+\nu+1\,)\sum_{m=0}^\infty f_P(m;\lambda)m + (\nu+2\,)(\nu+1)\sum_{m=0}^\infty f_P(m;\lambda)\end{aligned}$$

The second raw moment of the Poisson distribution is $\lambda^2 + \lambda$ (so that the variance is $\lambda$). The above is then

$$= \lambda^2 + \lambda + (2\nu + 3)\lambda + (\nu + 2)(\nu + 1)$$
$$= (\lambda + \nu + 1)^2 + 2\lambda + \nu + 1$$

Subtracting the square of the mean from this raw 2nd moment gives the variance of the density:

$$\mathrm{Var}[f_{\chi^2}] = (\lambda + \nu + 1)^2 + 2\lambda + \nu + 1 - (\lambda + \nu + 1)^2 = 2\lambda + \nu + 1.$$

Again, this is the variance of the full distribution if $\nu > -1$.

As for $f_{-\chi^2}$ (which includes the case of $f_{\chi^2}$ when $\nu = -1$), the $r^{th}$ raw moment is

$$\mathrm{E}_{x^r}[f_{-\chi^2}] = \sum_{m=0}^{\infty} f_\Gamma(\lambda; m - \nu + 1) \frac{\Gamma(r + m + 1)}{\Gamma(m + 1)}$$

For $r = 1$, we develop this as

$$= \sum_{m=0}^{\infty} f_\Gamma(\lambda; m - \nu + 1)(m - \nu) + \sum_{m=0}^{\infty} f_\Gamma(\lambda; m - \nu + 1)(\nu + 1) \tag{77}$$

We apply (59) to the second sum. And via the identity,

$$\alpha f_\Gamma(z; \alpha + 1) = z f_\Gamma(z; \alpha), \tag{78}$$

we rewrite the first sum in (77) as

$$\begin{aligned} \sum_{m=0}^{\infty} f_\Gamma(\lambda; m - \nu + 1)(m - \nu) &= \sum_{m=0}^{\infty} \lambda f_\Gamma(\lambda; m - \nu) \\ &= \lambda f_\Gamma(\lambda; -\nu) + \lambda \sum_{m=1}^{\infty} f_\Gamma(\lambda; m - \nu) \\ &= \lambda f_\Gamma(\lambda; -\nu) + \lambda \sum_{m=0}^{\infty} f_\Gamma(\lambda; m - \nu + 1) \\ &= \lambda f_\Gamma(\lambda; -\nu) + \lambda F_\Gamma(\lambda; -\nu). \end{aligned} \tag{79}$$

The development of $\mathrm{E}_x[f_{-\chi^2}]$ continues with

$$\begin{aligned} &= (\nu + 1) F_\Gamma(\lambda; -\nu) + \lambda f_\Gamma(\lambda; -\nu) + \lambda F_\Gamma(\lambda; -\nu) \\ &= \lambda f_\Gamma(\lambda; -\nu) + (\lambda + \nu + 1) F_\Gamma(\lambda; -\nu). \end{aligned} \tag{80}$$

This computation neglects the probability mass at 0. But the result is correct because the mass at 0 contributes 0 to the raw moment.

In the same vein, the 2nd raw moment of $f_{-\chi^2}$ is

$$\mathrm{E}_{x^2}[f_{-\chi^2}] = \sum_{m=0}^{\infty} f_\Gamma(\lambda; m - \nu + 1) \frac{\Gamma(m + 3)}{\Gamma(m + 1)} = \sum_{m=0}^{\infty} f_\Gamma(\lambda; m - \nu + 1)(m + 2)(m + 1) \tag{81}$$

A step in (77) was to recast $(m + 1)$ as $(m - \nu) + (\nu + 1)$; the second-degree equivalent is

$$(m + 1)(m + 2) = (m - \nu)(m - \nu - 1) + 2(\nu + 2)(m - \nu) + (\nu + 1)(\nu + 2). \tag{82}$$

Now,

$$\sum_{m=0}^{\infty} f_\Gamma(\lambda; m-\nu+1)(m-\nu)(m-\nu-1) = \lambda \sum_{m=0}^{\infty} f_\Gamma(\lambda; m-\nu)(m-\nu-1)$$

$$= \lambda^2 \sum_{m=0}^{\infty} f_\Gamma(\lambda; m-\nu-1)$$

$$= \lambda^2 f_\Gamma(\lambda; -\nu-1) + \lambda^2 f_\Gamma(\lambda; -\nu) + \lambda^2 \sum_{m=0}^{\infty} f_\Gamma(\lambda; m-\nu+1)$$

$$= \lambda(-\nu-1) f_\Gamma(\lambda; -\nu) + \lambda^2 f_\Gamma(\lambda; -\nu) + \lambda^2 F_\Gamma(\lambda; -\nu)$$

$$= \lambda(\lambda-\nu-1) f_\Gamma(\lambda; -\nu) + \lambda^2 F_\Gamma(\lambda; -\nu).$$

Substituting with this formula as well as (79) and (82) into (81), the 2nd raw moment is

$$= \sum_{m=0}^{\infty} f_\Gamma(\lambda; m-\nu+1)(m-\nu)(m-\nu-1) + 2(\nu+2) \sum_{m=0}^{\infty} f_\Gamma(\lambda; m-\nu+1)(m-\nu)$$

$$+ (\nu+1)(\nu+2) \sum_{m=0}^{\infty} f_\Gamma(\lambda; m-\nu+1)$$

$$= \lambda(\lambda-\nu-1) f_\Gamma(\lambda; -\nu) + \lambda^2 F_\Gamma(\lambda; -\nu) + 2(\nu+2)\big(\lambda f_\Gamma(\lambda; -\nu) + \lambda F_\Gamma(\lambda; -\nu)\big) + (\nu+1)(\nu+2) F_\Gamma(\lambda; -\nu)$$

$$= \lambda(\lambda+\nu+3) f_\Gamma(\lambda; -\nu) + \big(\lambda + (\lambda+\nu+1)(\lambda+\nu+2)\big) F_\Gamma(\lambda; -\nu)$$

Subtracting the square of the mean, in (80), then gives the full expression for the variance of the Feller distribution:

$$\mathrm{Var}\big[f_{-\chi^2}\big] = \lambda(\lambda+\nu+3) f_\Gamma(\lambda; -\nu) + \big(\lambda + (\lambda+\nu+1)(\lambda+\nu+2)\big) F_\Gamma(\lambda; -\nu) - [\lambda f_\Gamma(\lambda; -\nu) + (\lambda+\nu+1) F_\Gamma(\lambda; -\nu)]^2$$

To obtain cross-sectional means and variances of the *diffusions*, we divide corresponding formulas for the distributions by $dx/dX_t = e^{-bt}/a\tilde{t}$ and its square, respectively, in order to pull back through $X_t \mapsto x$ in (41) for fixed $t$.

Summary, after additional rearrangements:

$$\mathrm{E}_x\big[f_{\chi^2}\big] = \lambda + \nu + 1$$

$$\mathrm{Var}_x\big[f_{\chi^2}\big] = 2\lambda + \nu + 1$$

$$\mathrm{E}_x\big[f_{-\chi^2}\big] = \lambda f_\Gamma(\lambda; -\nu) + (\lambda+\nu+1) F_\Gamma(\lambda; -\nu) = \lambda F_\Gamma(\lambda; -\nu-1) + (\nu+1) F_\Gamma(\lambda; -\nu)$$

$$\mathrm{Var}_x\big[f_{-\chi^2}\big] = \lambda(\lambda+\nu+3) f_\Gamma(\lambda; -\nu) + \big(\lambda + (\lambda+\nu+1)(\lambda+\nu+2)\big) F_\Gamma(\lambda; -\nu) - [\lambda f_\Gamma(\lambda; -\nu) + (\lambda+\nu+1) F_\Gamma(\lambda; -\nu)]^2$$

$$\mathrm{E}_{X_t}\left[f^*_{\pm\chi^2}\right] = \frac{a\tilde{t}}{e^{-bt}} \mathrm{E}_x\big[f_{\pm\chi^2}\big]$$

$$\mathrm{Var}_{X_t}\left[f^*_{\pm\chi^2}\right] = \left(\frac{a\tilde{t}}{e^{-bt}}\right)^2 \mathrm{Var}\big[f_{\pm\chi^2}\big]$$

Moreover, since the excess kurtosis of $f_\Gamma(\cdot; \alpha)$ is $6/\alpha$, when $f_{\pm\chi^2}$ are valid distributions, they inherit the leptokurticity via the definitions (60) and (61).

## Moments of $\ln x$

If $x$ is gamma-distributed, the $r^{th}$ raw moment of its natural logarithm is

$$\mathrm{E}_{(\ln x)^r}[f_\Gamma] = \int_0^\infty (\ln x)^r f_\Gamma(x; \alpha) dx$$

$$= \int_0^\infty (\ln x)^r \frac{x^{\alpha-1}}{\Gamma(\alpha)} e^{-x} dx$$

$$= \frac{1}{\Gamma(\alpha)} \int_0^\infty \frac{\partial^r}{\partial t^r} x^{\alpha-1+t} \bigg|_{t=0} e^{-x} dx$$
$$= \frac{1}{\Gamma(\alpha)} \frac{\partial^r}{\partial t^r} \int_0^\infty x^{\alpha-1+t} e^{-x} dx \bigg|_{t=0}$$
$$= \frac{1}{\Gamma(\alpha)} \frac{\partial^r}{\partial t^r} \Gamma(\alpha+t) \bigg|_{t=0}$$
$$= \frac{1}{\Gamma(\alpha)} \frac{\partial^r}{\partial \alpha^r} \Gamma(\alpha)$$

To compute the derivates $\frac{\partial^r}{\partial \alpha^r}\Gamma(\alpha)$, we use Pascal's Triangle–type identities for the derivatives of a function in terms of the derivatives of its log—which here are the digamma function $\psi$ and the higher polygamma functions $\psi_1, \psi_2, \ldots$.

$$\begin{aligned}
\Gamma' &= \Gamma\psi \\
\Gamma'' &= \Gamma'\psi + \Gamma\psi_1 \\
&= \Gamma \cdot (\psi^2 + \psi_1) \\
\Gamma''' &= \Gamma''\psi + 2\Gamma'\psi_1 + \Gamma\psi_2 \\
&= (\Gamma\psi^2 + \Gamma\psi_1)\psi + 2\Gamma\psi\psi_1 + \Gamma\psi_2 \\
&= \Gamma \cdot (\psi^3 + 3\psi\psi_1 + \psi_2) \\
\Gamma'''' &= \Gamma'''\psi + 3\Gamma''\psi_1 + 3\Gamma'\psi_2 + \Gamma\psi_3 \\
&= \Gamma \cdot \big((\psi^3 + 3\psi\psi_1 + \psi_2)\psi + 3(\psi^2 + \psi_1)\psi_1 + 3\psi\psi_2 + \psi_3\big) \\
&= \Gamma \cdot (\psi^4 + 3\psi^2\psi_1 + \psi\psi_2 + 3(\psi^2\psi_1 + \psi_1^2) + 3\psi\psi_2 + \psi_3) \\
&= \Gamma \cdot (\psi^4 + 6\psi^2\psi_1 + 4\psi\psi_2 + 3\psi_1^2 + \psi_3)
\end{aligned}$$

We get

$$\mathrm{E}_{\ln x}[f_\Gamma] = \frac{1}{\Gamma(\alpha)} \frac{\partial}{\partial \alpha} \Gamma(\alpha) = \psi(\alpha)$$
$$\mathrm{E}_{(\ln x)^2}[f_\Gamma] = \frac{1}{\Gamma(\alpha)} \frac{\partial^2}{\partial \alpha^2} \Gamma(\alpha) = \psi(\alpha)^2 + \psi_1(\alpha)$$
$$\mathrm{Var}_{\ln x}[f_\Gamma] = \mathrm{E}_{(\ln x)^2}[f_\Gamma] - (\mathrm{E}_{\ln x}[f_\Gamma])^2 = \psi_1(\alpha)$$

The fourth central moment is

$$\begin{aligned}
\mathrm{E}_{(\ln x - \mathrm{E}_{\ln x}[f_\Gamma])^4}[f_\Gamma] &= \mathrm{E}_{(\ln x)^4}[f_\Gamma] - 4\,\mathrm{E}_{(\ln x)^3}[f_\Gamma]\,\mathrm{E}_{\ln x}[f_\Gamma] + 6\,\mathrm{E}_{(\ln x)^2}[f_\Gamma]\,\mathrm{E}_{\ln x}[f_\Gamma]^2 - 4\,\mathrm{E}_{\ln x}[f_\Gamma]\,\mathrm{E}_{\ln x}[f_\Gamma]^3 + \mathrm{E}_{\ln x}[f_\Gamma]^4 \\
&= \psi^4 + 6\psi^2\psi_1 + 4\psi\psi_2 + 3\psi_1^2 + \psi_3 - 4\psi(\psi^3 + 3\psi\psi_1 + \psi_2) + 6\psi^2(\psi^2 + \psi_1) - 3\psi^4 \\
&= 3\psi_1^2 + \psi_3
\end{aligned}$$

So the excess kurtosis of $\ln x$ under the standard gamma distribution is

$$\frac{\mathrm{E}_{(\ln x - \mathrm{E}_{\ln x}[f_\Gamma])^4}[f_\Gamma]}{(\mathrm{Var}_{\ln x}[f_\Gamma])^2} - 3 = \frac{3\psi_1^2(\alpha) + \psi_3(\alpha)}{\psi_1^2(\alpha)} - 3 = \frac{\psi_3(\alpha)}{\psi_1^2(\alpha)} > 0$$

If $\nu \geq -1$, so that $f_{\pm\chi^2}(x; \lambda, \nu)$ is a valid distribution, then the leptokurticity of $\ln x$ is also bequeathed to $f_{\chi^2}$. Under $f_{-\chi^2}$, $\ln x$ has infinite kurtosis on the low side because of the non-zero mass at $x = 0$. These properties transfer in the now-familiar way to $X_t$ and $X_t^\gamma$.